\documentclass[a4paper,11pt,oneside]{article}
\usepackage{a4wide}
\usepackage{color}
\usepackage[normalem]{ulem}

\usepackage{natbib}
\usepackage{bm}
\usepackage{graphicx}
\usepackage{url}

\def\jgr{\rm{J.~Geophys.~Res.}}    
\newcommand{\apj}{Astrophys.~J.}
\newcommand{\apjl}{Astrophys.~J.}
\newcommand{\mnras}{Mon.~Not.~R.~Astron.~Soc.}
\newcommand{\aap}{Astron.~Astrophys.}
\newcommand{\aj}{Astron.~J.}
\newcommand{\nat}{Nature}
\newcommand{\pre}{Phys.~Rev.~E}

\newcommand{\vc}[1]{\bm{#1}}
\newcommand{\vcs}[1]{\mbox{\boldmath{$\scriptstyle{#1}$}}}
\newcommand{\dpa}[0]{\partial}
\newcommand{\lepar}[1]{%
    \left( \raisebox{#1}{\hspace{-2pt}} \right.}
\newcommand{\ripar}[1]{%
    \left. \raisebox{#1}{\hspace{-2pt}} \right)}

\newcommand{\ii}{\mathrm i}
\newcommand{\de}{\mathrm d}

\DeclareMathSymbol{\varOmega}{\mathord}{letters}{"0A}
\DeclareMathSymbol{\varPhi}{\mathord}{letters}{"08}
\DeclareMathSymbol{\varSigma}{\mathord}{letters}{"06}
\DeclareMathSymbol{\varPsi}{\mathord}{letters}{"09}

\makeatletter
\def\thebibliography#1{\subsection*{Supplementary Notes\@mkboth
    {SUPPLEMENTARY NOTES}{SUPPLEMENTARY NOTES}}\list
  {[\arabic{enumi}]}{\settowidth\labelwidth{[#1]}\leftmargin\labelwidth
    \advance\leftmargin\labelsep
    \usecounter{enumi}}
  \def\newblock{\hskip .11em plus .33em minus .07em}
  \sloppy\clubpenalty4000\widowpenalty4000
  \sfcode`\.=1000\relax}
\makeatother

\begin{document}

\nocite{Safronov1969}
\nocite{Dominik+etal2007}
\nocite{Benz2000}
\nocite{Weidenschilling1977}
\nocite{GoldreichWard1973}
\nocite{YoudinShu2002}
\nocite{WeidenschillingCuzzi1993}
\nocite{Hartmann1998}
\nocite{JohansenKlahrHenning2006}
\nocite{YoudinGoodman2005}
\nocite{JohansenHenningKlahr2006}
\nocite{JohansenYoudin2007}
\nocite{Cuzzi+etal1993}
\nocite{BalbusHawley1998}
\nocite{BargeSommeria1995}
\nocite{FromangNelson2005}
\nocite{Rice+etal2006}
\nocite{Cuzzi+etal2001}
\nocite{HockneyEastwood1981}
\nocite{Gammie2001}
\nocite{Gammie1996}
\nocite{Tanga+etal2004}
\nocite{Salo1992}
\nocite{Boss1997}
\nocite{Mayer+etal2002}
\nocite{Weidenschilling1995}
\nocite{DullemondDominik2005}
\nocite{Weidenschilling1997}
\nocite{ThroopBally2005}
\nocite{HaghighipourBoss2003}

\chapter{{\bf {\huge Supplementary Information for\\
``Rapid planetesimal formation in turbulent circumstellar discs''}}}

\section*{Abstract}

This document contains refereed supplementary information for the paper ``Rapid
planetesimal formation in turbulent circumstellar discs''. It contains 15
sections (\S1.1 -- \S1.15) that address a number of subjects related to the
main paper. Some of the subjects are highlighted here in the abstract. We
describe in detail the Poisson solver used to find the self-potential of the
solid particles, including a linear and a non-linear test problem (\S1.3).
Dissipative collisions remove energy from the motion of the particles by
collisional cooling (\S1.4), an effect that allows gravitational collapse to
occur in somewhat less massive discs (\S1.7). A resolution study of the
gravitational collapse of the boulders is presented in \S1.6. We find that
gravitational collapse can occur in progressively less massive discs as the
grid resolution is increased, likely due to the decreased smoothing of the
particle-mesh self-gravity solver with increasing resolution. In \S1.10 we show
that it is in good agreement with the Goldreich \& Ward (1973) stability
analysis to form several-hundred-km-sized bodies, when the analysis is applied
to 5 AU and to regions of increased boulder column density. \S11 is devoted to
the measurement of random speeds and collision speeds between boulders. We find
good agreement between our measurements and analytical theory for the random
speeds, but the measured collision speeds are 3 times lower than expected from
analytical theory.  Higher resolution studies, and an improved analytical
theory of collision speeds that takes into account epicyclic motion, will be
needed to determine whether collision speeds have converged. In \S1.12 we
present models with no magnetic fields. The boulder layer still exhibits strong
clumping, due to the streaming instability, if the global solids-to-gas ratio
is increased by a factor 3. Gravitational collapse occurs as readily as in
magnetised discs.
\\ \\
{\it Authors}:
\\ \\
Anders Johansen$^1$, Jeffrey S. Oishi$^{2,3}$, Mordecai-Mark Mac
Low$^{2,1}$, Hubert Klahr$^1$, Thomas Henning$^1$ \& Andrew Youdin$^4$
\\ \\
{\it Affiliations}:
\begin{enumerate}
  \item Max-Planck-Institut f\"ur Astronomie, K\"onigstuhl 17, D-69117
    Heidelberg, Germany
  \item Department of Astrophysics, American Museum of Natural History, 79th
    Street at Central Park West, New York, NY 10024-5192, USA
  \item also Department of Astronomy, University of Virginia, Charlottesville,
    VA, USA
  \item Canadian Institute for Theoretical Astrophysics, University of
    Toronto, 60 St. George Street, Toronto, Ontario M5S 3H8, Canada
\end{enumerate}

\newpage

\section{Supplementary Discussion}

\subsection{Pencil Code}

The Pencil Code\cite{Brandenburg2003} is a finite difference code that uses
symmetric derivatives of sixth order in space and a third order Runge-Kutta
time-stepping scheme. The difference equations as implemented are formally
dissipation free, having phase errors but no amplitude error. Only a small
amount of numerical dissipation is introduced from time-stepping the advection
term. Therefore, one must explicitly add dissipation to the dynamical equations
to suppress numerically unstable modes near the grid scale and to dissipate the
turbulent energy that is (in our case) released from the Keplerian shear by the
Reynolds and Maxwell stresses.  For this purpose, we use sixth order
hyperdiffusivity operators, where the usual $\nabla^2$ diffusivity operator is
replaced with a $\nabla^6$
operator\cite{HaugenBrandenburg2004,JohansenKlahr2005}. Hyperdiffusivity
dissipates energy at high wave numbers -- at the smallest scales in the
simulation -- but preserves energy at low wave numbers. Hyperviscosity and
magnetic hyperresistivity have been used extensively to study the properties of
forced magnetohydrodynamic turbulence (see Brandenburg \&
Sarson\cite{BrandenburgSarson2002} and references therein). They are designed
to affect large scales as little as possible by dissipation, thus widening the
inertial range beyond what can be achieved with a regular viscosity operator
while still maintaining numerical stability. We have tested that particle
overdensities occur for both the stringent hyperviscosity scheme used by Haugen
\& Brandenburg\cite{HaugenBrandenburg2004} and for the simplified scheme
described in Johansen \& Klahr\cite{JohansenKlahr2005}.
 
Possible side effects of using hyperviscosity and hyperresistivity include an
artificial increase in the bottleneck effect\cite{BiskampMueller2000}, a
physical effect in turbulence where energy piles up around the dissipative
scale, and a higher saturation level for dynamo-generated magnetic fields in
helical flows compared to what is seen when using a regular viscosity
operator\cite{BrandenburgSarson2002}. The bottleneck effect is unlikely to be
relevant to the dynamics of boulders in turbulence, since marginally coupled
particles are mostly affected by turbulent structures at the largest scales of
our simulation where the bottleneck effect is unimportant. The saturation level
of turbulence driven by the magnetorotational instability (MRI)  is indeed
affected by the numerical scheme and by the dissipation scheme, but the Pencil
Code agrees well with other grid codes regarding the statistical properties of
MRI turbulence\cite{BalbusHawley1998,Sano+etal2004}.

\subsection{Drag force}\label{s:drag_force}

\subsubsection{Drag force algorithm}

The computation of drag forces between Lagrangian particles and an Eulerian
grid requires some care to avoid spurious accelerations and to ensure momentum
conservation. Small errors in the gas velocity can be dangerously amplified by
the subtraction of highly correlated particle velocities. Tests of our drag
force algorithm  are described in detail elsewhere\cite{YoudinJohansen2007}. It
involves three steps:
\begin{enumerate}
  \item Interpolate gas velocities at particle positions.
  \item Calculate the drag force on the particles from the gas in nearby cells.
  \item Assign the back-reaction force to the gas from particles in nearby
  cells.
\end{enumerate}
For the first step, interpolation, we begin with gas velocities,
$\vc{u}^{(\vcs{j})}$, defined on a uniform grid where the index $\vc{j}$ labels
the cells centred on positions $\vc{x}^{(\vcs{j})}$. We interpolate to the
particle positions, $\vc{x}^{(i)}$, using a weight function, $W_{\rm I}$, as
\begin{equation} \label{InterpGasVel}
  \overline{\vc{u}(\vc{x}^{(i)})} = \sum_{\vcs{j}} W_{\rm I}(\vc{x}^{(i)} -
  \vc{x}^{(\vcs{j})})\vc{u}^{(\vcs{j})} \, .
\end{equation}
The weight function is normalised as $\sum_{\vcs{j}} W_{\rm
I}(\vc{x}^{(i)}-\vc{x}^{(\vcs{j})}) = 1$, for any $\vc{x}^{(i)}$, and has
non-zero contributions only from the cells in the immediate vicinity of
$\vc{x}^{(\vcs{j})}$.

The second step, calculating the drag acceleration on particle $i$,
\begin{equation}\label{fpart}
  \vc{f}_{\rm p}^{(i)} = -\frac{1}{\tau_{\rm f}} \left[
      \vc{v}^{(i)} - \overline{\vc{u}(\vc{x}^{(i)})} \right] \, ,
\end{equation}
is trivial once the relevant quantities are defined. We assume that the drag
force is proportional to the velocity difference between gas and particles,
i.e.\ that the friction time $\tau_{\rm f}$ is independent of the velocity
difference.

Finally, we calculate the back-reaction drag force, $\vc{f}_{\rm
g}^{(\vcs{j})}$, on the gas in cell $\vc{j}$.   Assigning the particle
velocities to a grid risks violating momentum conservation.  Instead we use
Newton's third law to directly assign the force on the particles back to the
gas,
\begin{eqnarray} \label{eq:fgj}
  \vc{f}_{\rm g}^{(\vcs{j})} = -\frac{m_{\rm p}}{\rho_{\rm g}^{(\vcs{j})}
  V_{\rm cell}} \sum_iW_{\rm A}(\vc{x}^{(i)}-\vc{x}^{(\vcs{j})})\vc{f}_{\rm p}^{(i)}    \label{newtonIII}\, ,
\end{eqnarray}
where $m_{\rm p}$ is the mass of a superparticle, $\rho_{\rm g}^{(\vcs{j})}$ is
the gas density in cell $\vc{j}$, and $V_{\rm cell}$ is the volume of a grid
cell.  The assignment function $W_{\rm A}$ obeys the same conditions as $W_{\rm
I}$, so that only particles in a given cell or in that cell's nearby neighbours
contribute to the sum.  We opt for the second order Triangular Shaped Cloud
(TSC) assignment scheme\cite{HockneyEastwood1981}, which uses an identical
function for assignment and for interpolation, $W_{\rm A} = W_{\rm I}$. The TSC
spreads the influence of particles and grid points to three grid points in each
direction, for a total of 27 points in three-dimensional (3-D) simulations.

The interpolation errors associated with the TSC assignment is found by
considering a periodic function (of arbitrary phase) sampled at the grid
points\cite{HockneyEastwood1981,YoudinJohansen2007}. The result is that the
assigned amplitude of a single Fourier component at scale $k$ relative to the
actual amplitude is to second order $1- (\Delta k)^2/8$, where $\Delta$ is the
linear size of a grid cell. Thus already at 5 grid cells there is
significant smoothing. This smoothing is found to have an influence on the
gravitational collapse, especially at crude resolution, see
\S\ref{s:convergence}.

The time-step constraint set by the drag force is $\delta t_{\rm
drag}=\tau_{\rm f}/(1+\epsilon)$, where $\epsilon$ is the local solids-to-gas
ratio\cite{JohansenHenningKlahr2006}. As the solids-to-gas ratio increases, the
allowed time-step decreases. We have made sure that the friction time-step
never dominates over the Courant time-step of the code by artificially
increasing the friction time in regions of high solids-to-gas ratio
($\epsilon>100$). We have experimented with the threshold and found no improved
collapse for a higher threshold, presumably because inelastic collisions
dominates the kinetic energy dissipation at these densities anyway.

\subsubsection{Particle sizes and drag regimes}\label{s:drag_regime}

In our work, as in most theoretical work, we characterise particles in terms of
their dimensionless friction time $\varOmega_{\rm K} \tau_{\rm f}$ (where
$\varOmega_{\rm K}$ is the Keplerian orbital frequency). The translation to a
particle size depends on the assumed disc model. The smallest particles are
subject to Epstein drag, valid when the particle radius is smaller than the
mean free path of the gas (see below), with
\begin{equation} 
  \varOmega_{\rm K} \tau_{\rm f}^{\rm (Ep)} = \frac{\varOmega_{\rm K}
  \rho_\bullet a}{\rho_{\rm g} c_{\rm s}} = \sqrt{2\pi} \frac{\rho_\bullet a}{\varSigma_{\rm g}} \label{eq:tauEpstein}\, ,\\
\end{equation} 
where the second step applies in the disc midplane. Here $\rho_\bullet$ is the
material density of the solids, $a$ is the radius of a solid body, $\rho_{\rm
g}$ is the gas density, $c_{\rm s}$ is the sound speed, while $\varSigma_{\rm
g}$ is the column density of gas. Epstein drag depends on gas density, but in
practice the gas density fluctuations are negligible (order 1\%) in our
subsonic flows, so we ignore them. Solving for particle size gives
\begin{equation}
  a =\frac{\varOmega_{\rm K} \tau_{\rm f}^{\rm (Ep)} \varSigma_{\rm
  g}}{\sqrt{2\pi} \rho_\bullet}
  \approx 30\,{\rm cm}\,\varOmega_{\rm K} \tau_{\rm f}^{\rm (Ep)}
  \left( \frac{\varSigma_{\mathrm{g,}5}}{150\,{\rm g\,cm^{-2}}} \right)
  \left( \frac{\rho_\bullet}{2\,{\rm g\,cm^{-3}}}
  \right)^{-1}\left(\frac{r}{5~\rm AU}\right)^{-1.5}\, ,
\end{equation}
where the normalisation of the gas surface density at 5 AU,
$\varSigma_{\mathrm{g,}5}$, and the power law slope follows the minimum mass
solar nebula model\cite{Hayashi1981}. Applying a simulation with given
$\tau_{\rm f}$ values to different disc radii changes the relevant particle
sizes. Considering, as in the main text, $r=5\,{\rm AU}$ and $\varSigma_{\rm
g}=300\,{\rm g\,cm^{-2}}$ yields particle sizes of $a=60,45,30,15$ cm, for
$\varOmega_{\rm K}\tau_{\rm f}=1.0,0.75,0.5,0.25$, respectively. When the model
is applied to the outer solar nebula at $r=40\,{\rm AU}$ instead, the
corresponding particle sizes are as low as a few centimetres (per
$\varOmega_{\rm K}\tau_{\rm f}$).

The Epstein regime of free molecular (or Knudsen) flow ceases to apply once the
particle radius exceeds (9/4 of) the gas mean free path,
\begin{eqnarray}
  \lambda &=&  \frac{\mu}{\rho_{\rm g} \sigma_{\rm mol}} = \frac{\sqrt{2\pi} \mu H}{\varSigma_{\rm g} \sigma_{\rm mol}} \label{eq:mfp}\\
  &\approx & 1\, {\rm m} \left(\frac{\varSigma_{\mathrm{g,}5}}{150\,{\rm g\,cm^{-3}}} \right)^{-1}\frac{H/r}{0.04}\left(\frac{r}{5 \rm AU}\right)^{2.5}\, ,
\end{eqnarray}
where $\mu = 3.9 \times 10^{-24}\,{\rm g}$ is the mean molecular weight and
$\sigma_{\rm mol} = 2 \times 10^{-15} {\rm cm}^2$ is the molecular cross
section of molecular hydrogen\cite{Nakagawa+etal1986,ChapmanCowling1970}.  The
radial dependence does not include flaring of the aspect ratio, $H/r$.  Epstein
drag applies as long as
\begin{equation} 
  \varOmega_{\rm K} \tau_{\rm f}^{\rm (Ep)} <  \frac {9 \pi}{2}
  \frac{\rho_\bullet \mu H}{\varSigma_{\rm g}^2 \sigma_{\rm mol}} \approx 7
  \left(\frac{\rho_\bullet}{2\,{\rm g\,cm^{-3}}}\right)
  \left(\frac{\varSigma_{\mathrm{g,}5}}{150\,{\rm g\,cm^{-2}}}
  \right)^{-2}\left(\frac{H/r}{0.04}\right)\left(\frac{r}{5\,{\rm
  AU}}\right)^{4}\, .
\end{equation} 
Thus Epstein drag is the relevant regime for the application of our model to 5
AU.

Extrapolation of our model to higher column densities (or regions closer to the
star) requires consideration of Stokes drag, for which
\begin{equation} 
  \varOmega_{\rm K} \tau_{\rm f}^{\rm (St)} = \varOmega_{\rm K}\tau_{\rm f}^{\rm (Ep)}  \frac{4 a}{9 \lambda} = \frac{4 \rho_\bullet a^2 \sigma_{\rm mol}}{9 \mu H} \label{eq:tauStokes}\, .
\end{equation} 
Since Stokes drag is also linear in the relative velocity between gas and
solids, our model applies with a different scaling of particle size,
\begin{equation} 
  a = \left[ \frac{9 \varOmega_{\rm K} \tau_{\rm f}^{\rm (St)}\mu H}{4
  \rho_\bullet \sigma_{\rm mol}} \right]^{1/2}
    \approx 80\,{\rm cm}\left[ \varOmega_{\rm K}
  \tau_{\rm f}^{\rm (St)}\left(\frac{\rho_\bullet}{2\,{\rm
  g\,cm^{-3}}}\right)^{-1}\left(\frac{H/r}{0.04}\right)\left(\frac{r}{5 \rm
  AU}\right) \right]^{1/2} \, . \label{eq:aSt}
\end{equation} 
Actually since Stokes drag is independent of gas density (including
fluctuations), our model is more exact in this regime, but modelling particles
in the Stokes regime would require a different treatment of collisional cooling
(see \S\ref{s:collisional_cooling}).


\subsubsection{Energy dissipation by drag force}

The dissipation of kinetic energy caused by drag force can be calculated as
\begin{equation}
  \dot{e}_{\rm kin}=\vc{u} \cdot \left(\rho_{\rm g} \frac{\dpa
  \vc{u}}{\dpa t}\right)_{\rm drag} + \vc{w} 
\cdot \left(\rho_{\rm p} \frac{\dpa \vc{w}}{\dpa t}\right)_{\rm drag},
\end{equation} 
where $\vc{u}$ and $\vc{w}$ are the respective gas and particle velocity
fields. We refer to references \cite{Dobrovolskis+etal1999} and
\cite{Youdin2005a} for considerations of the effect of drag force damping on
the dynamics of particles.  Inserting the fluid expressions for the drag force
acceleration,
\begin{eqnarray}
  \frac{\dpa \vc{u}}{\dpa t} &=& -\frac{\rho_{\rm p}/\rho_{\rm g}}{\tau_{\rm
  f}} (\vc{u} - \vc{w}) \, , \\
  \frac{\dpa \vc{w}}{\dpa t} &=& -\frac{1}{\tau_{\rm
  f}} (\vc{w} - \vc{u}) \, ,
\end{eqnarray}
yields
\begin{equation}
  \dot{e}_{\rm kin}
      = -\frac{\rho_{\rm p}}{\tau_{\rm f}} |\vc{w}-\vc{u}|^2 \, .
  \label{eq:dekindt}
\end{equation}
Dissipation of kinetic energy by drag forces comes automatically when applying
the momentum-conserving drag force. We usually assume that the dissipated
energy can immediately radiate away efficiently, keeping the temperature of the
gas constant. We show, however, in \S\ref{s:heating} that even if all the
released energy remains locally as heat, the resulting temperature increase of
the gas is insignificant and has no influence on the initial stages of the
gravitational collapse.

As long as there is a velocity difference between the solid particles and the
surrounding gas, then drag force cooling is efficient according to equation
(\ref{eq:dekindt}). But the relative speed of gas and particles in drag force
equilibrium decreases with increasing solids-to-gas
ratio\cite{Nakagawa+etal1986}, as the solid particles entrain the gas. In that
case inelastic collisions take over as the dominant cooling mechanism (see
\S\ref{s:collisional_cooling}).

\subsection{Self-gravity solver}

We compute the gravitational potential $\varPhi$ of the particles by
determining a particle density on the mesh and solving the Poisson equation for
this assigned density field. The gravitational acceleration is then
interpolated back to the positions of the particles. The particles are assigned
to the mesh using the TSC scheme\cite{HockneyEastwood1981}, spreading each
particle over the 27 nearest grid points (in 3-D), and the gravitational
acceleration is added back to the particles using a second order spline
interpolation to avoid any risk of self-acceleration of the particles.

The Poisson equation,
\begin{equation}
  \nabla^2 \varPhi = 4 \pi G \rho_{\rm p} \, ,
  \label{eq:poisson}
\end{equation}
where $G$ is the gravitational constant and $\rho_{\rm p}$ is the assigned
particle density field, is solved using a Fourier method. The solution for a
single Fourier density mode of wave number $\vc{k}=(k_x,k_y,k_z)$ and complex
Fourier amplitude $\rho_{\vc{k}}$ is
\begin{equation}
  \varPhi_{\vc{k}} = -\frac{4 \pi G \rho_{\vc{k}}}{k^2}
  \label{eq:Phi}
\end{equation}
for $k=|\vc{k}|>0$.  The full potential is then
$\varPhi(\vc{x})=\sum_{\vc{k}}\varPhi_{\vc{k}}\exp[\ii \vc{k}\cdot\vc{x}]$. 

The radial $x$-direction in the shearing sheet is not strictly periodic, but is
rather shear periodic with the connected points at the inner and outer boundary
separated by the distance $\Delta y(t)={\rm mod}[(3/2)  \varOmega_{\rm K} L_x
t, L_y]$ in the $y$-direction. We follow the algorithm of
Gammie\cite{Gammie2001} to include shear-periodic boundaries in the Fourier
method, with the difference that we perform the necessary interpolation in
Fourier space rather than in real space, as suggested by C. McNally at
\url{http://imp.mcmaster.ca/~colinm/ism/rotfft.html}.  First we apply a
discrete Fast Fourier Transform (FFT) in the periodic $y$-direction. We then
shift the entire $y$-direction by the amount $\delta y(x)=\Delta y(t) x/L_x$ to
make the $x$-direction periodic, before proceeding with discrete FFTs along $x$
and then $z$. Performing the $\delta y(x)$ shift in Fourier space (essentially
using a Fourier interpolation method) has the advantage over polynomial
interpolation that it is continuous and smooth in all its derivatives.  After
solving the Poisson equation in Fourier space, using the algebraic solution
from equation (\ref{eq:Phi}), we transform the potential back to real space in
the opposite order, shifting the $y$-direction back to the sheared frame on the
way.

The gravitational constant $G$ in equation (\ref{eq:poisson}) must be defined
in code units. We adopt the unit system $\varOmega_{\rm K}=c_{\rm
s}=H=\rho_{\rm g,mid}=1$, where the parameters are respectively the Keplerian
frequency, isothermal sound speed, pressure scale height, and midplane gas
density of the disc. The non-dimensional form of the Poisson equation is
\begin{equation}
  (H \nabla)^2 \varPhi/c_{\rm s}^2 =
      \tilde{G} \frac{\rho_{\rm p}}{\rho_{\rm g,mid}}\, ,
\end{equation}
where
\begin{equation}
  \tilde{G}\equiv4\pi G \rho_{\rm g, mid} \varOmega_{\rm K}^{-2}
\end{equation}
is a self-gravity parameter which relates to the Toomre
parameter\cite{Toomre1964} $Q$ of self-gravitating gas discs as
\begin{equation}
  Q\approx1.6 \tilde{G}^{-1} \, .
\end{equation}
In case of vertical hydrostatic equilibrium the connection between the midplane
density $\rho_{\rm g,mid}$ and the column density $\varSigma_{\rm g}$ is
$\rho_{\rm g,mid}=\varSigma_{\rm g}/(\sqrt{2\pi}H)$. Introducing further the
dimensionless scale-height-to-radius parameter $H/r$ yields the power law
connection
\begin{equation}
  \varSigma_{\rm g} = 300 {\rm\,g\,cm^{-2}}
      \lepar{14pt} \frac{\tilde{G}}{0.1} \ripar{14pt}
      \lepar{14pt} \frac{H/r}{0.04} \ripar{14pt}
      \lepar{14pt} \frac{M_\star}{M_\odot} \ripar{14pt}
      \lepar{14pt} \frac{r}{5\, {\rm AU}} \ripar{14pt}^{-2} \, .
  \label{eq:gtilde}
\end{equation}
Here we have made use of the identity $\varOmega_{\rm K}^2=G M_\star/r^3$,
where $M_\star$ is the mass of the central gravity source and $r$ is the
orbital radius to a considered point in the disc.

\subsubsection{Testing the Poisson solver}
\label{s:codetest}

We validate here the FFT Poisson solver for a complex test case of a
self-gravitating, shearing particle wave with gas drag. This is not a standard
test problem, but we found it very useful for validating the particle-mesh
self-gravity scheme of the Pencil Code. The particle density, velocity and
self-potential are written as $\rho_{\rm p} = \rho_0 + \rho_{\rm p}'$, $\vc{w}
= \vc{w}_0 + \vc{w}'$, $\varPhi = \varPhi_0 + \varPhi'$. Here $\vc{w}_0 =
u_y^{(0)} \hat{\vc{y}} \equiv -(3/2) \varOmega_{\rm K} x \hat{\vc{y}}$ is the
Keplerian shear flow. The linearised continuity equation, equation of motion
and Poisson equation are
\begin{eqnarray}
  \partial_t \rho_{\rm p}' + u_y^{(0)}
      \partial_y \rho_{\rm p}' &=&
      -\rho_0 (\partial_x w_x' + \partial_y w_y') \, ,
  \label{eq:lin_rho} \\
  \partial_t w_x' + u_y^{(0)} \partial_y w_x' &=&
       2 \varOmega_{\rm K} w_y' -\partial_x \varPhi'
      -\frac{1}{\tau_{\rm f}} w_x',
  \label{eq:lin_wx} \\
  \partial_t w_y' + u_y^{(0)} \partial_y w_y' &=&
      -\frac{1}{2} \varOmega_{\rm K} w_x' -\partial_y \varPhi'
      -\frac{1}{\tau_{\rm f}} w_y' \, ,
  \label{eq:lin_wy} \\
  \nabla^2 \varPhi' &=& 4 \pi G \rho_{\rm p}'.
  \label{eq:lin_poisson}
\end{eqnarray}
Here we have assumed particle motion to take place in the radial-azimuthal
plane and zero gas velocity, $u_x=u_y=0$, in the friction terms.  Because of
the Keplerian shear flow, the coefficients of the linearised equations are not
independent of $t$, and thus an eigenmode analysis is not possible.  Instead,
we assume a separable solution $q(t,x,y) = \hat{q}(t) \exp[\ii(k_x(t) x + k_y
y)]$ for each dynamical variable. The time derivative of $q(t,x,y)$ is
\begin{equation}
  \dot{q}(t,x,y) = [\partial_t \hat{q}(t) + \hat{q}(t) i x \partial_t k_x(t)] 
      \exp[\ii(k_x(t) x + k_y y)] \, .
\end{equation}
By setting $x \partial_t k_x(t)= -u_y^{(0)} k_y$, we cancel the $u_y^{(0)}
\partial_y q$ terms in equations~(\ref{eq:lin_rho})--(\ref{eq:lin_wy}), leaving
a system of equations for $(\hat{\rho}_{\rm p}, \hat{w}_x, \hat{w}_y)$,
\begin{eqnarray}
  \frac{\de \hat{\rho}_{\rm p}}{\de t} &=&
      -\rho_0 \ii (k_x \hat{w}_x + k_y \hat{w}_y) \, ,
  \label{eq:lin_rho2} \\
  \frac{\de \hat{w}_x}{\de t} &=&
       2 \varOmega_{\rm K} \hat{w}_y
      +\frac{4 \pi \ii G k_x \hat{\rho}_{\rm p}}{k_x^2+k_y^2}
      -\frac{1}{\tau_{\rm f}} \hat{w}_x \, ,
  \label{eq:lin_wx2} \\
  \frac{\de \hat{w}_y}{\de t} &=&
      -\frac{1}{2} \varOmega_{\rm K} \hat{w}_x
      +\frac{4 \pi \ii G k_y \hat{\rho}_{\rm p}}{k_x^2+k_y^2}
      -\frac{1}{\tau_{\rm f}} \hat{w}_y \, ,
  \label{eq:lin_wy2}
\end{eqnarray}
together with $k_x(t)=k_x(t_0)+(3/2) \varOmega_{\rm K} t k_y$.

\begin{figure}
  \begin{center}
    \includegraphics[width=0.45\linewidth]{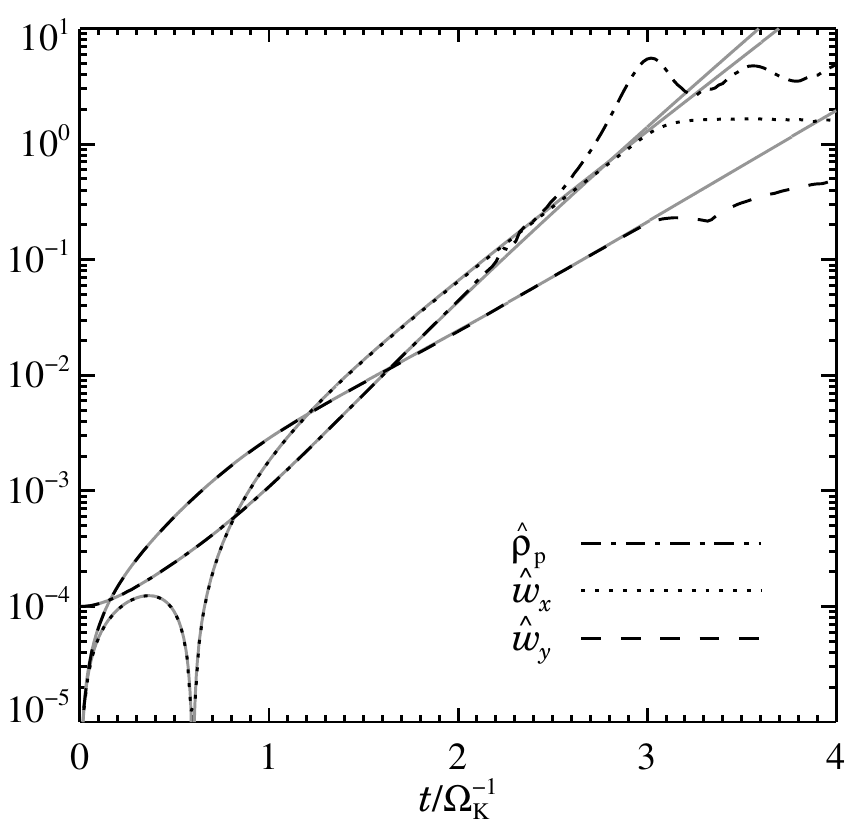}
  \end{center}
  \caption{The evolution of a self-gravitating shear wave of solid particles.
    The plot shows a comparison between the semi-analytic solution to the
    linearised equation system (grey lines) and the solution obtained with the
    full solver of the Pencil Code (dark dotted, dashed and dot-dashed lines)
    for the amplitude of the particle density $\hat{\rho}_{\rm p}$ and of the
    particle velocity components $\hat{w}_x$ and $\hat{w}_y$.  There is
    excellent agreement between the numerical and the analytical solutions in
    the entire linear range. After $\hat{\rho}_{\rm p}\sim0.1$ the analytical
    solution is no longer valid and the two solutions diverge.}
  \label{f:testmode}
\end{figure}

We solve this system of ordinary differential equations numerically using a
third order Runge-Kutta method to follow the temporal evolution of a
non-axisymmetric wave with the initial condition $k_x=-1, k_y=1,
\hat{w}_x=\hat{w}_y=0, \hat{\rho}_{\rm p}=10^{-4}$, and $G=\varOmega_{\rm
K}=\rho_0=\tau_{\rm f} = 1$.  This semi-analytic solution is then compared to
the evolution obtained with the full solver of the Pencil Code ($64^2$ grid
points and $4$ particles per cell).  Because we use particles to represent the
solids and a grid-based FFT solver to calculate the gravitational potential,
the comparison provides an excellent of test of all physics and numerical
schemes relevant to our investigation: numerical particles, grid-based FFT
self-gravity, shear and shear-periodic boundary conditions. We show the
evolution of the analytical solution and of the solution obtained with the
Pencil Code in Fig.\ \ref{f:testmode}. There is an excellent agreement between
the two solutions up to around $\hat{\rho}_{\rm p}=0.1$ where the problem
enters the nonlinear regime and the analytical solution loses its validity (but
the Pencil Code solution does not).

Next we derive an analytical solution to a fully non-linear one-dimensional
gravitational collapse problem, following Spitzer\cite{Spitzer1942}, to test
the Poisson solver outside the linear regime.

Consider the equation system governing a self-gravitating gas in an infinitely
extended one-dimensional space along the $z$-axis:
\begin{eqnarray}
  \frac{\dpa u_z}{\dpa t} + u_z \frac{\dpa u_z}{\dpa z} &=& -
  \frac{\dpa \varPhi}{\dpa z} - c_{\rm s}^2 \frac{\dpa \ln \rho}{\dpa z} \, ,\\
  \frac{\dpa \ln \rho}{\dpa t} + u_z \frac{\dpa \ln \rho}{\dpa z} &=& -
  \frac{\dpa u_z}{\dpa z} \, , \\
  \frac{\dpa^2 \varPhi}{\dpa z^2} &=& 4 \pi G \rho \, .
\end{eqnarray}
We search for stationary solutions with $\dot{u}_z=\dot{\rho}=u_z=0$. This
yields a second order differential equation in $\rho$,
\begin{equation}
  c_{\rm s}^2 \frac{\dpa^2 \ln \rho}{\dpa z^2} + 4 \pi G \rho = 0 \, .
\end{equation}
The general solution is
\begin{equation}
  \rho(z) = \frac{C_1 c_{\rm s}^2}{8 \pi G}
            \left[ 1- {\rm tanh}\left( \frac{1}{2} \sqrt{C_1} |z-z_0| \right)^2 \right]
\end{equation}
where
\begin{equation}
  C_1 = \left( \frac{2 \pi G \varSigma_{\rm g}}{c_{\rm s}^2} \right)^2
\end{equation}
is given by the column density of the gas and $z_0$ is an arbitrary constant.
\begin{figure}
  \begin{center}
    \includegraphics[width=0.45\linewidth]{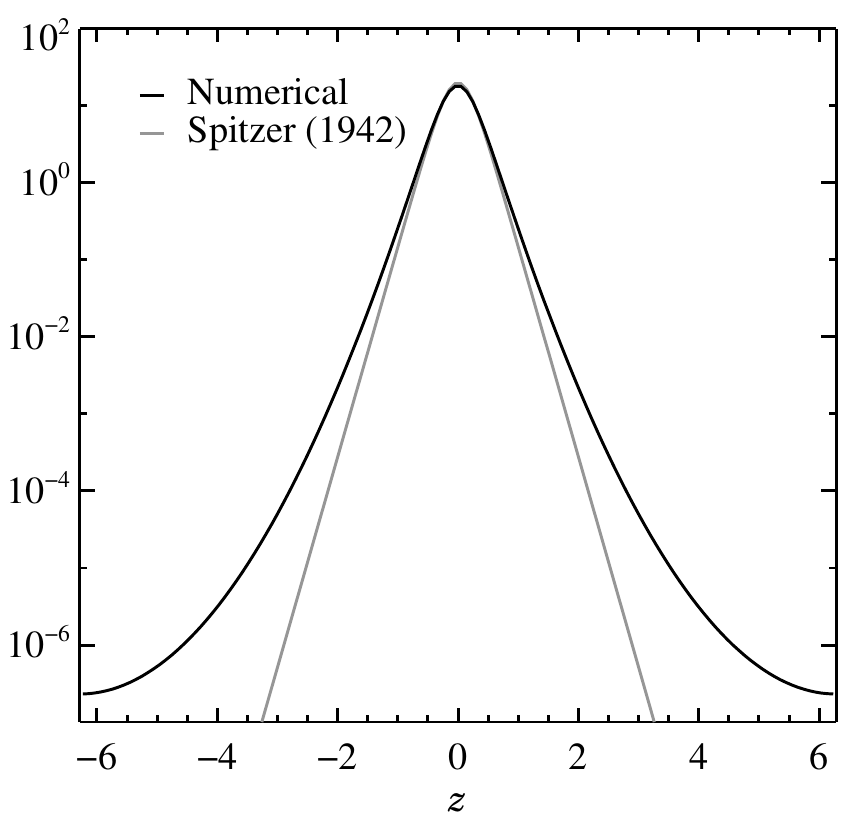}
  \end{center}
  \caption{Comparison of numerical and analytical solution to the
    Spitzer\cite{Spitzer1942} one-dimensional, non-linear collapse problem.
    The agreement is very good in high density regions, but there is
    disagreement in the underdense regions because the analytical solution
    assumes infinite space rather than a periodic domain.}
  \label{f:spitzer}
\end{figure}
This is an exact solution to the full non-linear equation system and thus
complements the linear test problem described in the previous section. We let
the Pencil Code start with a Jeans unstable mode of wavenumber $k_z=0.5$ in a
periodic $z$-space of length $L_z=4\pi$ covered by $64$ grid points (we used
shock viscosity to keep the code stable during the initial collapse).  We show
in Fig.\ \ref{f:spitzer} a comparison between the equilibrium state found by
direct time integration of the equation system by the Pencil Code and the
analytical equilibrium solution.  There is an excellent agreement between the
two solutions in regions of high and moderate density, whereas the underdense
regions contain too much mass in the Pencil Code solution. This is however just
an artefact of the comparison: the analytical solution works in infinitely
extended space, whereas we assumed periodic boundary conditions in the
numerical solution.

\subsection{Collisional cooling}\label{s:collisional_cooling}

In the models that include collisional cooling we let unresolved collisions
damp the velocity dispersion of the particles within each grid cell. The friction time in
the Epstein\cite{Epstein1924} regime is
\begin{equation}
  \tau_{\rm f} = \frac{a_\bullet \rho_\bullet}{c_{\rm s}\rho_{\rm g}} \, ,
  \label{eq:epstein}
\end{equation}
where we follow the nomenclature of equation (\ref{eq:tauEpstein}). The
collisional time-scale is assumed to follow a simple scaling with the Epstein
friction time,
\begin{equation} \label{eq:t_coll}
  \rho_{\rm g} c_{\rm s}\tau_{\rm f} =
      \rho_{\rm p} c_{\rm p} \tau_{\rm coll} \, ,
\end{equation}
where $\rho_{\rm p}$ is the bulk mass density of solids and $c_{\rm p}$ is the
velocity dispersion of the boulders in a grid cell. We ignore for simplicity
any proportionality factors of order unity (for the complete analytical
expression of the cooling time-scale, see Garz\'o \&
Dufty\cite{GarzoDufty1999}). If particles are in the Stokes drag regime rather
than Epstein (see \S\ref{s:drag_regime}), then the collisional cooling
time-scale actually decreases relative to the friction time, but for simplicity
we do not model this regime here. Equation~(\ref{eq:t_coll}) allows us to
calculate the collisional time-scale from the friction time as
\begin{equation}
  \tau_{\rm coll} =
    \frac{\tau_{\rm f}}{(c_{\rm p}/c_{\rm s}) (\rho_{\rm p}/\rho_{\rm g})}
  \, ,
  \label{eq:tcoll}
\end{equation}
without any reference to the radius and solid density of the particles.  We
discuss below how this expression changes when particles of several sizes
collide.

Taking a velocity dispersion of $c_{\rm p}/c_{\rm s} \sim 1\times10^{-2}$ for
the random motion of particles within a grid cell gives a threshold
solids-to-gas ratio of $\epsilon \sim 100$ where collisions between boulders
becomes as important as collisions between boulders and gas molecules.  The
collisional cooling is implemented as a simple term that reduces the velocity
of a particle relative to the mean velocity of the particles in a grid cell on
a collisional time-scale,
\begin{equation}
  \frac{\partial \vc{v}^{(i)}}{\partial t} = -\frac{1-C_{\rm res}}{\tau_{\rm coll}}
  (\vc{v}^{(i)}-\overline{\vc{v}}^{(\vc{k})}) \, ,
\end{equation}
where $\overline{\vc{v}}^{(\vc{k})}$ is the average particle velocity in cell
$\vc{k}$ in which particle $i$ is situated.  The coefficient of restitution
$C_{\rm res}$ is a measure of the degree of energy conservation in the
impacts.  We take $C_{\rm res}=0.1$, corresponding to a loss of 90\% of the
relative velocity in each collision, a reasonable value for collisions between
macroscopic solids\cite{Hartmann1985}, giving a collisional cooling time-scale
that is approximately the same as the collisional time-scale

For collisions between particles of multiple sizes we replace equation
(\ref{eq:tcoll}) with the equation
\begin{equation}
  \tau_{\rm coll} =
      \frac{\tau_{\rm f}^{*}}{(c_{\rm p}/c_{\rm s}) \sum_i (\rho_i/\rho_{\rm g})}\, ,
\end{equation}
where $\rho_i$ is the local bulk density of solids of species $i$.  The
size-modified friction time $\tau_{\rm f}^{*}$ is an average of the friction
times of the individual species weighted with the local bulk density of the
individual species,
\begin{equation}
 \tau_{\rm f}^{*} =
      \frac{\sum_i(\rho_i/\rho_{\rm g})}{\sum_i(\rho_i/\rho_{\rm g})\tau_i^{-1}}
      \, .
\end{equation}
Using $\tau_{\rm f}^{*}$ as the effective friction time, we can define a single
cooling time-scale for all particle size bins. This ensures that the
parameterised collisional cooling conserves momentum. For a single particle
size the effective friction time reduces to the friction time, $\tau_{\rm
f}^{*}=\tau_{\rm f}$. 

In the numerical simulations we calculate the collision speed of
particles in a given grid cell $i$ by first calculating the mean velocity in
the cell $\overline{\vc{v}}^{(i)}$. We then calculate the average particle
collision speed as
\begin{equation}
  c_{\rm p}^{(i)} = \overline{| \vc{v}^{(i)}-\overline{\vc{v}}^{(i)} |} \, .
\end{equation}
This value is then plugged into equation~(\ref{eq:tcoll}) to get the
collisional time-scale.  This approach must be taken when using superparticles,
as collisions cannot directly be followed as could, in principle, be done for
individual particles.

\subsection{Initial condition}\label{s:initial_condition}

Our simulations are initialised with a uniform density gas, $\rho_{\rm g} = 1$,
with isothermal sound speed $c_s=1$.  The superparticles are randomly
distributed in space, and the mass of each superparticle is set such that the
column density of solids is $\varSigma_{\rm p} = 0.01 \sqrt{2\pi} H \rho_{\rm
g,mid}$.  Here we have assumed that prior to the start of the simulation,
sufficient sedimentation has occurred to bring all the solids within the
computational domain.  The particles are given random velocities with a
Gaussian distribution of width $\delta v=1\times 10^{-3} c_{\rm s}$.  The
domain is cubic with $L_x=L_y=L_z=1.32 H$, where $H$ is the pressure scale
height of the disc. We use periodic boundary conditions in the azimuthal
$y$-direction and in the vertical $z$-direction and shearing periodic boundary
conditions in the radial $x$-direction\cite{Hawley+etal1995}. The Keplerian
rotation frequency at the central radius of the grid is $\varOmega_{\rm K} =
1$. Assuming an orbital distance of $r=5\,{\rm AU}$, a gas column density of
$\varSigma_{\rm g}=300\,{\rm g\,cm^{-2}}$ and a scale-height-to-radius ratio of
$H/r=0.04$ then gives $H=3 \times 10^{12}\,{\rm cm}$, $\rho_{\rm g}=4 \times
10^{-11}\,{\rm g\,cm^{-3}}$ and $c_{\rm s}=500\,{\rm m\,s^{-1}}$ as physical
properties of the disc. The column density of solids is $\varSigma_{\rm
p}=3\,{\rm g\,cm^{-2}}$ for a global solids-to-gas ratio of $\epsilon_0=0.01$.

\subsubsection{Radial pressure support}\label{s:radial_drift}

The assumed radial pressure gradient, $\dpa \ln P/\dpa \ln r$, causes radial
drift of the boulders following the expression\cite{Nakagawa+etal1986}
\begin{equation}
  v_x = \frac{(\dpa \ln P/\dpa \ln r) (H/r)}{\varOmega_{\rm K} \tau_{\rm
  f}+(\varOmega_{\rm K} \tau_{\rm f})^{-1}} c_{\rm s} = \frac{2 \Delta
  v}{\varOmega_{\rm K} \tau_{\rm
    f}+(\varOmega_{\rm K} \tau_{\rm f})^{-1}}
\end{equation}
in the absence of collective drag force effects on the gas (i.e.\ vanishing
solids-to-gas ratio).  With $(\dpa \ln P/\dpa \ln r) (H/r)=-0.04$ this gives
$v_x = \Delta v = -0.02 c_{\rm s}$ for marginally coupled boulders with
$\varOmega_{\rm K} \tau_{\rm f}=1$. We show in \S\ref{s:vary_drift} the effect
of less and more radial pressure support.

\subsubsection{Magnetic fields}

The Pencil Code solves the induction equation for the magnetic vector potential
$\vc{A}$, keeping the magnetic field $\vc{B}=\vc{\nabla} \times \vc{A}$
solenoidal (i.e.\ $\vc{\nabla}\cdot\vc{B}=0$) per construction. We initialise
magnetic fields by setting the initial vector potential
\begin{equation}
  \vc{A} = A_0 \cos\left(\frac{2\pi}{L_x} x\right)
               \cos\left(\frac{2\pi}{L_y} y\right)
               \cos\left(\frac{2\pi}{L_z} z\right) \hat{\vc{y}},
\end{equation}
with $A_0 = 0.04$, which leads to an initial zero-net flux field with a maximum
thermal to magnetic pressure ratio $\beta \simeq 55$. For the $256^3$ run we
add a very weak external vertical magnetic field of constant $\beta_{\rm
ext}=2\times10^5$. We found this latter field to be necessary to maintain a
constant turbulent viscosity of $\alpha=10^{-3}$ throughout the simulation.

\subsubsection{Disc mass}\label{s:disc_mass}

We set for the $256^3$ self-gravity simulation presented in the main text a
value of $\tilde{G}=0.1$, corresponding to a disc with Toomre\cite{Toomre1964}
parameter $Q \approx 16$, an order of magnitude below the limit for global
gravitational instability of the gas at $\tilde{G}=1$ (at very large scales in
an infinitely extended disc). For this reason we do not let the gas feel
self-gravity, nor do we let the gas contribute to the gravitational potential
in which the particles move. We have performed simulations that included the
self-gravity of the gas, but found no significant gas overdensity in the
gravitationally contracting particle clusters, due to the strong pressure
support of the gas. Our choice of $\tilde{G}$ gives a gas surface density of
$\varSigma_{\rm g} = 300\,{\rm g\,cm^{-2}}$ at $r=5\,{\rm AU}$ (see eq.\
\ref{eq:gtilde}), approximately two times the minimum mass solar nebula
(MMSN\cite{Hayashi1981}). With moderate radial pressure support $\Delta v=-0.05 c_{\rm s}$
the column density limit is twice as high as for $\Delta v=-0.02 c_{\rm s}$,
see \S\ref{s:vary_drift}.  These are not unrealistic values for the actual
column density of gas in the solar nebula, since the MMSN is a very
conservative estimate of the column density that is calculated by adding up the
heavy elements in the present Solar System and then dividing by the assumed
interstellar solids-to-gas ratio. Any loss of material that occurred due to
radial drift will then lead to an underestimate of the actual column density.
Radial drift can also enhance the solids-to-gas ratio in the inner 10 AU by as
much as an order of
magnitude\cite{StepinskiValageas1996,YoudinShu2002,YoudinChiang2004}.

\subsubsection{Particle sizes}

We assume in all the models described here a global solids-to-gas ratio of
$\epsilon_0=0.01$. There may be twice as much solid material present due to the
condensation of volatiles into ice beyond the snow line at a few AU from the
protostar.  Coagulation models generally yield a particle mass distribution
that spans two orders of magnitude in
radius\cite{Weidenschilling1997,DullemondDominik2005}.  Our particle size
distribution has size bin boundaries spanning an order of magnitude from
$\varOmega_{\rm K}\tau_{\rm f}=0.125$ to $\varOmega_{\rm K}\tau_{\rm
f}=1.125$. The assumption is then that the simulated boulders represent the
upper half of the actual size range and that the lower half is present in
smaller particles that are not effectively concentrated by transient high
pressures and streaming instability and also are unable to participate in the
collapse due to their strong coupling to the gas.

\subsection{Resolution study}\label{s:convergence}

\subsubsection{Power spectra}


\begin{figure}[!t]
  \begin{center}
    \includegraphics[width=\linewidth]{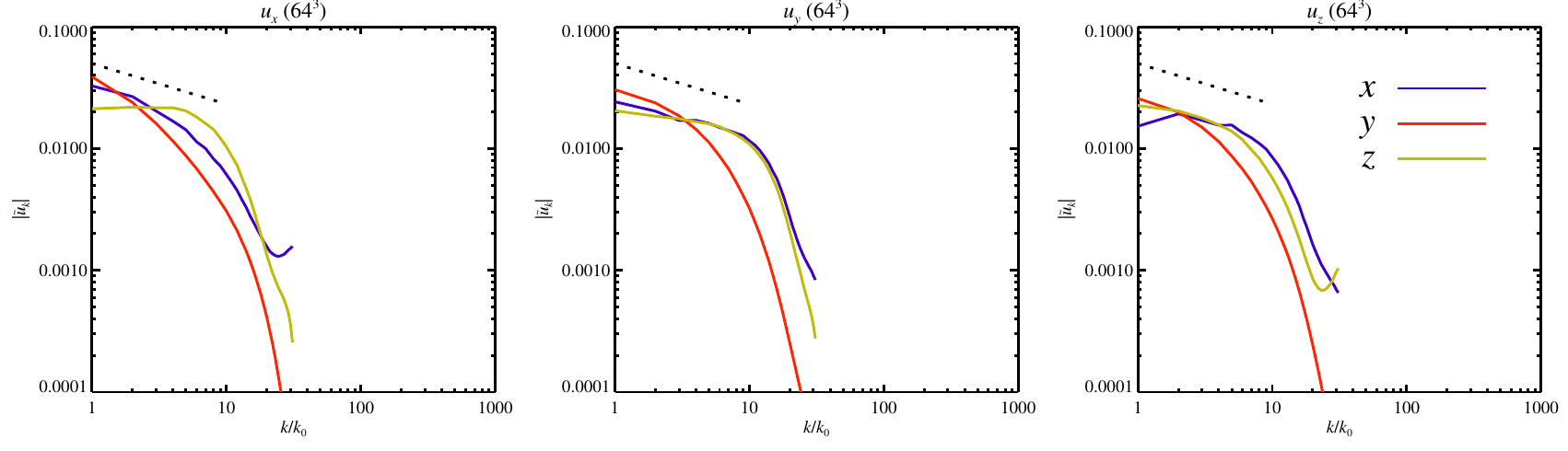}\\
    \includegraphics[width=\linewidth]{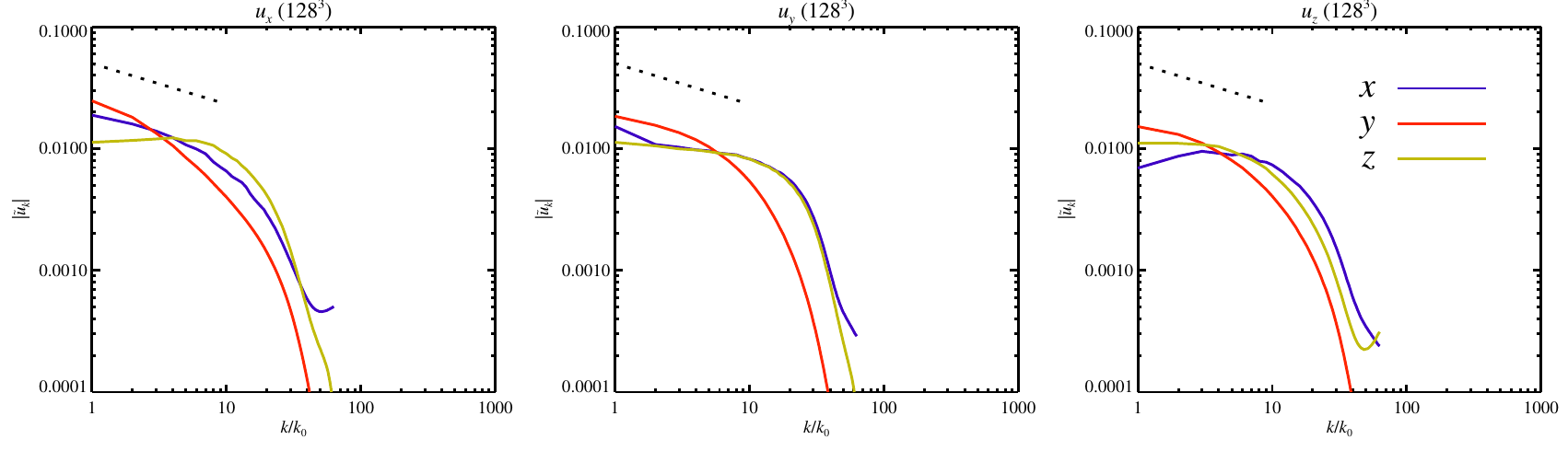}\\
    \includegraphics[width=\linewidth]{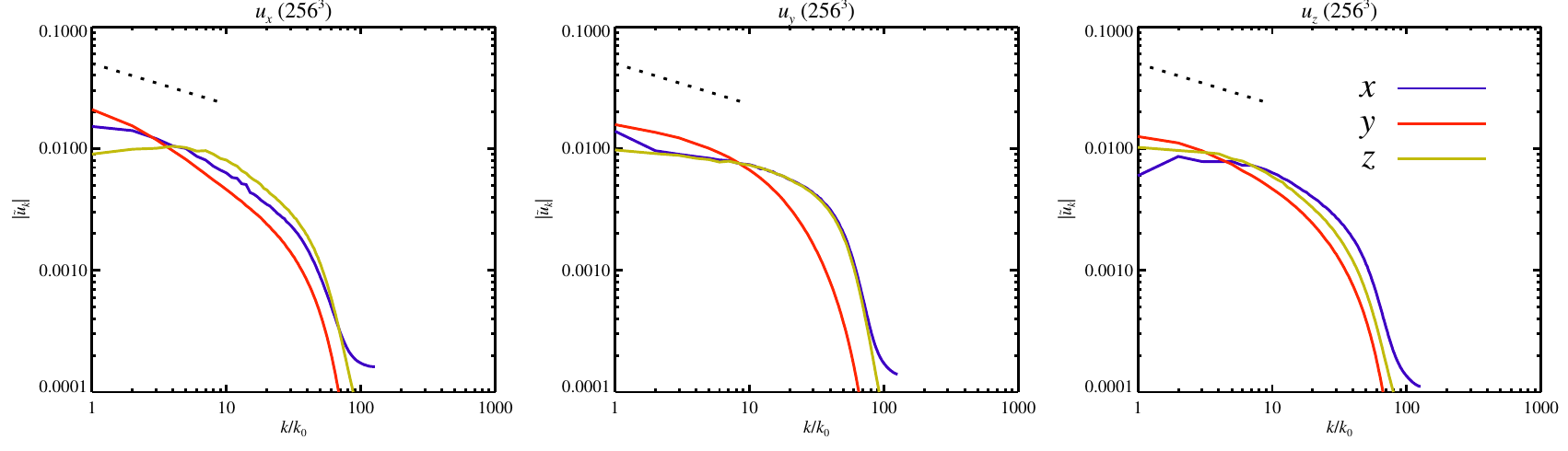}
  \end{center}
  \caption{1-D spectra of the gas velocity components with three different resolutions along
    the rows, the three different velocity components along the columns, and
    spectra taken in three different directions shown in each panel, in
    simulations with no particles. The different colours show power along the
    three coordinate directions, while the dotted line indicates a $k^{-1/3}$
    Kolmogorov law.}
  \label{f:power}
\end{figure}
Magnetorotational turbulence is known to produce a Kolmogorov-like power
spectrum\cite{Hawley+etal1995} with velocity amplitude on scale $k$ going as
$u_k \propto k^{-1/3}$ in the inertial range where numerical viscosity is
insignificant. Increasing resolution leads to an increase in the inertial
range. We show in Fig.\ \ref{f:power} the modulus of the velocity amplitude as
a function of wavenumber $k$ normalised with the wavenumber of the largest
scale of the box. A Kolmogorov $k^{-1/3}$ power law is indicated with a dotted
line for reference. The turbulence is notably weaker in the $z$-direction than
in the $x$- and $y$-directions. It also clear from Fig.\ \ref{f:power} that the
inertial range expands as expected with increasing resolution. Resolving the
turbulent scales is important for resolving the local velocity dispersion
(``temperature'') of the particles and thus for the gravitational collapse (see
\S\ref{s:collision_speeds}).

\subsubsection{Accretion}

\begin{figure}[!t]
  \begin{center}
    \includegraphics[width=0.5\linewidth]{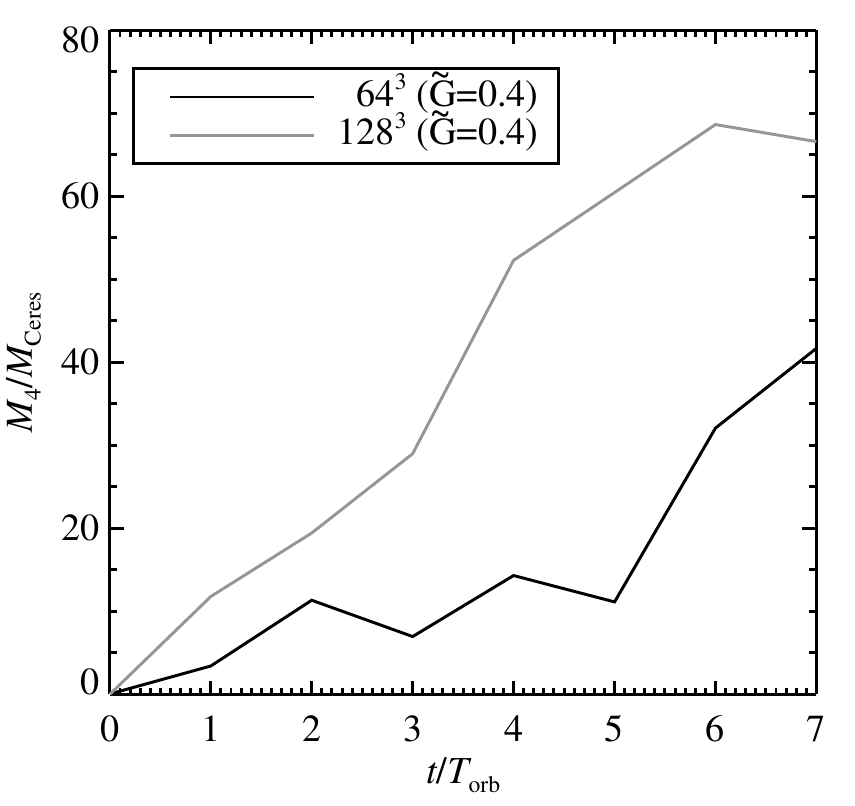}
  \end{center}
  \caption{Total mass in the four most massive gravitationally bound clusters,
    in units of Ceres masses, plotted versus time since self-gravity was turned
    on, for two resolutions but a fixed self-gravity parameter $\tilde{G}$ that
    allows gravitational collapse at both $64^3$ and at $128^3$. The higher
    resolution run accretes at a 50\% higher rate onto four clusters condensed
    out of the midplane layer, while only two clusters formed in the $64^3$
    run.}
  \label{f:mclumptot_t_fixed}
\end{figure}

\begin{figure}[!t]
  \begin{center}
    \includegraphics[width=0.45\linewidth]{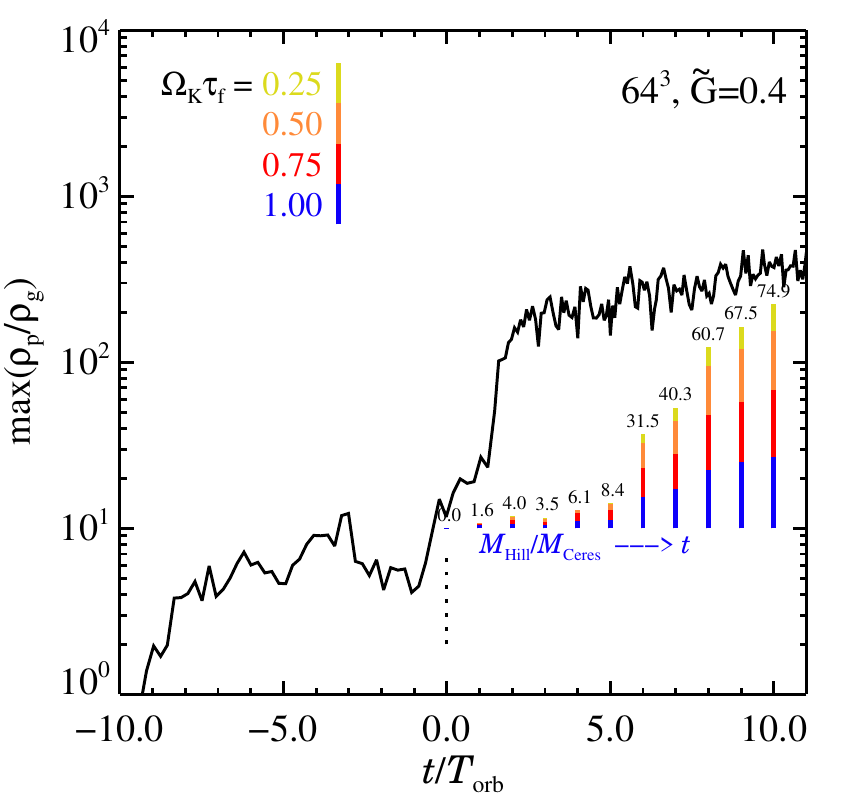}
    \includegraphics[type=pdf,ext=.pdf,read=.pdf,width=0.45\linewidth]{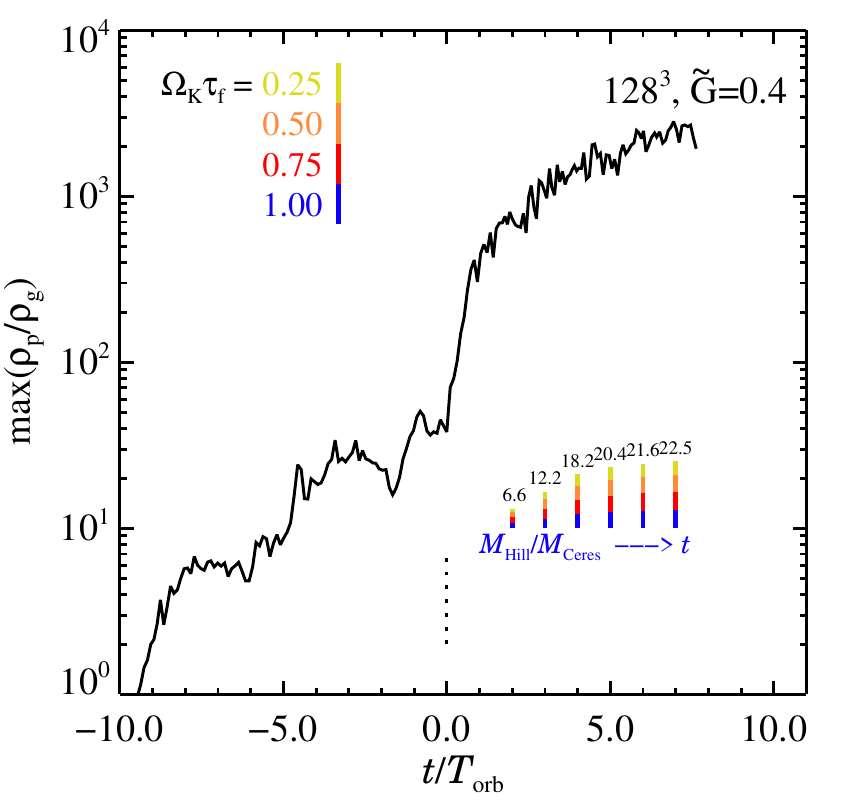}
  \end{center}
  \caption{The maximum bulk solid density and the mass of the most massive bound
    cluster as a function of time for a fixed value $\tilde{G}=0.4$ at $64^3$
    (left, $1.25 \times 10^5$ particles) and at $128^3$ (right, $1 \times 10^6$
    particles). The $64^3$ run produces only 2 clusters (see
    Table~\ref{t:convergence_fixG}), but the
    most massive one accretes lots of solid material (7.5 $M_{\rm Ceres}$ per
    orbit), whereas at the increased resolution of $128^3$ four clusters form
    and they must compete for particles to accrete, thus the accretion rate of
    the most massive cluster decreases. The maximum bulk density of particles
    increases for higher resolution because the particle density field is
    better resolved by the TSC assignment scheme (see \S\ref{s:drag_force}).
    Fig.\ \ref{f:mclumptot_t_fixed} shows that the {\em total} mass accretion
    onto bound clusters in the $128^3$ simulation is actually twice as high as
    in the $64^3$ simulation.}
  \label{f:max_pdensity_fixG}
\end{figure}

\begin{table}[t!]
  \begin{center}
    \begin{tabular}{cccc}
      \hline
      \hline
      Resolution & $N_{\rm par}/10^6$ & $\Delta t_{\rm grav}$ & $\alpha$ \\
      (1) & (2) & (3) & (4) \\
      \hline
      \,\,\,$64^3$  & \,\,\,\,\,\,$0.125$ & $10.0$ & $0.002$ \\
      $128^3$ & $1.0$    & $7.0$ & $0.001$ \\
      \hline
    \end{tabular}
    \begin{tabular}{ccccccc}
      \hline
      \hline
      Resolution & $\tilde{G}$ & $Q$
                 & $N_{\rm clusters}$ & $\dot{M}_{\rm cluster}$ 
                 & $\dot{M}_{\rm cluster}/\tilde{G}_{0.1}$
                 & $M_4$ \\
      (5) & (6) & (7) & (8) & (9) & (10) & (11) \\
      \hline
      \,\,\,$64^3$ & $0.4$ & $4.0$ & $2$ & $7.5$ & $1.9$ & $42$ \\
           $128^3$ & $0.4$ & $4.0$ & $4$ & $2.9$ & $0.7$ & $65$ \\
      \hline
    \end{tabular}
    \caption{Resolution study for fixed $\tilde{G}$.
      Col.\ (1): Mesh resolution.
      Col.\ (2): Number of superparticles in millions.
      Col.\ (3): Number of orbits with self-gravity.
      Col.\ (4): Measured turbulent viscosity.
      Col.\ (6): Self-gravity parameter.
      Col.\ (7): Corresponding Toomre $Q\approx1.6\tilde{G}^{-1}$.
      Col.\ (8): Number of clusters at the end of the simulation.
      Col.\ (9): Accretion rate of the most massive cluster in Ceres masses per
                 orbit.
      Col.\ (10): Accretion rate normalised with $\tilde{G}_{0.1} \equiv
                  \tilde{G}/0.1$.
      Col.\ (11): Total mass in the four most massive gravitationally bound
                  clusters at 7 orbits after self-gravity is turned on.
}
  \label{t:convergence_fixG}
  \end{center}
\end{table}
We check the convergence of the accreted mass using runs described in
Table~\ref{t:convergence_fixG}.
We compare runs at $64^3$ and $128^3$ zone resolution, using a column density
(parameterised by the self-gravity parameter $\tilde{G}$) that allows the $64^3$
simulation to undergo gravitational collapse into discrete clusters. We find
that the $128^3$ run forms four clusters while the $64^3$ run only forms two.
However, the most massive cluster accretes faster in the low resolution case,
because it does not have to compete for particles with other clusters. In Fig.\
\ref{f:mclumptot_t_fixed} we plot the mass in the four most massive clusters as
a function of time. The clusters in the $128^3$ run accrete at approximately
50\% higher rate than in the $64^3$ run, because there are more clusters upon
which particles can accrete. This is also evident from the last column of
Table~\ref{t:convergence_fixG}
where we write the total mass of the four most massive, bound clusters at a
time of 7 orbits after self-gravity is turned on. In Fig.\
\ref{f:max_pdensity_fixG} we plot the maximum bulk density of particles and the
mass of the most massive bound cluster as a function of time (similar to Fig.\
3 of the main text). It is clear that the most massive cluster contains more
mass and accretes faster in the $64^3$ simulation. At the same time the maximum
density is higher in the $128^3$ simulation, because the increased resolution
leads to less smoothing in the particle-mesh scheme (see \S\ref{s:drag_force}).

We emphasise that the gravitational collapse of overdense seeds in turbulence
is inherently a stochastic process, because of the temporal variability of the
overdensities, so that convergence is not expected in the strict meaning of the
term. Only convergence to mean values can be expected over a large ensemble of
simulations. The limited number of tests that we have been able to perform
within available computational resources should therefore be taken as
qualitatively suggestive rather than conclusive. 

\subsubsection{Gravitational collapse}

We have also done a resolution study of the column density required for
gravitational collapse of the particle layer to occur, with results given in
Table~\ref{t:multiG}.  The midplane layer in all runs presented here is
resolved with an initial average of around four particles per grid cell, enough
for the TSC scheme to define meaningful particle densities for the self-gravity
solver. Collapsing regions have many more particles in each cell, of course.
The column density limit for forming gravitationally bound clusters (see column
6 of Table~\ref{t:multiG}) decreases with increasing grid resolution at all
resolutions we have been able to study to date, as determined by varying the
column density in increments of $\Delta \varSigma_{\rm g} = 300$~g~cm$^{-3}$,
equivalent to $\Delta \tilde{G}=0.1$.

The observed decrease in limiting column density appears to be partly because
of a decrease in the strength of the magnetorotational turbulence with
increasing resolution, at least at lower resolution. The strength of the
turbulence can be parameterised by the effective turbulent viscosity, which
drops from $\alpha=0.002$ at $64^3$ to $\alpha=0.001$ at $128^3$ and $256^3$,
allowing a slightly denser midplane layer to form.

\begin{figure}
  \begin{center}
    \includegraphics[width=0.5\linewidth]{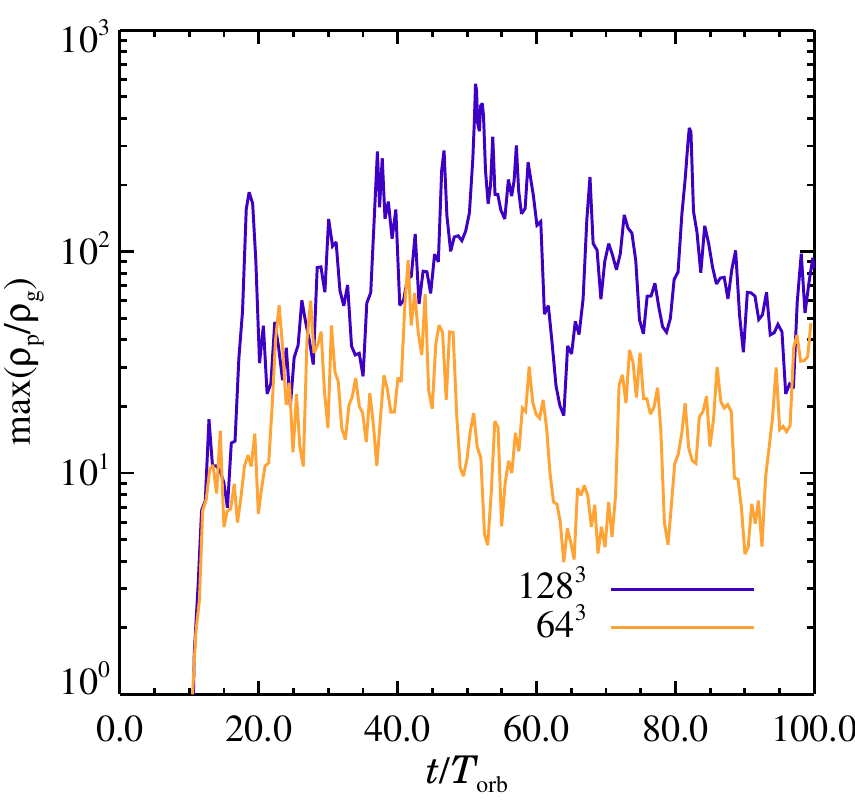}
  \end{center}
  \caption{Maximum particle density on the grid, assigned from particle
    positions with the TSC scheme\cite{HockneyEastwood1981,YoudinJohansen2007},
    versus time for runs with no self-gravity and no collisional cooling. The
    maximum density goes up by a factor of 3-4 for the $128^3$ simulation, in
    part because the density field assigned with the TSC scheme on the grid has
    less smoothing at higher resolution.}
  \label{f:rhopmax_t_convergence_nosg}
\end{figure}

A more important factor, though, is that the particle density field becomes
better resolved on the grid at higher resolution.  The TSC assignment scheme
underestimates the amplitude of density modes with wavelength near the grid
size\cite{HockneyEastwood1981}, so the self-gravity solver underestimates the
gravitational acceleration caused by structures at those small scales.  This
can be seen in Fig.\ \ref{f:rhopmax_t_convergence_nosg}, where we show the
maximum particle density produced by streaming instability and concentration in
transient high pressures in models with neither self-gravity nor collisional
cooling.  Increasing linear resolution by a factor of two increases maximum
density by around a factor four. Thus only higher resolution models reach the
Toomre criterion for collapse at low column densities. 

We write in columns 12 and 13 of Table~\ref{t:multiG} the mass in the four most
massive, bound clusters and the same parameter divided by $\tilde{G}_{0.1}
\equiv \tilde{G}/0.1$. There is less (normalised) mass accretion at $256^3$
than at lower resolutions. This is readily explained because collapse can take
place in smaller regions when the resolution is increased. The clusters thus
contain less mass and accrete slower, at least to begin with, although the
$256^3$ has not run long enough to reveal the behaviour over longer times.

Another important effect is the resolution of the initial gravitational
contraction. Fig.\ 2 in the main text shows that self-gravity results first in
a radial contraction of the overdense seed bands. This contraction continues
until the bands reach the Hill density where self-gravity dominates over tidal
force, allowing a full non-axisymmetric collapse into bound clusters occurs.
However, the self-gravity solver cannot follow collapse below a few grid cells,
setting a resolution-dependent minimum width on the radial bands. Increasing
the resolution decreases the final width of the bands, increasing their
density, and thus allowing full collapse to occur in models with lower initial
column densities. After collapse sets in, higher resolution allows collapse to
higher densities, again resulting in higher peak densities in higher resolution
models.
\begin{table}[t!]
  \begin{center}
    \begin{tabular}{cccc}
      \hline
      \hline
      Resolution & $N_{\rm par}/10^6$ & $\Delta t_{\rm grav}$ & $\alpha$ \\
      (1) & (2) & (3) & (4) \\
      \hline
      \,\,\,$64^3$  & \,\,\,\,\,\,$0.125$ & $10.0$ & $0.002$ \\
      $128^3$ & $1.0$    & $10.0$ & $0.001$ \\
      $256^3$ & $8.0$    & \,\,\,$ 7.0$ & $0.001$ \\
      \hline
    \end{tabular}
    \begin{tabular}{ccccccccc}
      \hline
      \hline
      Resolution & $\tilde{G}$
                 & $Q$
                 & $\tilde{G}^{\rm (no\,\,coll)}$
                 & $N_{\rm clusters}$ & $\dot{M}_{\rm cluster}$ 
                 & $\dot{M}_{\rm cluster}/\tilde{G}_{0.1}$
                 & $M_4$ & $M_4/\tilde{G}_{0.1}$\\
      (5) & (6) & (7) & (8) & (9) & (10) & (11) & (12) & (13) \\
      \hline
      \,\,\,$64^3$
          & $0.4$ & $4.0$ & $0.4$
          & $2$ & $7.5$ & $1.9$ & $42$ & $10.5$ \\
      $128^3$
          & $0.2$ & $8.0$ & $0.3$
          & $2$ & $1.8$ & $0.9$ & $19$ & $9.5$ \\
      $256^3$
          & $0.1$ & $16.0$ & --
          & $4$ & $0.5$ & $0.5$ & $6$ & $6$\\
      \hline
    \end{tabular}
    \caption{Resolution study for gravitational collapse.
      Col.\ (1): Mesh resolution.
      Col.\ (2): Number of superparticles in millions.
      Col.\ (3): Number of orbits with self-gravity.
      Col.\ (4): Measured turbulent viscosity.
      Col.\ (6): Minimum self-gravity parameter where gravitationally bound
                 clusters form (MMSN has $\tilde{G}\approx0.05$ at $r=5$ AU).
      Col.\ (7): Corresponding Toomre $Q\approx1.6\tilde{G}^{-1}$.
      Col.\ (8): Minimum self-gravity parameter where gravitationally bound
                 clusters form without collisional cooling.
      Col.\ (9): Number of clusters at the end of the simulation.
      Col.\ (10): Accretion rate of the most massive cluster in Ceres masses
                  per orbit.
      Col.\ (11): Accretion rate normalised with $\tilde{G}_{0.1} \equiv
                  \tilde{G}/0.1$.
      Col.\ (12): Total mass in the four most massive gravitationally bound
                  clusters at 7 orbits after self-gravity is turned on.
      Col.\ (13): Total cluster mass normalised with $\tilde{G}_{0.1} \equiv
                  \tilde{G}/0.1$.}
  \label{t:multiG}
  \end{center}
\end{table}

Poisson noise due to the discrete nature of the superparticles on the other
hand, appears to be quite insignificant for gravitational collapse. We have
done tests with particle numbers as high as thirty particles per cell (in the
mid-plane layer) in the 64$^3$ runs ($10^6$ particles) and eight particles per
cell in the 128$^3$ runs ($2 \times 10^6$ particles) and found the column
density for collapse to be unchanged from runs with the same grid resolution
and only four particles per cell. The overdense seeds of the gravitational
instability are under all circumstances resolved with several hundreds or
thousands of particles, explaining this result.

Although we have clearly not yet obtained convergence in the gravitational
collapse of the particles, the consistent trend towards higher peak densities
and gravitational instability at lower average column density with increasing
resolution lends strong support to our hypothesis that this mechanism can drive
planetesimal formation at gas column densities characteristic of the minimum
mass solar nebula.

To compensate for underresolving the peak densities, one could manually sharpen
the underestimated density modes\cite{HockneyEastwood1981}, but we prefer not
to solve for the gravitational acceleration better than the pressure term, to
avoid any risk of artificial fragmentation\cite{BateBurkert1997}. On the other
hand, studies that use the adaptive mesh refinement
method\cite{Truelove+etal1998} in combination with a kinetic particle code with
collision detections\cite{WisdomTremaine1988} could be adapted to fully resolve
the gas concentration of particles, and then follow the collapse and
gravitational fragmentation of the boulder layer all the way up to solid
density, allowing determination of the minimum column density for planetesimal
formation and the final multiplicity of the self-gravitating boulder clusters
that we model.

\subsection{Collisional cooling}

\begin{figure}
  \begin{center}
    \includegraphics[width=0.4\linewidth]{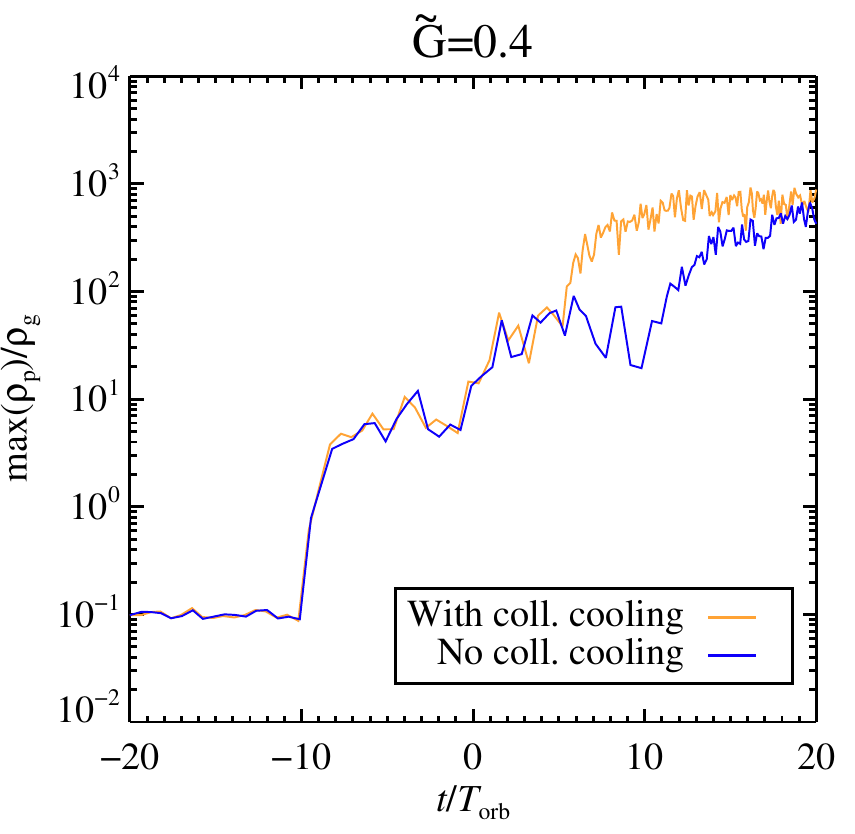}
    \includegraphics[width=0.4\linewidth]{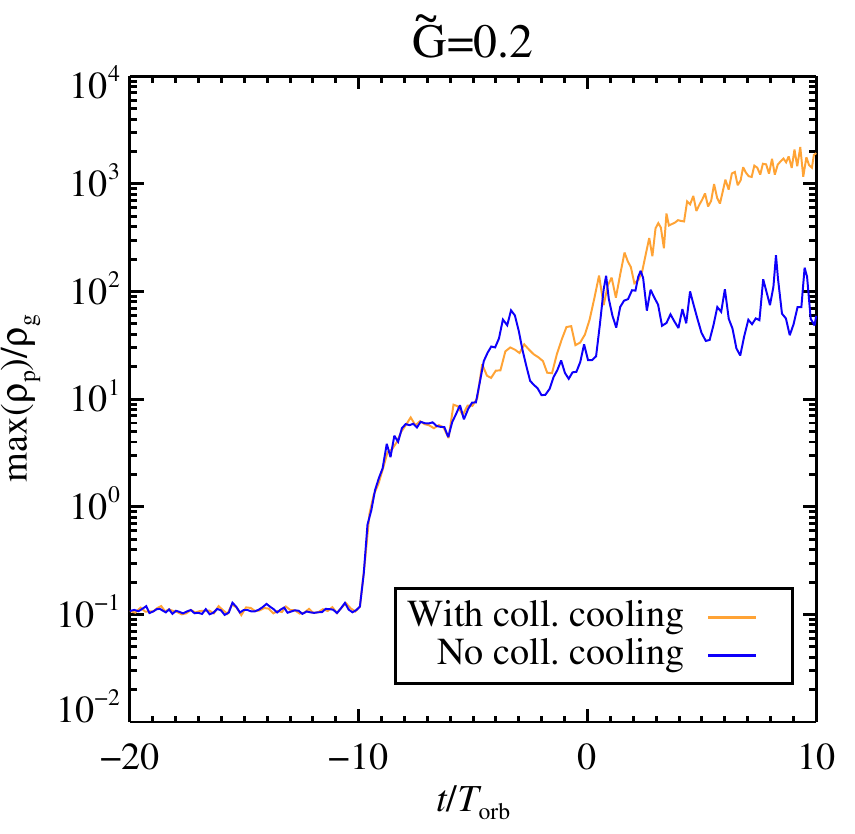}
  \end{center}
  \caption{Comparison of runs with (bright/yellow) and without (dark/blue)
    collisional cooling (self-gravity and collisional cooling are turned on at
    $t=0$). At
    $64^3$ resolution, we find that collapse occurs at the same gas column
    density, independent of collisional cooling, although collisional cooling
    lets the collapse happen faster. At $128^3$ collapse with collisional
    cooling occurs at $\tilde{G}=0.2$, while $\tilde{G}=0.3$ is needed when
    collisional cooling is not applied.}
  \label{f:rhopmax_t_coll_compare}
\end{figure}

In the runs presented in the main text, we include the effect of collisional
cooling on the collapse of the mid-plane layer into gravitationally bound
clusters.  However, the collisional cooling only marginally changes the column
density at which collapse initially occurs.  To demonstrate this, we have run
simulations without collisional cooling to quantify its importance for
gravitational collapse (see column 8 of Table~\ref{t:multiG}).  We show in
Fig.\ \ref{f:rhopmax_t_coll_compare} the maximum bulk density of solids in
simulations with and without collisional cooling, for $64^3$ (left plot) and
$128^3$ (right plot) grid simulations. At $64^3$ resolution collapse occurs at
$\tilde{G}=0.4$ both with and without collisional cooling, although its
inclusion seems to cause collapse to happen somewhat more quickly after
self-gravity is turned on at $t=0$. At a resolution of $128^3$ zones, collapse
occurs in the presence of collisional cooling at $\tilde{G}=0.2$, whereas the
column density limit without collisional cooling is roughly 50\% higher.
Collisional cooling is thus not a prerequisite of the collapse, but rather
allows collapse to occur in somewhat lighter discs.

\subsection{Jeans and Toomre criterion}

The Jeans criterion\cite{BateBurkert1997,Truelove+etal1997} states that the
local Jeans length for gravitational collapse of gas must be resolved by at
least a few grid points, as otherwise artificial fragmentation may occur in
marginally stable structures because of numerical discretisation errors. This
is particularly a problem in the presence of
rotation\cite{Truelove+etal1997,Li+etal2005}. A corresponding
Toomre\cite{Toomre1964} criterion\cite{Nelson2006} is appropriate for accretion
discs. We show in \S\ref{s:sizes} that the initial radial collapse phase is
resolved by around 20 grid points, but gravitationally bound clusters
eventually reach the size of a few grid cells, for which the unstable
wavelengths that would lead to further fragmentation are not resolved.  It is
not snown whether artificial fragmentation is an issue for solid particles
moving on a fixed mesh. The TSC scheme inherently smooths out small scale power
in the assigned particle density\cite{HockneyEastwood1981,YoudinJohansen2007}
(see also \S\ref{s:drag_force}), so the contribution of the small scales to the
gravitational potential is underestimated, whereas the particle velocity
dispersion that counteracts contraction is resolved equally well at all
scales.  It is thus more likely that our numerical scheme suppresses
fragmentation at the small scales and at crude resolutions; this is also
supported by the resolution tests where gravitational collapse occurs at lower
column density in higher resolution models.  We have no reason to believe that
the collapse of the particle layer should stop anywhere near the grid scale:
the velocity dispersion in the overdense clusters is around 1\% of the sound
speed, so the pressure support is negligible\cite{JohansenKlahrHenning2006},
and drag force and inelastic collisions are highly efficient at dissipating
kinetic energy.  Instead we consider the condensed particle clusters to be
equivalent to (numerically expensive) sink particles, a good approximation as
long as there is no feedback from the unresolved scales back to the large
scales\cite{Bate+etal1995}.

\subsection{Gammie cut-off}

Gammie\cite{Gammie2001} argues that one must exclude wave numbers above a limit
from the solution to the Poisson equation in the shearing box in order to keep
the gravitational acceleration approximately isotropic at small scales in the
presence of shear. The cut-off is at $k_{\rm max}=\sqrt{2} k_{\rm Ny}$ where
$k_{\rm Ny}$ is the Nyquist wavenumber.  This cut-off has been applied in all
our simulations. We have run test simulations without the cut-off and found no
significant differences in the resulting gravitational fragmentation, probably
because the TSC assignment scheme itself underestimates the potential at these
small scales.

\subsection{Size of the forming bodies}\label{s:sizes}

The linear stability analysis of Goldreich \& Ward\cite{GoldreichWard1973}
gives the largest wavelength unstable to radial self-gravity as
\begin{equation}
  \lambda_{\rm GW} = \frac{4\pi^2 G \varSigma_{\rm p}}{\varOmega_{\rm K}^2} \, ,
\end{equation}
where $G$ is the gravity constant, $\varSigma_{\rm p}$ is the column density of
solids and $\varOmega_{\rm K}$ is the Keplerian orbital frequency. This
expression is formally valid in the limit of vanishing particle velocity
dispersion and ignores the potentially important effect of drag forces on the
collapse\cite{Youdin2005a,Youdin2005b}.  Inserting nominal values at $r=5\,{\rm
AU}$ in the protosolar nebula gives
\begin{equation}
  \lambda_{\rm GW} \approx 1.4 \times 10^{10}\,{\rm cm}
  \left( \frac{M_\star}{M_\odot} \right)^{-1} \left( \frac{\varSigma_{\rm
  p}}{1.5\,{\rm g\,cm^{-2}}} \right) \left( \frac{r}{5\,{\rm AU}} \right)^3 \, .
\end{equation}
Using a scale height of $H=3\times10^{12}\,{\rm cm}$ (see
\S\ref{s:initial_condition}) and $\varSigma_{\rm p}=15\,{\rm g\,cm^{-2}}$ in
the radial overdensities of our simulations yields $\lambda_{\rm GW}/H \approx
0.1$, which is initially well-resolved (the grid size is $\delta x=\delta
y=\delta z=0.005 H$ in the $256^3$ simulation). Assuming that all the solid
mass within a single unstable wavelength collapses to a gravitationally bound
solid object, the radius of the object is
\begin{equation}
  R \approx \left( \frac{\varSigma_{\rm p} \xi^2 \lambda_{\rm GW}^2}{\rho_\bullet} \right)^{1/3}
    \approx 50\,{\rm km} \, \xi^{2/3} \left( \frac{M_\star}{M_\odot} \right)^{-2/3} \left(
    \frac{\varSigma_{\rm p}}{1.5\,{\rm g\,cm^{-2}}} \right) \left(
    \frac{r}{5\,{\rm AU}} \right)^2
    \left( \frac{\rho_\bullet}{2\,{\rm g\,cm^{-3}}}\right)^{-1/3} \, ,
  \label{eq:radius_GW}
\end{equation}
where $\xi<1$ is a parametrisation of the most unstable wavelength relative to
the largest unstable wavelength. A reasonable value is $\xi=1/2$.  

At $r=1\,{\rm AU}$ with $\varSigma_{\rm p}=15\,{\rm g\,cm^{-2}}$ one recovers
the 10 km planetesimals found by Goldreich \& Ward\cite{GoldreichWard1973}. The
size of planetesimals falls to less than 1 km if the velocity dispersion of the
particles is included, but the derivation ignores the effect of drag forces on
energy dissipation\cite{Ward1976,Ward2000,Youdin2005a,Youdin2005b}.
Considering, on the other hand, an orbital radius $r=5\,{\rm AU}$, with
$\varSigma_{\rm p}=30\,{\rm g\,cm^{-2}}$ in the radial overdensities
found in the current work, the planetesimals reach radii of around 600 km,
because the unstable wavelengths contain far more mass. These sizes agree
roughly with the results of our numerical models.

\subsection{Collision speeds}\label{s:collision_speeds}

\begin{figure}
  \begin{center}
    \includegraphics[width=\linewidth]{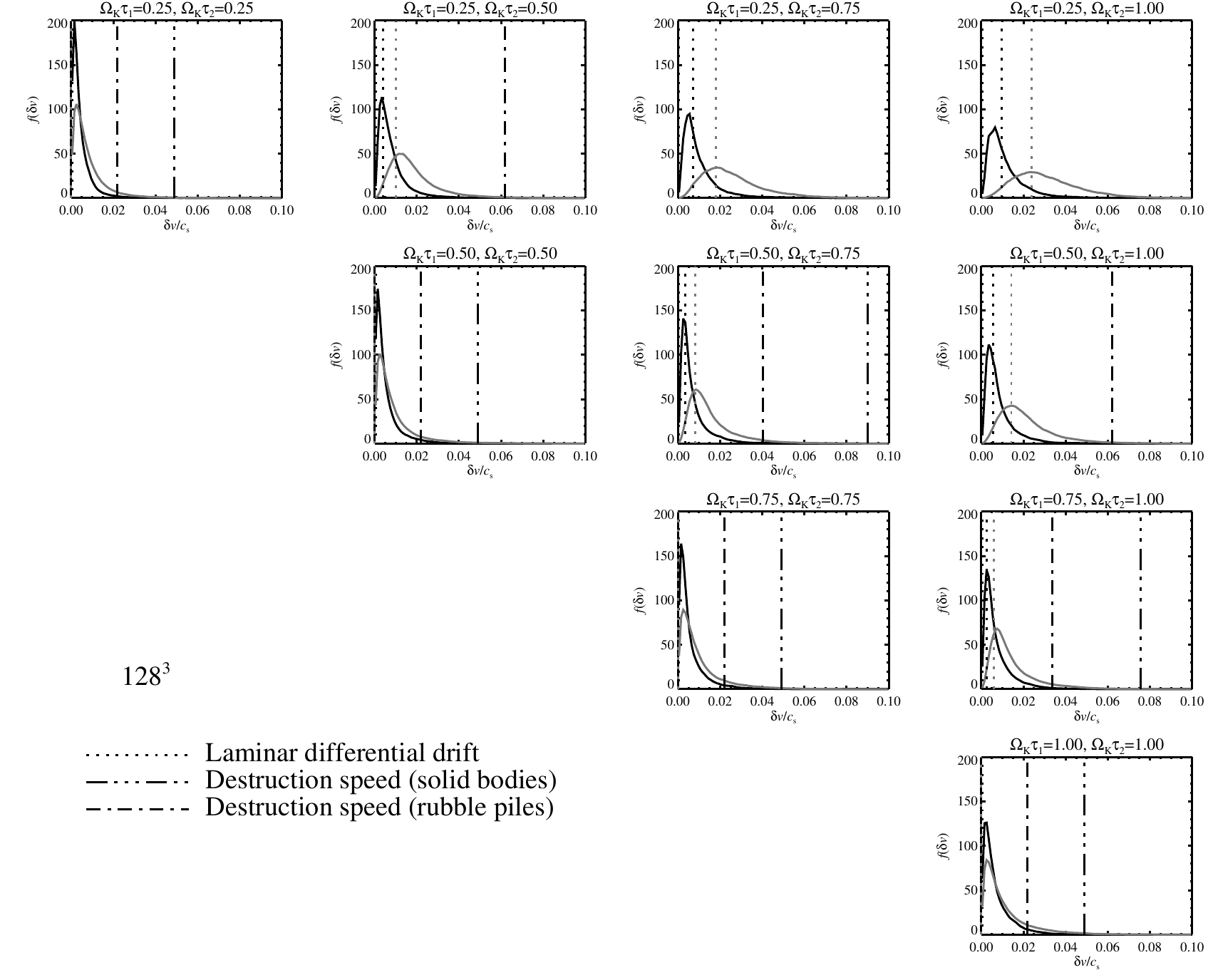}
  \end{center}
  \caption{Distribution of collision speeds between boulders of different sizes
    for a radial pressure support of $\Delta v=-0.05 c_{\rm s}$ (grey line) and
    $\Delta v=-0.02 c_{\rm s}$ (dark line), both for $128^3$ grid points and
    1,000,000 particles {\em without} either collisional cooling or
    self-gravity. The laminar solution for the relative speed due to
    differential drift (in radial and azimuthal velocity) is shown with a
    dotted line (with an assumed solids-to-gas ratio of $\epsilon=0.5$). This
    is the major contribution for different-sized bodies, whereas equal-sized
    bodies only obtain collision speed due to turbulent motions in the gas.
    The threshold for destruction is shown for solid bodies (dot-dot-dot-dashed
    line) and for rubble piles (dot-dashed line) following
    Benz\cite{Benz2000}.  Equal-sized bodies have the lowest destruction
    threshold, but that is still in the far wing of the collision distribution.
    Small bodies and large bodies can survive collisions at much higher speeds,
    sometimes even beyond the range shown along the $\delta v$ axis (i.e.
    $>0.1c_{\rm s}$, where $c_{\rm s}$=500 m/s).  The collision speed between
    equal-sized bodies is relatively unchanged when increasing the pressure
    support to $\Delta v=-0.05 c_{\rm s}$. The increased differential drift,
    however, manifests itself in higher collision speeds between
    different-sized bodies.}
  \label{f:vcoll_128}
\end{figure}

We show in Fig.\ \ref{f:vcoll_128} the collision speeds between bodies of
different sizes in a $128^3$ simulation without self-gravity or collisional
cooling (from a snapshot taken at $t=20T_{\rm orb}$). We consider a million
particles of four different particle sizes at the same time, $\varOmega_{\rm K}
\tau_{\rm f}=0.25,0.50,0.75,1.00$ and two values of the radial pressure
support, $\Delta v=-0.05 c_{\rm s}$ and $\Delta v=-0.02 c_{\rm s}$.

\begin{figure}
  \begin{center}
    \includegraphics[width=\linewidth]{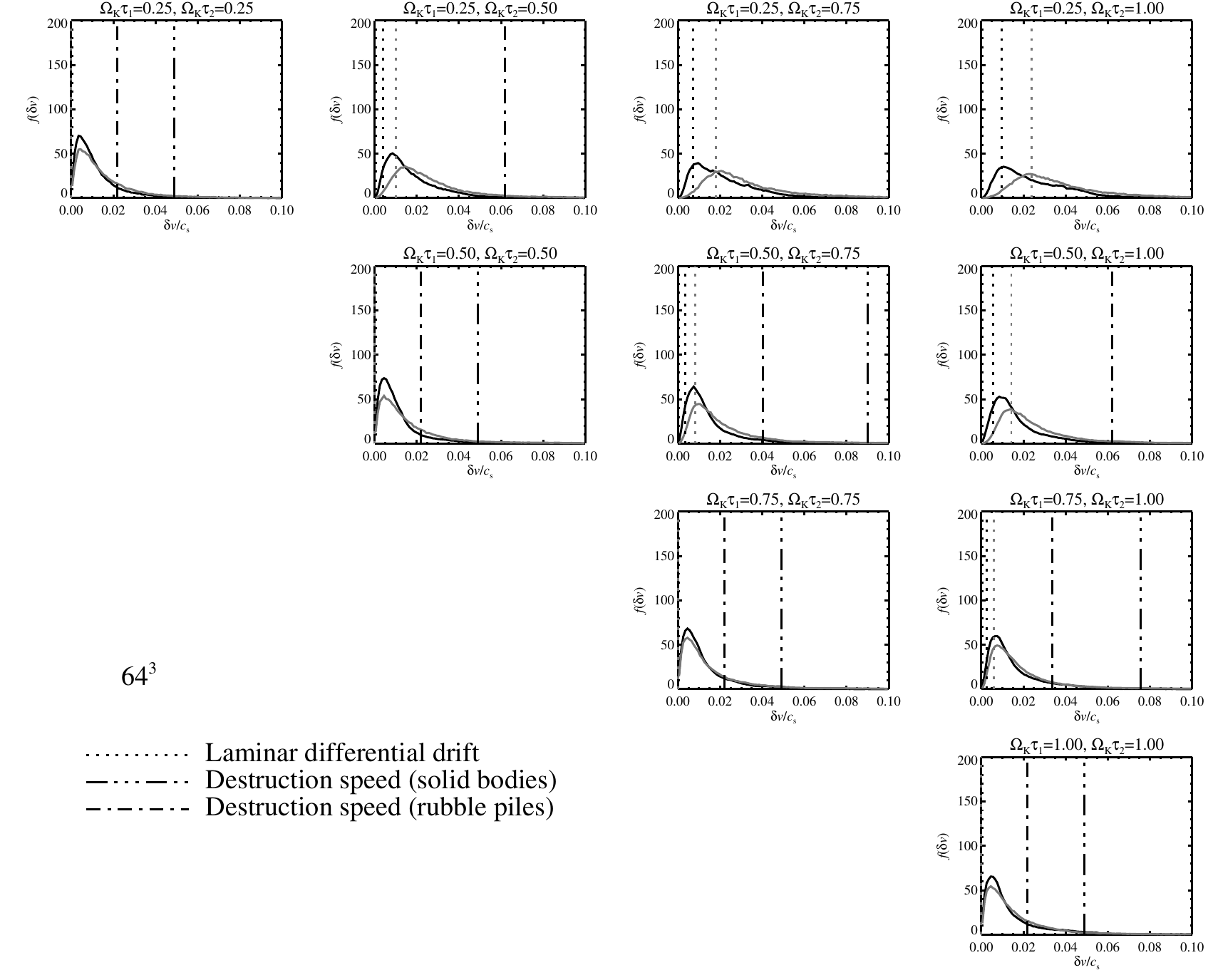}
  \end{center}
  \caption{Same as Fig.\ \ref{f:vcoll_128}, but for $64^3$ grid points with
    125,000 particles. The overall Mach number of the turbulent flow is 30-40\%
    higher than in the higher resolution runs, hence collision speeds are
    somewhat higher.}
  \label{f:vcoll_64}
\end{figure}
\begin{figure}
  \begin{center}
    \includegraphics[width=\linewidth]{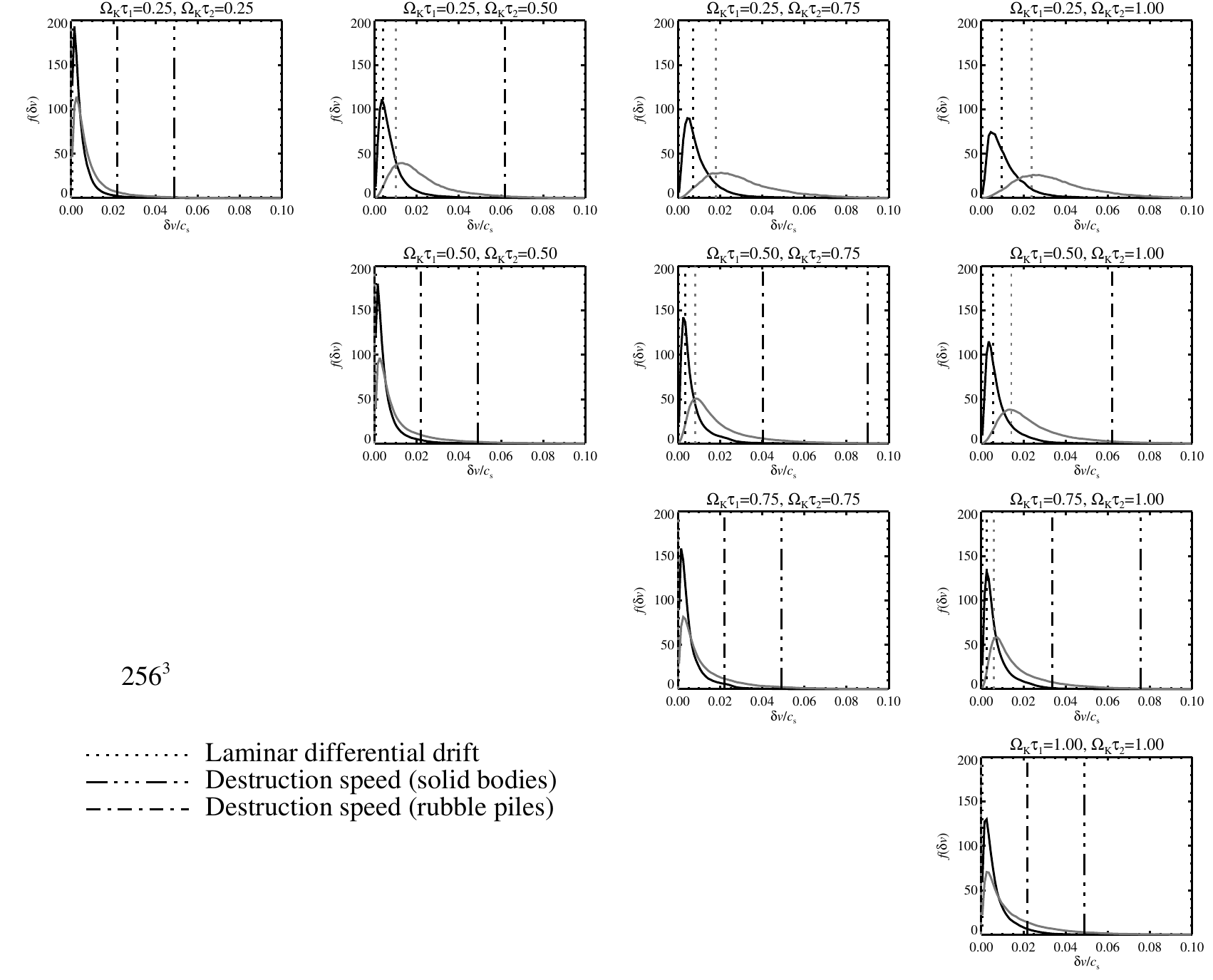}
  \end{center}
  \caption{Same as Fig.\ \ref{f:vcoll_128}, but for $256^3$ grid points with
    8,000,000 particles. Here we considered for simplicity only 1/8 of the
    immense amount of possible collisions. The collision speeds agree well with
    those of Fig.\ \ref{f:vcoll_128}.}
  \label{f:vcoll_256}
\end{figure}
We have considered collisions between all particles within the same grid cell 
by measuring the relative speed of every single pair of particles, giving
$(1/2)N(N-1)$ unique collisions in a grid cell with $N$ particles. We note that
this may underestimate the actual rate of fast collisions since in reality high
speed collisions would happen more frequently. We also note that grid points
with many particles get a high weight in the overall picture, and if these very
dense grid points have a lower collision speed due to feedback shielding from
the gas (see Figs.\ \ref{f:urms_z} and \ref{f:uk_z}), this will also
lead to a generally lower collision speed. All collisions were considered for
simplicity to be head on. The more realistic case of random impact parameters
would lead to a reduction of the measured collision speeds by more than 50\%.
We show for comparison the collision speeds measured at $t=20 T_{\rm orb}$ for
$64^3$ and $256^3$ simulations in Figs.\ \ref{f:vcoll_64}--\ref{f:vcoll_256}.
For the $64^3$ the Mach number of the gas is overall larger, ${\rm
Ma}\approx0.07$ compared to ${\rm Ma}\approx0.05$ at $128^3$, and collision
speeds are also larger than in Fig.\ \ref{f:vcoll_128}. There is very little
difference between the $128^3$ and the $256^3$ collision speeds, but the data
is based on only one snapshot. More statistically significant convergence
tests, given in \S\ref{s:vcoll}, show a slight increase in collision speeds
(relative to large scale rms speeds) with increasing resolution.

Equal-sized bodies have no differential drift, so turbulent motions dominate
the distribution of collision speeds, while differential drift (dotted lines in
Fig.\ \ref{f:vcoll_128}) dominates the distribution of collisions between
bodies of different sizes.

Following Benz\cite{Benz2000} we calculate the threshold for destructive
collisions. We use the specific incoming kinetic energy
\begin{equation}
  {\cal K} = (1/2) m v^2 / M \, ,
  \label{eq:Qimpact}
\end{equation}
where $v$ is the collision speed and $m$ and $M$ are the impactor and target
mass, respectively, with $M>m$. Collisional fragmentation occurs at a limit
${\cal K}^\star \approx 3\times10^6\,{\rm erg\,g^{-1}}$ for metre-sized solid
bodies and ${\cal K}^\star \approx 6 \times 10^5\,{\rm erg\,g^{-1}}$ for
metre-sized rubble piles\cite{Benz2000}.  This allows us to calculate the
destructive collision threshold from equation (\ref{eq:Qimpact}) for any pair
of bodies (see Fig.\ \ref{f:vcoll_128}).  We have ignored any shape effects that
could lead to differential drift of equal mass bodies\cite{Benz2000}.  More
collisions exceed the destruction threshold for equal-sized bodies than other
cases, but destructive collisions still only form a minority of the total,
although the fraction of destructive collisions depends on poorly known
material properties of the boulders. Note that our bodies are in fact smaller
than the 1 m considered by Benz, so the destruction threshold speeds may in
reality be higher since smaller bodies are expected to be able to withstand
higher collision speeds.  Higher radial pressure support does not produce
markedly more collisional fragmentation, as can be seen by comparing the grey
and dark lines in Fig.\ \ref{f:vcoll_128}. The collisions for which the relative
speed increases with the drift speed are between different-sized bodies, and
those collisions can withstand much higher relative speeds, whereas higher
radial pressure support does not give equal-sized bodies a higher collision
speed.  A third possibility for the internal makeup of the bodies is that they
are icy, porous structures. It is shown by Ryan et al.\cite{Ryan+etal1999} that
such objects are generally as strong as the silicates considered by
Benz\cite{Benz2000}. We ignore in the present analysis other effects of
collisions, such as erosion that may deplete boulders of mass even if the
collision speed is below the threshold for destruction.

\begin{figure}[!t]
  \begin{center}
    \includegraphics[width=15.0cm]{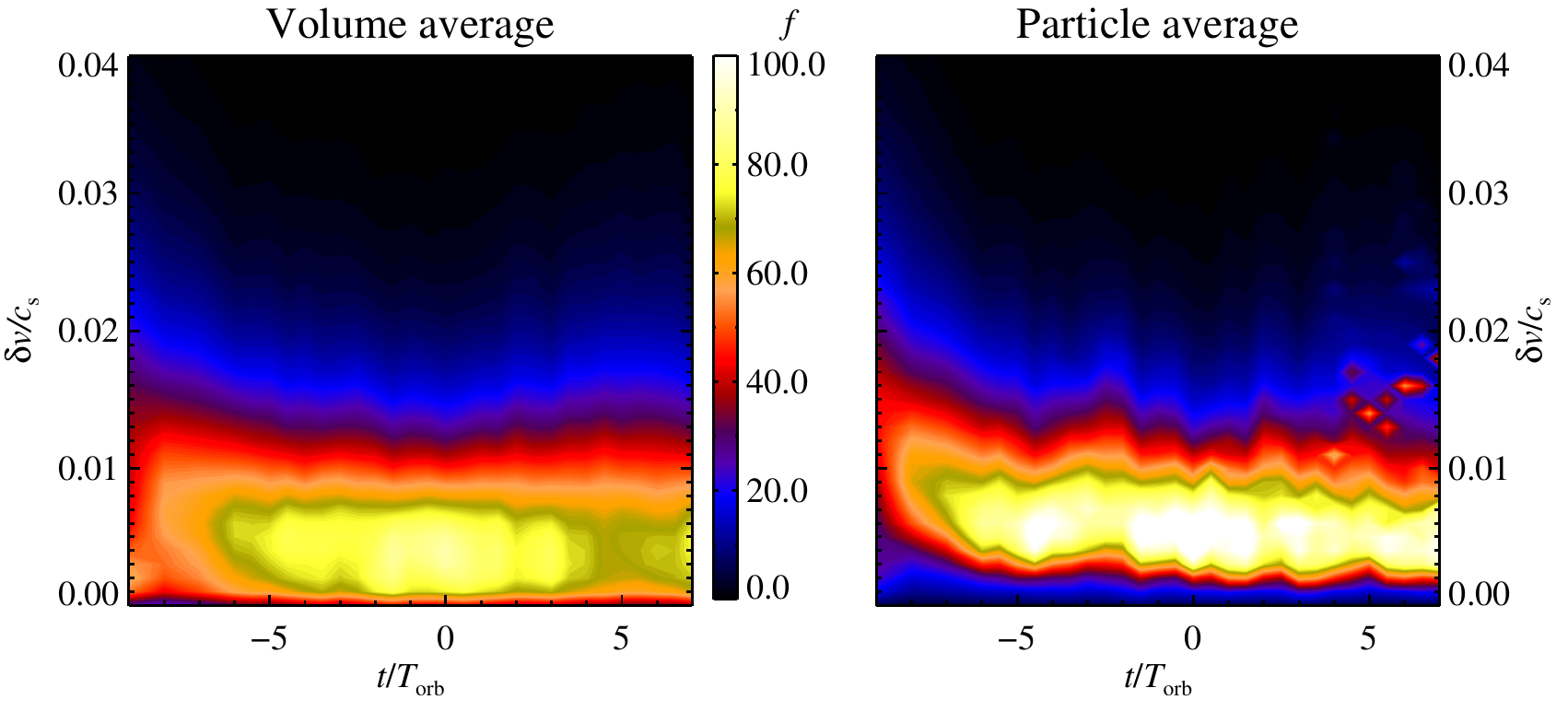}
  \end{center}
  \caption{The distribution of particle collision speeds (volume average, left
    plot) in units of sound speed, also shown weighted with number of particles
    in the grid cell (particle average, right plot), for $\Delta v=-0.02 c_{\rm
    s}$ and four different particle sizes $\varOmega_{\rm
    K}=0.25,0.50,0.75,1.00$. The distribution function $f$ is normalised such
    that $\int_0^\infty f(t,\delta v) \de (\delta v/c_{\rm s})=1$ for all $t$.
    Collisional cooling and self-gravity are turned on at $t=0$, halfway
    through these runs, with little effect on average collision speeds until
    late times. Collisional cooling is only efficient in the few regions where
    the solids-to-gas ratio is very high, so the average collision speeds are
    not much affected. Accretion onto the condensed out clusters is visible in
    the right panel at late times.}
  \label{f:collision_speeds}
\end{figure}

A quite serious issue related to collisional fragmentation is that it could
reduce the number of particles large enough to take part in gravitational
collapse prior to the collapse to as little as 20\% of the total mass of the
solids\cite{Weidenschilling1997,DullemondDominik2005}. This is particularly
relevant in the inner solar nebula where the sound speed is high. In that case
we speculate that an augmentation of the solids-to-gas ratio by up to a factor
5 would be needed for gravitational collapse to occur in the midplane layer.
Note that we already assume that half of the solid mass is present in small
grains too well-coupled to the gas to take place in the collapse (assuming that
the total abundance of solids and ices is $0.017$ times that of the
gas\cite{Hayashi1981}). A more detailed modelling of collisional fragmentation
is needed to quantify its actual effect on the gravitational collapse.

We show in Fig.\ \ref{f:collision_speeds} the time evolution of the average
collision speed of the particles for a radial pressure support of $\Delta
v=-0.02 c_{\rm s}$. The volume average collision speed distribution has one
speed associated with each grid cell whereas the particle average collision
speed distribution has the local collision speed weighted by the number of
particles in the grid cell.  Collisional cooling, turned on together with
self-gravity at $t=0$ in Fig.\ \ref{f:collision_speeds}, does not reduce
collision speeds broadly, since only regions with very high solids-to-gas
ratios have sufficiently high collision rates for cooling to be effective.
Accretion with high relative speeds onto condensed clusters is visible at late
times in the particle average distribution. Some of the high-speed collisions
that occur during the accretion phase, especially among boulders that have
already been deformed in earlier collision events, may in reality lead to
erosion or destruction of the colliding bodies\cite{Benz2000}.  However, the
fragments remain within the Hill sphere of the gravitationally bound cluster,
so their dynamics is dominated by the gravity of the nearby cluster rather than
by the tidal force of the protostar.  Furthermore, the mean free path of the
fragments, $\ell/H \approx \varOmega_{\rm K} \tau_{\rm f}/(\rho_{\rm
p}/\rho_{\rm g}) \sim 10^{-3}$, is shorter than the size of the bound clusters,
so they will undergo collisions with the remaining boulders, leading only to
the further growth of the remaining boulders as they sweep up the fragments.
All in all collisional destruction of boulders (or, rather, the lack of
widespread collisional destruction) is important for obtaining an initial
condition with a sufficient amount of the solid material bound in boulders,
whereas collisional destruction during the gravitational collapse phase is
probably only of secondary importance.

\subsubsection{Comparison of rms and collision speeds with analytic values}
\label{s:vcoll}

\begin{table}
  \begin{center}
    Without feedback ($128^3$):
    \vspace{0.3cm}
    \begin{tabular}{cccccccccc}
      \hline
      \hline
      $\varOmega_{\rm K} \tau_{\rm f}$ &
      $\sigma^{\rm (p)}_x$ & $\sigma^{\rm (p)}_y$ & $\sigma^{\rm (p)}_z$&
      $\sigma^{\rm (g)}_x$ & $\sigma^{\rm (g)}_y$ & $\sigma^{\rm (g)}_z$&
      $\sigma^{\rm (p)}_x/\sigma^{\rm (g)}_x$ &
      $\sigma^{\rm (p)}_y/\sigma^{\rm (g)}_y$ &
      $\sigma^{\rm (p)}_z/\sigma^{\rm (g)}_z$ \\
      \hline
      $0.20$ & $0.0307$ & $0.0230$ & $0.0205$ & $0.0317$ & $0.0294$ & $0.0240$ &
               $0.9686$ & $0.7844$ & $0.8515$ \\
      $0.50$ & $0.0313$ & $0.0191$ & $0.0172$ & $0.0321$ & $0.0300$ & $0.0244$ &
               $0.9762$ & $0.6367$ & $0.7049$ \\
      $1.00$ & $0.0305$ & $0.0149$ & $0.0138$ & $0.0326$ & $0.0300$ & $0.0246$ &
               $0.9353$ & $0.4967$ & $0.5599$ \\
      $2.00$ & $0.0271$ & $0.0120$ & $0.0113$ & $0.0331$ & $0.0312$ & $0.0251$ &
               $0.8192$ & $0.3834$ & $0.4510$ \\
      $5.00$ & $0.0186$ & $0.0082$ & $0.0074$ & $0.0329$ & $0.0311$ & $0.0249$ &
               $0.5644$ & $0.2648$ & $0.2961$ \\
      \hline
    \end{tabular}
    Without feedback ($64^3$):
    \vspace{0.3cm}
    \begin{tabular}{cccccccccc}
      \hline
      \hline
      $\varOmega_{\rm K} \tau_{\rm f}$ &
      $\sigma^{\rm (p)}_x$ & $\sigma^{\rm (p)}_y$ & $\sigma^{\rm (p)}_z$&
      $\sigma^{\rm (g)}_x$ & $\sigma^{\rm (g)}_y$ & $\sigma^{\rm (g)}_z$&
      $\sigma^{\rm (p)}_x/\sigma^{\rm (g)}_x$ &
      $\sigma^{\rm (p)}_y/\sigma^{\rm (g)}_y$ &
      $\sigma^{\rm (p)}_z/\sigma^{\rm (g)}_z$ \\
      \hline
      $0.20$ & $0.0428$ & $0.0322$ & $0.0289$ & $0.0448$ & $0.0422$ & $0.0344$ &
               $0.9554$ & $0.7628$ & $0.8394$ \\
      $0.50$ & $0.0429$ & $0.0263$ & $0.0248$ & $0.0448$ & $0.0422$ & $0.0344$ &
               $0.9573$ & $0.6234$ & $0.7201$ \\
      $1.00$ & $0.0404$ & $0.0207$ & $0.0198$ & $0.0448$ & $0.0422$ & $0.0344$ &
               $0.9019$ & $0.4909$ & $0.5764$ \\
      $2.00$ & $0.0340$ & $0.0161$ & $0.0152$ & $0.0448$ & $0.0422$ & $0.0344$ &
               $0.7590$ & $0.3815$ & $0.4408$ \\
      $5.00$ & $0.0237$ & $0.0115$ & $0.0104$ & $0.0448$ & $0.0422$ & $0.0344$ &
               $0.5283$ & $0.2731$ & $0.3025$ \\
      \hline
    \end{tabular}
    With feedback ($128^3$):
    \vspace{0.3cm}
    \begin{tabular}{cccccccccc}
      \hline
      \hline
      $\varOmega_{\rm K} \tau_{\rm f}$ &
      $\sigma^{\rm (p)}_x$ & $\sigma^{\rm (p)}_y$ & $\sigma^{\rm (p)}_z$&
      $\sigma^{\rm (g)}_x$ & $\sigma^{\rm (g)}_y$ & $\sigma^{\rm (g)}_z$&
      $\sigma^{\rm (p)}_x/\sigma^{\rm (g)}_x$ &
      $\sigma^{\rm (p)}_y/\sigma^{\rm (g)}_y$ &
      $\sigma^{\rm (p)}_z/\sigma^{\rm (g)}_z$ \\
      \hline
      $0.20$ & $0.0276$ & $0.0194$ & $0.0172$ & $0.0318$ & $0.0307$ & $0.0244$ &
               $0.8655$ & $0.6315$ & $0.7046$ \\
      $0.50$ & $0.0279$ & $0.0166$ & $0.0150$ & $0.0341$ & $0.0341$ & $0.0263$ &
               $0.8199$ & $0.4858$ & $0.5698$ \\
      $1.00$ & $0.0220$ & $0.0111$ & $0.0106$ & $0.0337$ & $0.0331$ & $0.0257$ &
               $0.6520$ & $0.3350$ & $0.4105$ \\
      \hline
    \end{tabular}
    With feedback ($64^3$):
    \begin{tabular}{cccccccccc}
      \hline
      \hline
      $\varOmega_{\rm K} \tau_{\rm f}$ &
      $\sigma^{\rm (p)}_x$ & $\sigma^{\rm (p)}_y$ & $\sigma^{\rm (p)}_z$&
      $\sigma^{\rm (g)}_x$ & $\sigma^{\rm (g)}_y$ & $\sigma^{\rm (g)}_z$&
      $\sigma^{\rm (p)}_x/\sigma^{\rm (g)}_x$ &
      $\sigma^{\rm (p)}_y/\sigma^{\rm (g)}_y$ &
      $\sigma^{\rm (p)}_z/\sigma^{\rm (g)}_z$ \\
      \hline
      $0.20$ & $0.0423$ & $0.0291$ & $0.0265$ & $0.0462$ & $0.0434$ & $0.0355$ &
               $0.9147$ & $0.6719$ & $0.7450$ \\
      $0.50$ & $0.0372$ & $0.0224$ & $0.0217$ & $0.0479$ & $0.0459$ & $0.0363$ &
               $0.7769$ & $0.4881$ & $0.5991$ \\
      $1.00$ & $0.0388$ & $0.0182$ & $0.0184$ & $0.0465$ & $0.0432$ & $0.0358$ &
               $0.8335$ & $0.4218$ & $0.5134$ \\
      \hline
    \end{tabular}
  \end{center}
  \caption{Particle and gas rms speeds and their ratios, for different values
    of the friction time. The top part of the table shows results for passive
    particles (i.e.\ with no friction force on the gas), whereas the bottom
    part shows the result when gas is allowed to feel friction.  The gas Mach
    number is around ${\rm Ma}\approx0.05$ for $128^3$ grid points, while ${\rm
    Ma}\approx0.07$ for $64^3$ grid points.  Small particles show the expected
    trend towards equal particle and gas rms speeds (the ratios are shown in
    the last three columns). The azimuthal particle rms speed falls quickly
    with increasing friction time as the loosely coupled particles go on
    elliptic orbits, and the ratio between radial and azimuthal particle rms
    speed approaches the expected value of 0.5. Particle rms speeds fall by a
    significant factor when gas is allowed to feel the friction from the
    particles -- for $\varOmega_{\rm K} \tau_{\rm f}=1.0$ the reduction is
    around 25\% when adding the components quadratically.}
  \label{t:vrms}
\end{table}

We now aim to explain why the relative (i.e.\ collisional) speeds are only
$\delta v \approx 0.01 c_{\rm s}$ for marginally coupled particles despite
turbulent motions of Mach number $\mathcal{M} = u_{\rm t}/c_{\rm s} =  0.05$.
We summarise the four main effects: (1) due to finite friction times particles
are not fully accelerated by the turbulence, so rms particle speeds are reduced
below that of the gas; (2) particle drag on gas tends to reduce both random and
relative speeds; (3) particles are well enough coupled to eddies that collision
speeds are slower than rms speeds due to entrainment by the same extended
eddies; and (4) even though we have a fairly long inertial range in the $256^3$
simulations, particles gain relative speed from scales near the onset of the
dissipative subrange, so it is not certain that all the relevant scales are
resolved. We present extensive documentation of effects (1) and (2) in this
section, while point (4) is given some support by a 10-20\% increase of
collision speeds relative to rms speeds when going from $64^3$ to $128^3$ grid
points (see Figure \ref{f:vcoll_over_vrms}).

We emphasise that there is no complete theory describing all these effects, so
the current numerical study deepens our understanding of these issues. Notably
an analytical theory for collision speeds that includes orbital dynamics is, to
our knowledge, missing (even without drag of particles on gas), so verification
of point (3) above is difficult and the point is left as speculation in this
work. We also emphasise that our results show that the collision speed
increases slightly with increasing resolution and that higher resolution
studies will be needed to see if the collision speeds have converged. The
important issue of collision speeds of marginally coupled solids will need more
consideration in the future.

First a word on nomenclature and on calculation procedure:
\begin{itemize}
  \item With {\it rms speeds} we refer to the overall root-mean-square speed of
    either gas or particles. This value is directly obtained from calculating
    the root-mean-square of all values of a chosen component of gas or particle
    velocity.
  \item {\it Collision speed} refers to the local velocity differences
    of particles within a single grid cell (similar to the sound speed or mean
    thermal speed of the gas). We calculate the mean collision speed of
    particles by following 1000 particles from the greater ensemble and
    measuring the velocity components of each particle relative to a random
    other particle in its grid cell, calculating in the end the rms of each
    velocity component from the distribution of relative velocities. We make
    sure that the particle has a neighbour by excluding grid points with
    only 1 particle. We also exclude a neighbour particle if it is one of
    the 1000 chosen particles. This method is computationally much cheaper
    than the full tabulation of all collisions used in Fig.\ \ref{f:vcoll_128}.
\end{itemize}
For all measurements presented in Tables \ref{t:vrms} and \ref{t:vrms2} we have
averaged over snapshots taken equidistantly 10 orbits apart, starting at $t=20
T_{\rm orb}$ where the sedimentary mid-plane layer has already had 10 orbits to
form in balance with the turbulent diffusion.

\begin{figure}[!t]
  \begin{center}
    \includegraphics[width=0.495\linewidth]{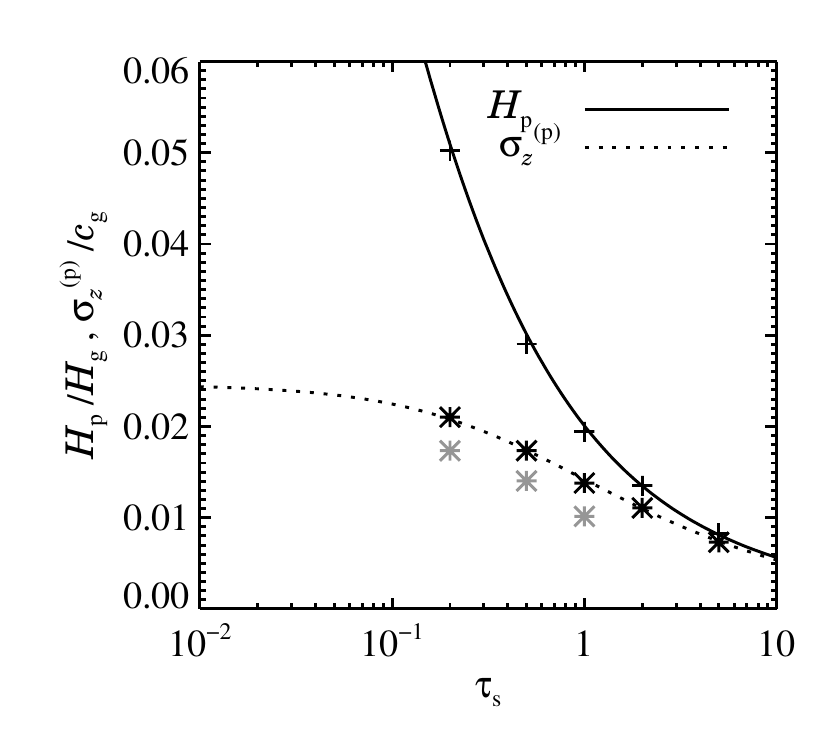}
    \includegraphics[width=0.495\linewidth]{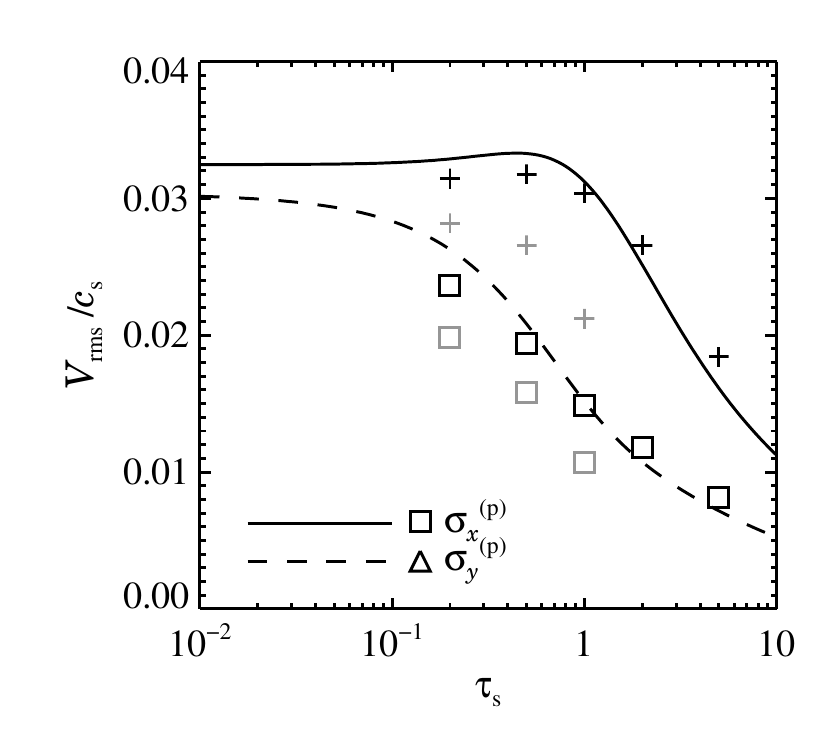}
  \end{center}
  \caption{Comparison of simulated particle rms speeds and scale-heights to a
    simple analytic model of turbulent forcing that includes orbital dynamics,
    but no feedback\cite{YoudinLithwick2007}.  Symbols are the measured rms
    particle speeds from single particle size simulations $\tau_{\rm s} =
    \varOmega_{\rm K} \tau_{\rm f} = 0.2, 0.5, 1.0, 2.0, 5.0$ ($128^3$ grid
    points and 1,000,000 particles) either without ({\it black symbols}) or
    with ({\it grey symbols}) particle feedback. {\it Left plot:} Particle
    scale-height ({\it crosses}) and vertical rms speeds ({\it asterisks})
    compared to analytical curves which use the measured vertical rms gas
    speed, $u_z = 0.024 c_{\rm s}$ and an eddy time, $t_{\rm eddy} = 1.0
    \varOmega_{\rm K}^{-1}$ as input. {\it Right plot:} Measured radial ({\it
    crosses}) and azimuthal ({\it squares}) particle rms speeds compared to
    analytic curves using the measured gas velocities (radial, $u_x = 0.032
    c_{\rm s}$; azimuthal $u_y = 0.030 c_{\rm s}$; and correlated $\langle u_x
    u_y \rangle^{1/2} = 0.016 c_{\rm s}$) and $t_{\rm eddy} = 1.0
    \varOmega_{\rm K}^{-1}$ as input. See text for further discussion.}
  \label{f:rmsfit}
\end{figure}
We show in Table~\ref{t:vrms} the rms speeds of gas $\sigma^{\rm (g)}$ and
particles $\sigma^{\rm (p)}$ for different values of the friction time (from
$128^3$ simulations with single-sized particles). The top part shows the
results of simulations where drag force from particles on gas was not applied,
whereas this back-reaction drag force is included in the bottom part of
Table~\ref{t:vrms}. The turbulent rms speed of the gas has in all cases a Mach
number of around ${\rm Ma} = u_{\rm t}/c_{\rm s}=0.05$ for $128^3$ grid points
(seen by adding up the three components quadratically), while ${\rm Ma}=0.07$
for $64^3$ grid points. The smallest particles ($\varOmega_{\rm K} \tau_{\rm
f}=0.2$) have comparable rms speeds, but the rms speeds fall with stopping time
as expected (see  Fig.\ \ref{f:rmsfit} and explanation below).

To understand the simulated rms speeds of particles in MRI turbulence, we
compare to Youdin \& Lithwick (hereafter YL)\cite{YoudinLithwick2007}.   YL
consider the rms response of particles to stochastic turbulent forcing (with a
temporal Kolmogorov spectrum) with orbital dynamics -- in particular epicyclic
motions -- included, which previous works have ignored . The inputs for the
simple analytic theory include the measured rms speeds of the turbulence
(radial, azimuthal, vertical and the radial-azimuthal correlations), the
friction time of the particles, and the eddy time, i.e.\ the correlation time
of turbulent velocity fluctuations. Since the eddy time is a measured quantity,
we treat it as a free parameter and find the reasonable value $t_{\rm eddy}
\approx \varOmega_{\rm K}^{-1}$.
\begin{figure}[!t]
  \begin{center}
    \includegraphics[width=\linewidth]{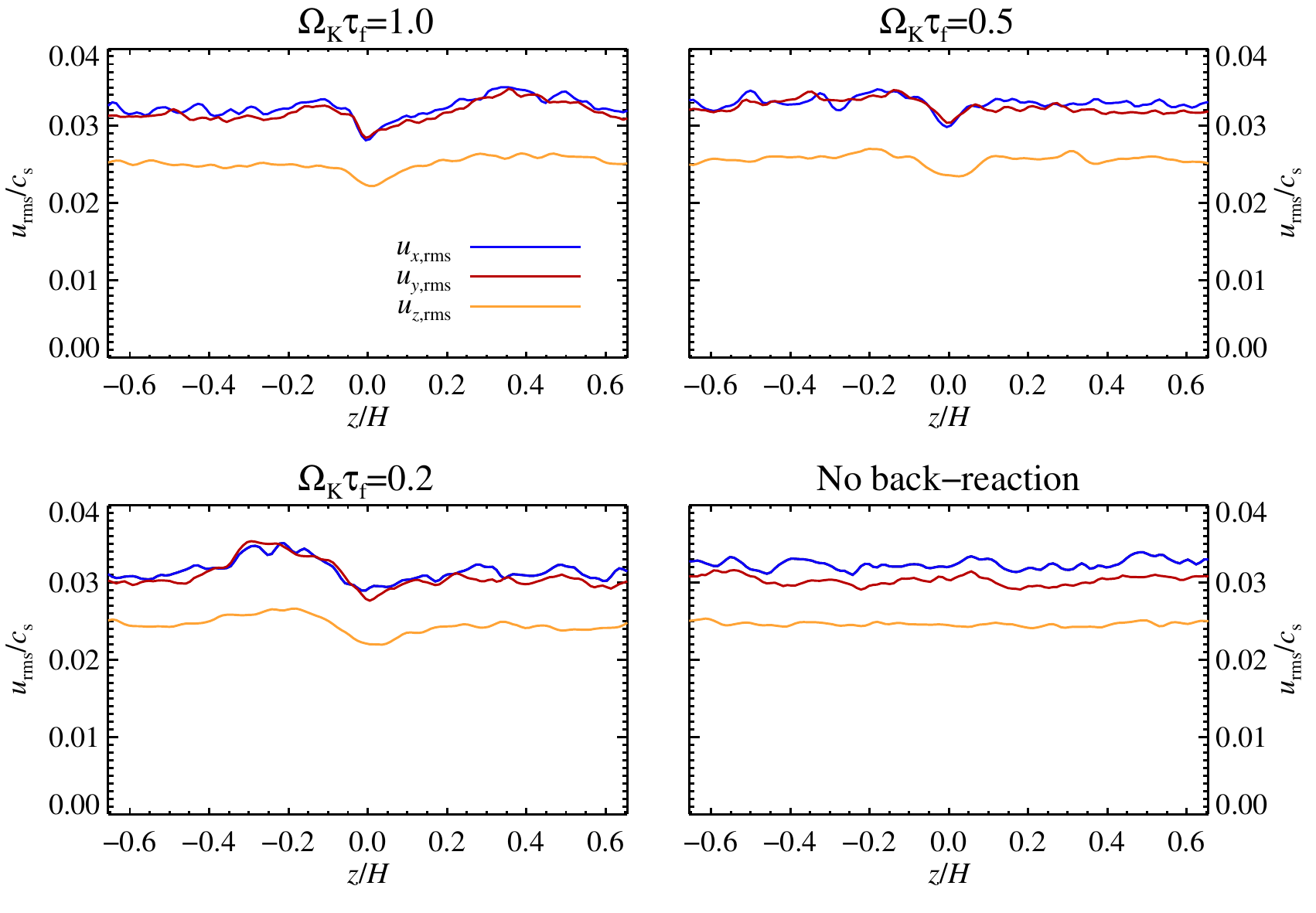}
  \end{center}
  \caption{Gas rms speed as a function of height over the mid-plane in
    simulations with $128^3$ grid points and 1,000,000 particles. At each
    $z$-value we have calculated the dispersion of the three velocity
    components.  There is a clear dip in the gas velocity dispersion around the
    mid-plane where the solids-to-gas ratio is high. This dip is not present in
    the run with no back-reaction friction force (lower right panel).}
  \label{f:urms_z}
\end{figure}

The left panel of Fig.\ \ref{f:rmsfit} shows that vertical speeds and scale
heights of particles agree very well with the analytic theory of YL. Since
vertical particle motions are decoupled from in-plane motions, the analytic
description is fairly simple.  When feedback is included ({\it grey symbols})
the random particle velocities drop, by around 25\% for $\tau_{\rm s} = 1.0$
particles.  This is not due to a decrease in the total turbulent Mach number
(which actually goes up slightly), but likely comes from the increased particle
inertia in dense clumps.

\begin{figure}[!t]
  \begin{center}
    \includegraphics[width=\linewidth]{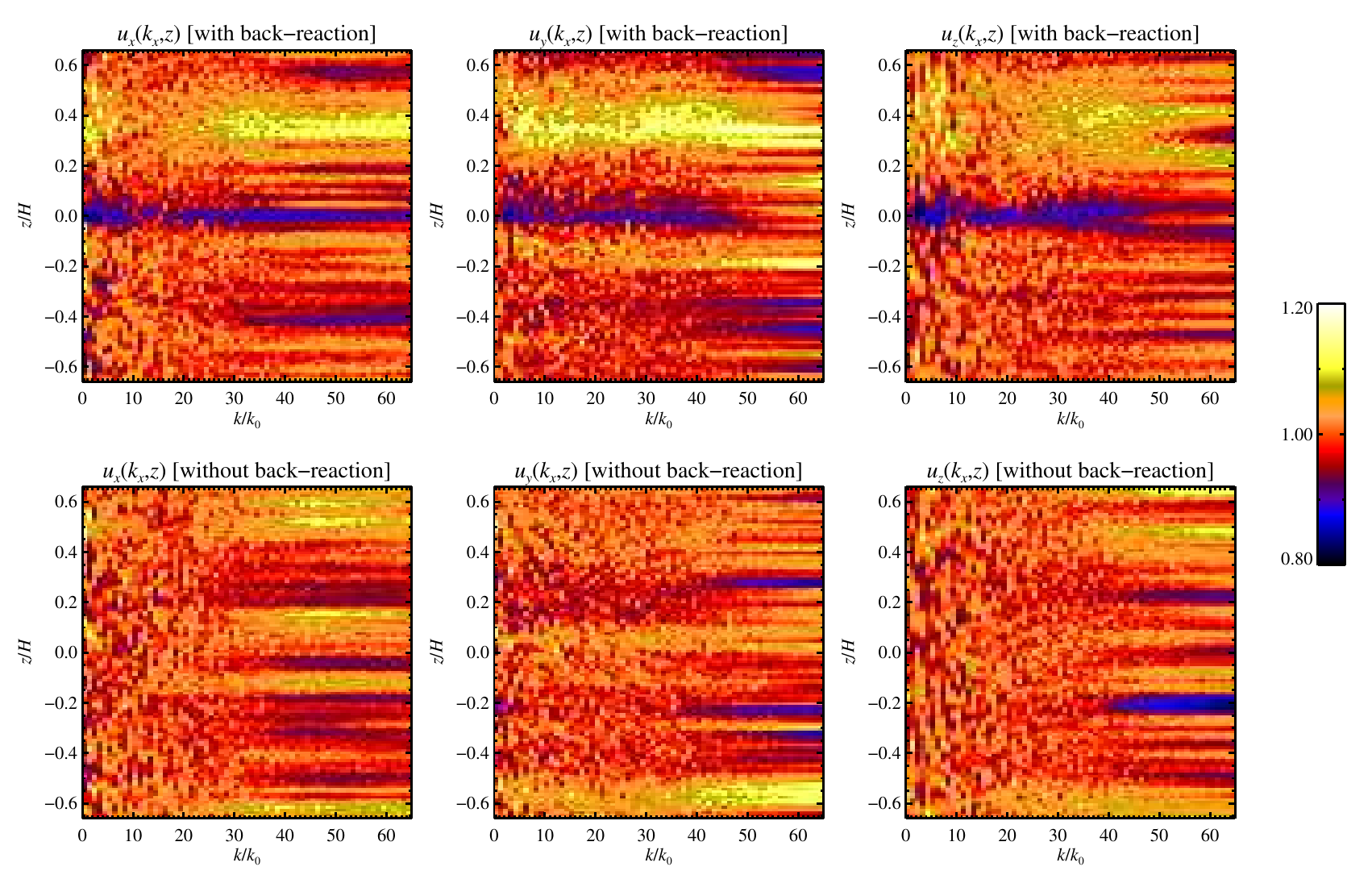}
  \end{center}
  \caption{Normalised Fourier amplitude of the gas velocity as a function of
    wavenumber $k$ (in units if the largest scale in the box $k_0$) and height
    over the mid-plane $z$. Shown here is a simulation with single-sized
    $\varOmega_{\rm K} \tau_{\rm f}=1$ particles (no self-gravity or
    collisional cooling). We have done 1-D Fourier transforms along the
    $x$-direction and averaged this over all values of azimuthal coordinate $y$
    and time $t$ for each given height $z$. The power spectrum has been
    normalised with the average value at each wavenumber $k$. The colours
    represent values of $\pm 20\%$. Feedback drag force from particles on the
    gas reduces the velocity amplitude in the mid-plane by around 20\%,
    explaining a similar reduction in rms speeds visible in Table \ref{t:vrms}.}
  \label{f:uk_z}
\end{figure}
The right panel of Fig.\ \ref{f:rmsfit} shows in-plane particle velocities
compared to the YL model. The agreement is good, if not as perfect as the
vertical case, due to more complex dynamics.  Notably, the prediction that
azimuthal velocities fall before radial speeds (as $\tau_{\rm s}$ increases)
due to epicyclic motion is confirmed.   
Drag feedback ({\it grey symbols}) again produces a decrease in rms speeds, by
about 30\% in the radial speeds for $\tau_{\rm s} = 1.0$.

It may be surprising that the eddy time-scale is so short, $t_{\rm
eddy}\approx1 \varOmega_{\rm K}^{-1}$ (a similar result for the eddy time was
found recently by Johansen, Klahr, \& Mee\cite{JohansenKlahrMee2006} by
considering the relation between turbulent diffusion and gas rms speeds).
Simple mixing length estimates based on Fig.\ \ref{f:power} yield an eddy time
at least an order of magnitude higher than our best fit value, i.e.\ $t_k =
1/(k u _k) \sim 100 \varOmega_{\rm K}^{-1}$ on the largest scales. But the
large scales of the flow are dominated by Coriolis force and centrifugal terms
rather than non-linear advection, and thus the correlation time of these
important structures is forced to be similar to the Keplerian
frequency\cite{Shakura+etal1978,Cuzzi+etal2001}. Mixing length theory gives a
good estimate of the life times of large scale structures, but their
correlation time, which is the important quantity for mixing purposes, is never
longer than a Keplerian shear time $\varOmega_{\rm K}^{-1}$.

\begin{figure}[!t]
  \begin{center}
    \includegraphics[width=0.5\linewidth]{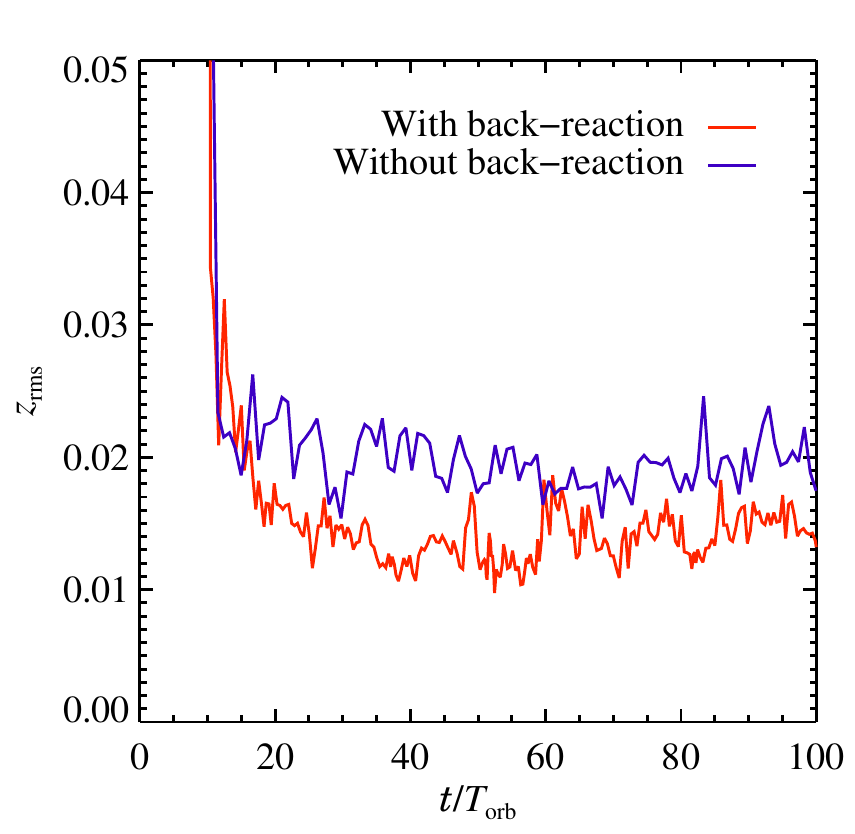}
  \end{center}
  \caption{The root-mean-square of the particle $z$-coordinate as a function of
    time, with and without feedback, for a $128^3$ simulation with single-sized
    $\varOmega_{\rm K} \tau_{\rm f}=1$ particles. The scale height of the
    particles falls by around 20\% when feedback is applied, in accordance with
    Fig.\ \ref{f:uk_z} that shows that feedback damps MRI turbulence by around
    20\% in the sedimented mid-plane layer (the reduction in scale height is
    also seen in the left panel of Fig.\ \ref{f:rmsfit}).}
  \label{f:zprms_t}
\end{figure}

\begin{figure}[!t]
  \begin{center}
    \includegraphics[width=0.5\linewidth]{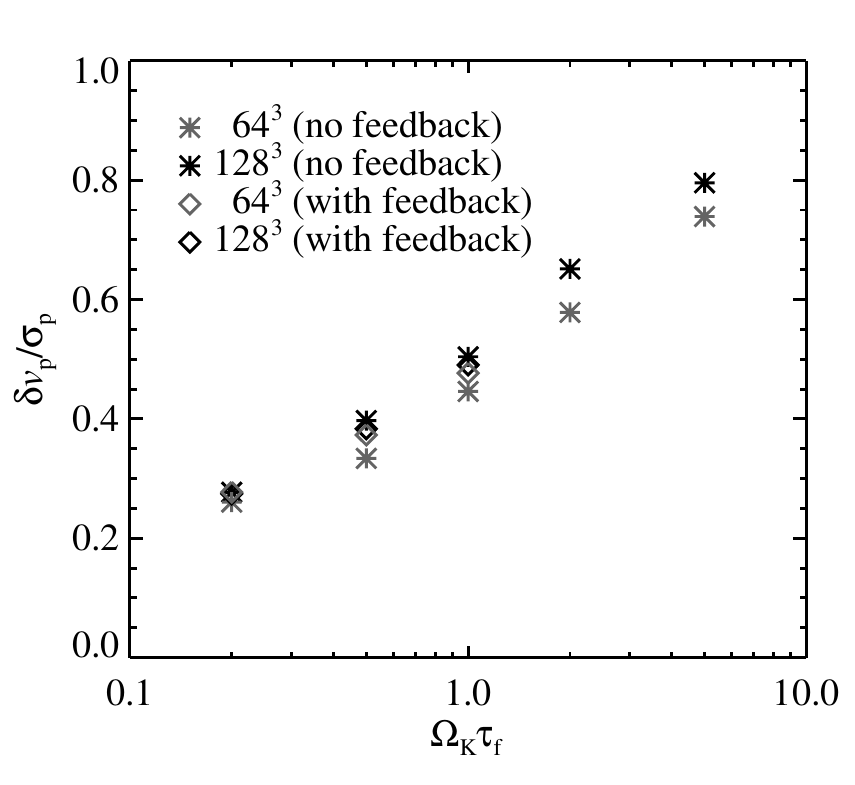}
  \end{center}
  \caption{The ratio of particle collision speed $\delta v_{\rm p}$ to large
    scale particle rms speed $\sigma_{\rm p}$, both calculated as quadratic
    sums of the directional components given in Table \ref{t:vrms2} and shown
    for two resolutions and with and without particle feedback on the gas. The
    collision speeds increase as expected with increasing friction time as
    particles decouple from fast large scale eddies. Collision speeds increase
    by 10-20\% when increasing resolution from $64^3$ to $128^3$, likely
    because the inertial range of the gas turbulence expands to smaller scales.
    Feedback has only little influence on the ratio of collision speed to rms
    speed, but the overall rms speed of particles is reduced by feedback (see
    Fig.\ \ref{f:rmsfit}). Higher resolution studies will be needed to
    determine whether $\varOmega_{\rm K}\tau_{\rm f}=1$ particles will
    eventually obtain the fully mixed $\delta v_{\rm p}/\sigma_{\rm
    p}=\sqrt{2}$.}
  \label{f:vcoll_over_vrms}
\end{figure}

The net effect of decoupling and feedback gives a total (radial, azimuthal and
vertical) rms speed of $0.027 c_{\rm s}$ for $\tau_{\rm s} = 1.0$, nearly half
the gas rms speed of $v_t = 0.05 c_{\rm s}$.  Regions of high particle density
damp out the gas velocity on the time-scale $\tau_{\rm g}=\tau_{\rm
f}/(1+\epsilon)$ where $\tau_{\rm f}$ is the friction time of the particles and
$\epsilon$ is the solids-to-gas ratio. It was shown by Dobrovolskis et
al.\cite{Dobrovolskis+etal1999} that this damping can lead to a minor reduction
of turbulence around the mid-plane layer. We show in Fig.\ \ref{f:urms_z} the
gas velocity dispersion as a function of height over the mid-plane. There is a
clear, though rather modest, dip around the mid-plane where the solids-to-gas
ratio is also high.  The dip is only visible in simulations that include the
back-reaction friction force. In Fig.\ \ref{f:uk_z} we show the magnitude of
the different Fourier components of the gas velocity as a function of height
over the mid-plane. Here we have Fourier-transformed along the $x$-direction
and averaged over every single $y$ coordinate at each height $z$. The spectrum
has afterwards been normalised with the average spectrum so that differences
appear more clearly. There is again a clear dip, of around 20\%, in the
strength of the velocity fluctuations around the mid-plane. The dip occurs at
all scales, although the small scales of $u_y$ are less affected than the large
scales. In Fig.\ \ref{f:zprms_t} we show the particle rms $z$-coordinate as a
function of time. With particle feedback on the gas there is a drop by around
20\% in the scale height of the particles, in agreement with what is expected
for a $1-0.2^2=36\%$ drop in the diffusion coefficient\cite{Dubrulle+etal1995}.

\begin{table}
  \begin{center}
    Without feedback ($128^3$):
    \vspace{0.3cm}
    \begin{tabular}{cccccccccc}
      \hline
      \hline
      $\varOmega_{\rm K} \tau_{\rm f}$ &
      $\sigma^{\rm (p)}_x$ & $\sigma^{\rm (p)}_y$ & $\sigma^{\rm (p)}_z$ &
      $\delta v _x$ & $\delta v_y$ & $\delta v_z$ &
      $\delta v_x/\sigma^{\rm (p)}_x$ & $\delta v_y/\sigma^{\rm (p)}_y$ &
      $\delta v_z/\sigma^{\rm (p)}_z$ \\
      \hline
      $0.20$ & $0.0307$ & $0.0230$ & $0.0205$ & $0.0077$ & $0.0078$ & $0.0050$ &
               $0.2503$ & $0.3363$ & $0.2432$ \\
      $0.50$ & $0.0313$ & $0.0191$ & $0.0172$ & $0.0126$ & $0.0076$ & $0.0065$ &
               $0.4020$ & $0.3942$ & $0.3767$ \\
      $1.00$ & $0.0305$ & $0.0149$ & $0.0138$ & $0.0154$ & $0.0068$ & $0.0076$ &
               $0.5018$ & $0.4557$ & $0.5491$ \\
      $2.00$ & $0.0271$ & $0.0120$ & $0.0113$ & $0.0170$ & $0.0073$ & $0.0092$ &
               $0.6225$ & $0.6101$ & $0.8098$ \\
      $5.00$ & $0.0186$ & $0.0082$ & $0.0074$ & $0.0143$ & $0.0062$ & $0.0073$ &
               $0.7634$ & $0.7549$ & $0.9836$ \\
      \hline
    \end{tabular}
    Without feedback ($64^3$):
    \vspace{0.3cm}
    \begin{tabular}{cccccccccc}
      \hline
      \hline
      $\varOmega_{\rm K} \tau_{\rm f}$ &
      $\sigma^{\rm (p)}_x$ & $\sigma^{\rm (p)}_y$ & $\sigma^{\rm (p)}_z$ &
      $\delta v _x$ & $\delta v_y$ & $\delta v_z$ &
      $\delta v_x/\sigma^{\rm (p)}_x$ & $\delta v_y/\sigma^{\rm (p)}_y$ &
      $\delta v_z/\sigma^{\rm (p)}_z$ \\
      \hline

      $0.20$ & $0.0428$ & $0.0322$ & $0.0289$ & $0.0105$ & $0.0093$ & $0.0074$ &
               $0.2461$ & $0.2907$ & $0.2588$ \\
      $0.50$ & $0.0429$ & $0.0263$ & $0.0248$ & $0.0147$ & $0.0079$ & $0.0085$ &
               $0.3432$ & $0.3013$ & $0.3462$ \\
      $1.00$ & $0.0404$ & $0.0207$ & $0.0198$ & $0.0183$ & $0.0078$ & $0.0096$ &
               $0.4533$ & $0.3814$ & $0.4879$ \\
      $2.00$ & $0.0340$ & $0.0161$ & $0.0152$ & $0.0191$ & $0.0082$ & $0.0109$ &
               $0.5572$ & $0.5099$ & $0.7199$ \\
      $5.00$ & $0.0237$ & $0.0115$ & $0.0104$ & $0.0170$ & $0.0078$ & $0.0094$ &
               $0.7163$ & $0.6847$ & $0.9072$ \\

      \hline
    \end{tabular}
    With feedback ($128^3$):
    \vspace{0.3cm}
    \begin{tabular}{cccccccccc}
      \hline
      \hline
      $\varOmega_{\rm K} \tau_{\rm f}$ &
      $\sigma^{\rm (p)}_x$ & $\sigma^{\rm (p)}_y$ & $\sigma^{\rm (p)}_z$ &
      $\delta v _x$ & $\delta v_y$ & $\delta v_z$ &
      $\delta v_x/\sigma^{\rm (p)}_x$ & $\delta v_y/\sigma^{\rm (p)}_y$ &
      $\delta v_z/\sigma^{\rm (p)}_z$ \\
      \hline
      $0.20$ & $0.0276$ & $0.0194$ & $0.0172$ & $0.0068$ & $0.0065$ & $0.0044$ &
               $0.2463$ & $0.3376$ & $0.2555$ \\
      $0.50$ & $0.0279$ & $0.0166$ & $0.0150$ & $0.0108$ & $0.0065$ & $0.0054$ &
               $0.3845$ & $0.3855$ & $0.3569$ \\
      $1.00$ & $0.0220$ & $0.0111$ & $0.0106$ & $0.0109$ & $0.0049$ & $0.0055$ &
               $0.4982$ & $0.4501$ & $0.5224$ \\
      \hline
    \end{tabular}
    With feedback ($64^3$):
    \vspace{0.3cm}
    \begin{tabular}{cccccccccc}
      \hline
      \hline
      $\varOmega_{\rm K} \tau_{\rm f}$ &
      $\sigma^{\rm (p)}_x$ & $\sigma^{\rm (p)}_y$ & $\sigma^{\rm (p)}_z$ &
      $\delta v _x$ & $\delta v_y$ & $\delta v_z$ &
      $\delta v_x/\sigma^{\rm (p)}_x$ & $\delta v_y/\sigma^{\rm (p)}_y$ &
      $\delta v_z/\sigma^{\rm (p)}_z$ \\
      \hline

      $0.20$ & $0.0423$ & $0.0291$ & $0.0265$ & $0.0108$ & $0.0094$ & $0.0072$ &
               $0.2579$ & $0.3224$ & $0.2715$ \\
      $0.50$ & $0.0372$ & $0.0224$ & $0.0217$ & $0.0144$ & $0.0082$ & $0.0073$ &
               $0.3892$ & $0.3622$ & $0.3364$ \\
      $1.00$ & $0.0388$ & $0.0182$ & $0.0184$ & $0.0187$ & $0.0082$ & $0.0088$ &
               $0.4855$ & $0.4361$ & $0.4798$ \\

      \hline
    \end{tabular}
  \end{center}
  \caption{Particle rms speeds ($\sigma^{\rm (p)}_{x|y|z}$), collision speeds
    ($\delta v_{x|y|z}$) and their ratios. The top part of the table shows
    results without drag force on the gas, the bottom part includes drag force
    on the gas.  The collision speeds are very low for small particles, but the
    ratio between collision speed and rms speed increases with increasing
    friction time as the equal-sized marginally coupled particles are big
    enough to have significantly different histories when they meet.
    Back-reaction friction force on the gas has very little influence on the
    ratio of collision speeds to rms speeds. There is a 10-20\% increase in the
    ratio of collision to rms speed when going from $64^3$ to $128^3$ grid
    points, possibly due an expansion of the inertial range of the turbulence.
    Higher resolution studies will be needed to determine whether this trend
    continues of whether the collision speeds have already converged.}
  \label{t:vrms2}
\end{table}

\subsubsection{Why are collision speeds so low?}

Table~\ref{t:vrms2} shows the collision speed $\delta v$ of the particles for
comparison with the particle rms speeds. Here we have chosen a thousand
particles throughout the simulation and calculated their relative speed with a
random different particle in the same grid cell. The collision speed of
marginally coupled particles (with $\varOmega_{\rm K} \tau_{\rm f}=1$) is
around 50\% of the overall rms speed.  This factor 2 decrease in the relative
speeds on small scales compared to rms speeds remains to be explained. An exact
analytic prediction of this effect is difficult, and has not to our knowledge
been done with either orbital dynamics or feedback, let alone both effects.
Higher resolution simulations will be needed to determine if small scale eddies
can change this result. Fig.\ \ref{f:power} indicates that over an order of
magnitude in the inertial range is resolved, but scales around the onset of the
dissipative subrange may be important for collision speeds.
\begin{figure}[!t]
  \begin{center}
    \includegraphics[width=\linewidth]{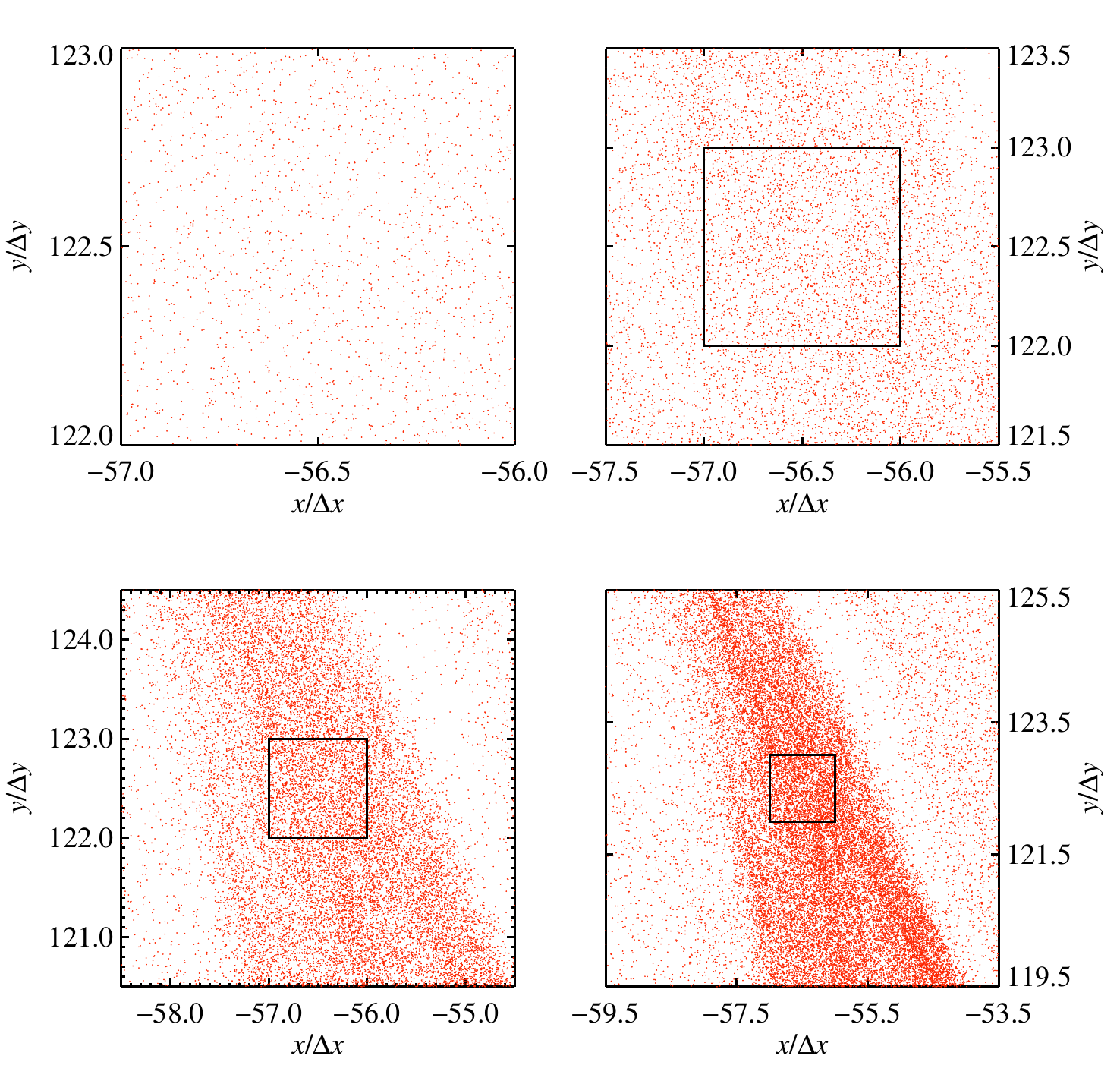}
  \end{center}
  \caption{Particles in the densest grid cell at $t=20T_{\rm orb}$ for
    the $256^3$ simulation presented in the main paper (this is before
    collisional cooling and self-gravity are turned on). The panels show a
    gradual zoom out on the surrounding grid cells (still in the same
    $z$-plane). No spurious structure is evident either at the subgrid scale or
    at the interfaces between grid cells, since the high order particle-mesh
    drag force scheme samples 27 nearby grid points when evaluating the gas
    velocity at the position of a particle. The particles are shown in velocity
    space in the left panel of Fig.\ \ref{f:vpx_vpy}.}
  \label{f:particles}
\end{figure}
\begin{figure}[!t]
  \begin{center}
    \includegraphics[width=\linewidth]{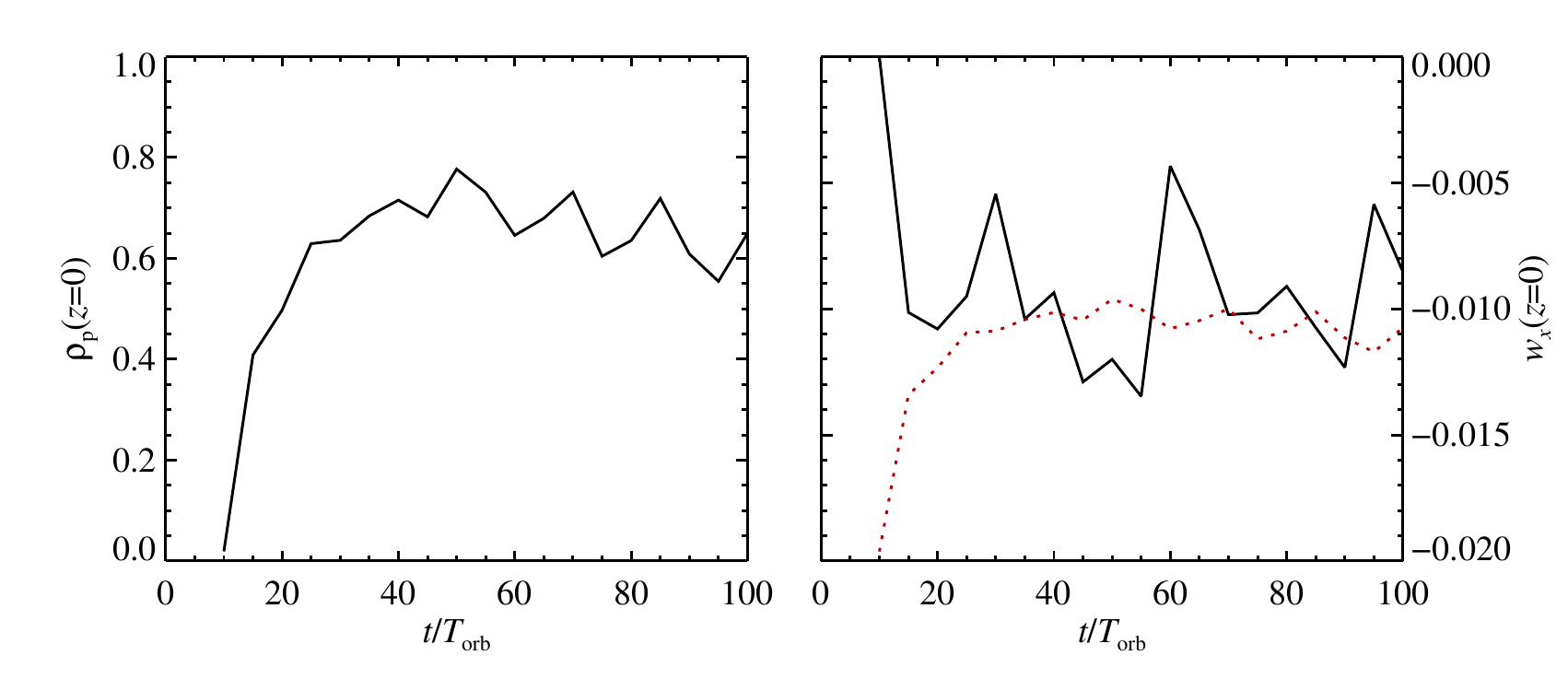}
  \end{center}
  \caption{The mean particle density in the mid-plane (left panel) for a
    simulation with single-sized $\varOmega_{\rm K} \tau_{\rm f}=1$ particles
    and feedback on the gas. The right panel shows the radial velocity $w_x$ of
    the particles in the mid-plane. Isolated particles would have $w_x=-0.02
    c_{\rm s}$, but the increased inertia in the mid-plane has decreased the
    radial drift by a factor two there. The dashed line shows the expected
    radial velocity calculated from the mid-plane density shown in the left
    plot. There is relatively good agreement, an indication that drag forces
    are calculated correctly in the code (we note that a thorough test of the
    particle-mesh drag force scheme was performed by Youdin \&
    Johansen\cite{YoudinJohansen2007}). Radial drift is a bit lower than
    expected, but that may be explained by large scale gas density fluctuations
    that modify the radial pressure gradient enough to slow the radial drift of
    marginally coupled particles\cite{JohansenKlahrHenning2006}.}
  \label{f:vpx_z}
\end{figure}

We can think of two possible reasons for the low collision speeds. The first is
insufficient numerical resolution. The numerical dissipation of the code
results in an underestimate of the amplitude of turbulent velocities at scales
smaller than around 6 grid points. At the same scales, the particle-mesh drag
force (\S\ref{s:drag_force}) begins to be
underestimated\cite{YoudinJohansen2007}. Taking the stopping length of the
boulders to be $\ell_{\rm stop}=\tau_{\rm f} v_{\rm rms}$, where $v_{\rm rms}$
potentially depends on $\tau_{\rm f}$, this yields for marginally coupled
particles a stopping length of $\ell_{\rm stop}=0.025 H$, when reduction of the
rms speed due to marginal coupling and feedback is taken into account. The
resolution is either $\delta x=0.01$, for $128^3$ grid cells, or $\delta
x=0.005$ for $256^3$ grid cells. If we assume that relative collision speed
between particles of stopping length $\ell_{\rm stop}$ is typically given by
turbulent eddies of wavelength $2\ell_{\rm stop}$, then these eddies are of
size 5-10 grid cells, depending on resolution.  Thus there is potentially a
damping of the collision speeds, as some of the scales that are important for
relative motion are just at the onset of the dissipative subrange. Table
\ref{t:vrms2} and Fig.\ \ref{f:vcoll_over_vrms} show that the ratio of
collision speeds to rms speeds does increase by 10-20\% when going from $64^3$
to $128^3$ grid points, although the increase is much more modest when feedback
is included.  An increase in grid resolution, implying both an increased
inertial range and a sharpening of the particle-mesh scheme at small scales,
will be needed to determine whether the collision speeds will continue to go up
with increasing resolution.

A second possible explanation is missing orbital dynamics in analytical
estimates.  The collision speeds predicted by Voelk et al.\cite{Voelk+etal1980}
rely on a Kolmogorov spectrum with eddy turn over times from mixing length
theory. This approach does not take into account epicyclic motion, nor the
upper limit to structure correlation times given by the Coriolis force. YL
already showed that orbital dynamics are important for particle rms speeds, but
there is no similar theory for collision speeds that takes into account orbital
dynamics. Such analytical work should be a high priority for future research
because of its importance to the dynamics and growth of boulders across the
m-sized barrier.

We note that the particle-mesh drag force scheme of the Pencil Code has been
extensively tested in the recent paper by Youdin \&
Johansen\cite{YoudinJohansen2007}. The same TSC assignment/interpolation scheme
is also used for self-gravity, for which we present a test problem in
\S\ref{s:codetest}.

We proceed with a few reality checks of particle positions in physical space
and in velocity space. While we do not claim that any of these tests are
exhaustive, they are meant to serve the purpose to exclude that spurious
particle structure at the grid scale is the source of the low collision
speeds.  We show in Fig.\ \ref{f:particles} particle positions around the
densest grid point at $t=20T_{\rm orb}$ for the $256^3$ simulation presented in
the main paper, before collisional cooling and self-gravity are turned on. No
spurious structure is evident at subgrid scales or at the interfaces between
the grid cells. In Fig.\ \ref{f:vpx_z} we show the mean particle density in the
mid-plane as a function of time (left panel), for a simulation with
single-sized $\varOmega_{\rm K} \tau_{\rm f}=1$ particles. The right panel of
Fig.\ \ref{f:vpx_z} shows the mean radial drift velocity in the mid-plane.
Particles drift slower than the expected $-0.02c_{\rm s}$ due to increased
particle inertia in the mid-plane\cite{Nakagawa+etal1986}. The dashed line
indicates the expected drift velocity given the mid-plane density shown in the
left panel. It is in relatively good agreement with the measured drift, another
indication that the code applies drag forces correctly. A slight decrease in
the drift speed may be due to large scale pressure gradients in the gas slowing
down the radial drift\cite{JohansenKlahrHenning2006}.

\begin{figure}[!t]
  \begin{center}
    \includegraphics[width=0.49\linewidth]{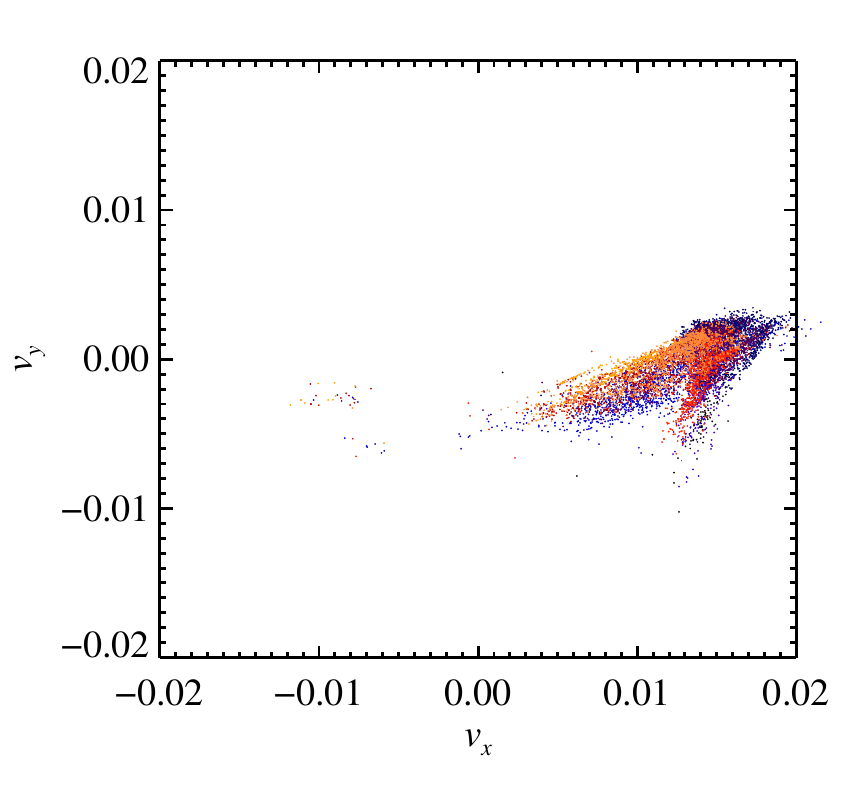}
    \includegraphics[width=0.49\linewidth]{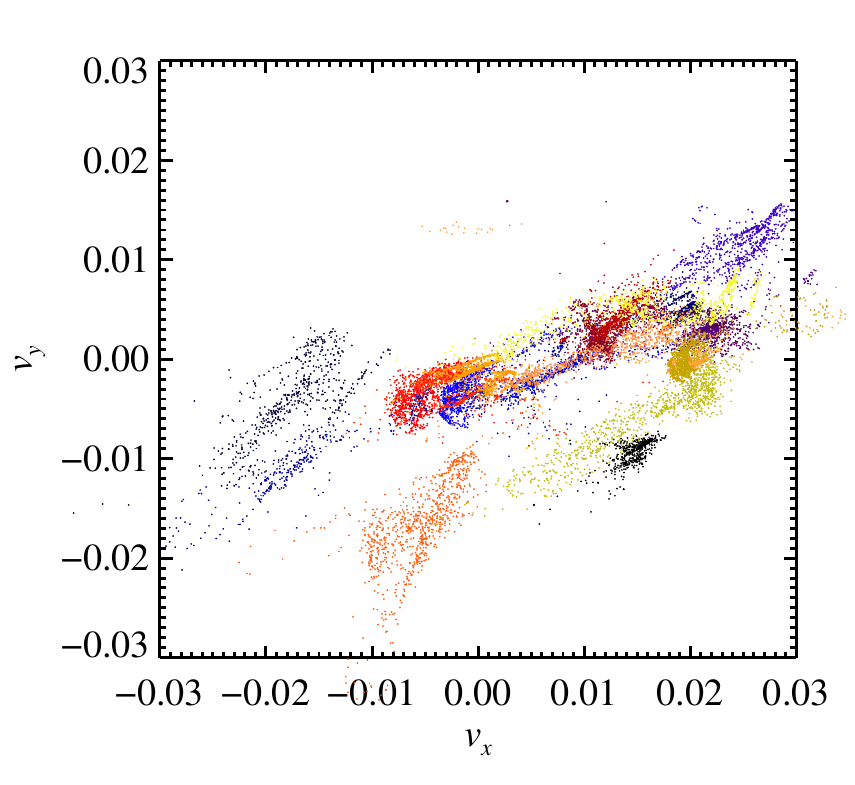}
  \end{center}
  \caption{Particles shown in velocity space $(v_x,v_y)$. The left panel shows
    the particles from Fig.\ \ref{f:particles} -- the colours refer to the 16
    individual grid points in the lower right panel of Fig.\ \ref{f:particles}.
    The right panel shows a $128^3$ simulation with single-sized
    $\varOmega_{\rm K} \tau_{\rm f}=1$ particles. Notice the higher axis range.
    Here we have chosen 20 random grid points containing between 500 and 1000
    particles. The overall spread of particle velocities is large, comparable
    to $0.02c_{\rm s}$ in each direction, whereas the spread within individual
    grid cells is much smaller. The positive correlation between the two
    velocity components arises from the positive Reynolds stress of the MRI
    turbulence.}
  \label{f:vpx_vpy}
\end{figure}

Finally we show in Fig.\ \ref{f:vpx_vpy} particle positions in velocity space
$(v_x,v_y)$. While the overall spread in velocities is large, the individual
grid cells (shown with individual colours) show a much more coherent velocity
structure. We hope to explain this difference in future publications, both with
increased numerical resolution and with improved analytical models that include
orbital dynamics on the calculation of the collision speeds.

\subsection{Low ionisation discs}

A high enough ionisation fraction is needed for the magnetorotational
instability to operate\cite{BlaesBalbus1994}.  This threshold may be reached in
hot parts of the disc and in parts where the column density is low enough that
cosmic rays
penetrate\cite{Fromang+etal2002,Semenov+etal2004,IlgnerNelson2006}.  Typically
a penetration depth of $\varSigma_{\rm g} \approx 100\,{\rm g\,cm^{-2}}$ is
found. Other sources of ionisation are decay of radioactive elements and
chemical reactions.  Small dust grains, on the other hand, soak up electrons
and reduce the conductivity\cite{Sano+etal2000}. This has lead to the concept
of a dead zone in the midplane of the disc where ionisation is insufficient for
MRI \cite{Gammie1996}. Several mechanisms have been proposed for reviving the
dead zone: growth of the peak dust grain size by an order of
magnitude\cite{Sano+etal2000}, using the turbulent kinetic energy to ionise
molecules, thus maintaining any turbulence that is already
present\cite{InutsukaSano2005}, and X-ray flares from the central star that
would ionise the disc periodically\cite{IlgnerNelson2006c}.

We investigate the limit in which the MRI completely fails to influence the
midplane by modelling the same problem as in the main text, but absent magnetic
fields.  A more advanced model would still have magnetically active surface
layers that could influence the ``dead zone''
dynamically\cite{FlemingStone2003,TurnerSanoDziourkevitch2007,OishiMacLowMenou2007}.
We do not, in general, expect density fluctuations in the gas similar to what
happens in the MRI. This is because the Mach number of Kelvin-Helmholtz
turbulence and streaming turbulence is generally smaller than for the
MRI\cite{JohansenHenningKlahr2006,JohansenYoudin2007}. This situation may
change in a more advanced model with active surface layers that can influence
the magnetically dead mid-plane, as the surface layers send density waves
through the dead zone\cite{FlemingStone2003,OishiMacLowMenou2007}.

\subsubsection{Low radial pressure support}

\begin{figure}[!t]
  \begin{center}
    \includegraphics[width=\linewidth]{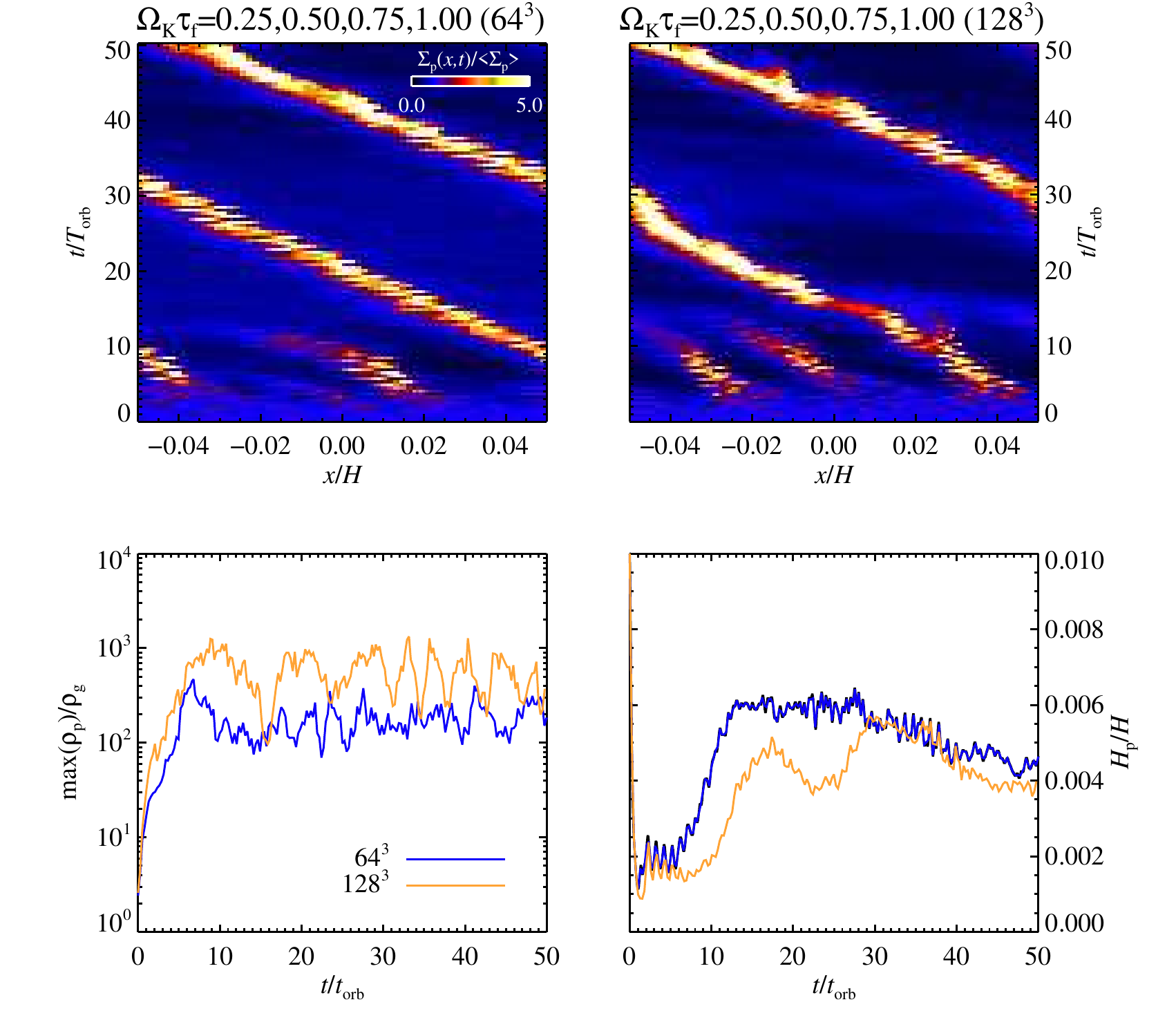}
  \end{center}
  \caption{Particle column density for $64^3$ and $128^3$ simulation with no
    magnetic fields and the usual range of boulder sizes. Particles quickly end
    up in a single azimuthally extended roll and slowly drift in together for
    at least 50 orbits. The maximum bulk density of particles reaches 1000
    times the ambient gas density.}
  \label{f:noMRI_lowdrift}
\end{figure}

We show in Fig.\ \ref{f:noMRI_lowdrift} the evolution of the sedimented
mid-plane layer for the usual range of particle sizes and a radial pressure
support of $\Delta v=-0.02 c_{\rm s}$. The box size was set to $L_x=L_y=L_z=0.1
H$, 1.32 times smaller than in the magnetic runs, in order to resolve the thin
mid-plane layer that forms in the absence of global MRI turbulence. A turbulent
state develops anyway since vertical shear in the radial and azimuthal gas flow
is unstable to the Kelvin-Helmholtz
instability\cite{GoldreichWard1973,Weidenschilling1980,WeidenschillingCuzzi1993,Cuzzi+etal1993,GomezOstriker2005,JohansenHenningKlahr2006}.
The streaming instability is also a source of non-linear dynamics and
turbulence\cite{YoudinGoodman2005,YoudinJohansen2007,JohansenYoudin2007}.

It is evident from Fig.\ \ref{f:noMRI_lowdrift} that the particles quickly
gather at a single radial location and drift radially together for the full
duration of the simulation (50 orbits, the Courant time-step in these small
boxes limits the possible simulation times compared to the magnetic runs). This
can be understood as an effect of the streaming instability. Any isolated
particle drifts quickly into the overdense band where the dynamics is so
dominated by particles that the radial drift is significantly reduced. Very
high particle densities are reached within the band, with peaks at more than
three orders of magnitude times the gas density.

The high particle densities are very susceptible to gravitational collapse. We
show in Table~\ref{t:lowdv_noMRI} that gravitational collapse is possible in
discs with as low mass as one half of the MMSN. The accretion rate of the most
massive bound cluster is generally 3 orders of magnitude lower than in the
magnetic runs, which follows the expected trend of
equation~(\ref{eq:radius_GW}) that the mass of the bound object scales with the
cube of the column density.  The solid size of the most massive bound cluster
in the $128^3$ run of Table~\ref{t:lowdv_noMRI} is 150 km, smaller than in the
runs with magnetorotational turbulence presented in the main text, but still
larger than the classical view of planetesimals.
\begin{table}[t!]
  \begin{center}
    \begin{tabular}{cccccccc}
      \hline
      \hline
      Resolution & $N_{\rm par}/10^6$ & $\Delta t_{\rm grav}$
                 & $\tilde{G}$ & $Q$
                 & $N_{\rm clusters}$ & $\dot{M}_{\rm cluster}$  \\
      (1) & (2) & (3) & (4) & (5) & (6) & (7) \\
      \hline
      \,\,\,$64^3$  & \,\,\,\,\,\,$0.125$ & $5.0$ & $0.1$ & $16.0$ & $1$ &
      $0.0050$ \\
      $128^3$ & $1.0$    & $5.0$ & $0.025$ & $64.0$ & $2$ & $0.0007$ \\
      \hline
    \end{tabular}
    \caption{Resolution study for runs with no magnetic fields and $\Delta v=-0.02 c_{\rm s}$.
      Col.\ (1): Mesh resolution.
      Col.\ (2): Number of superparticles in millions.
      Col.\ (3): Number of orbits with self-gravity.
      Col.\ (4): Minimum self-gravity parameter where gravitationally bound
                 clusters form (MMSN has $\tilde{G}\approx0.05$ at $r=5$ AU).
      Col.\ (5): Corresponding Toomre $Q\approx1.6 \tilde{G}^{-1}$.
      Col.\ (6): Number of clusters at the end of the simulation.
      Col.\ (7): Accretion rate of the most massive cluster in Ceres masses per
                 orbit.}
  \label{t:lowdv_noMRI}
  \end{center}
\end{table}

\subsubsection{Moderate radial pressure support}

\begin{figure}[!t]
  \begin{center}
    \includegraphics[width=\linewidth]{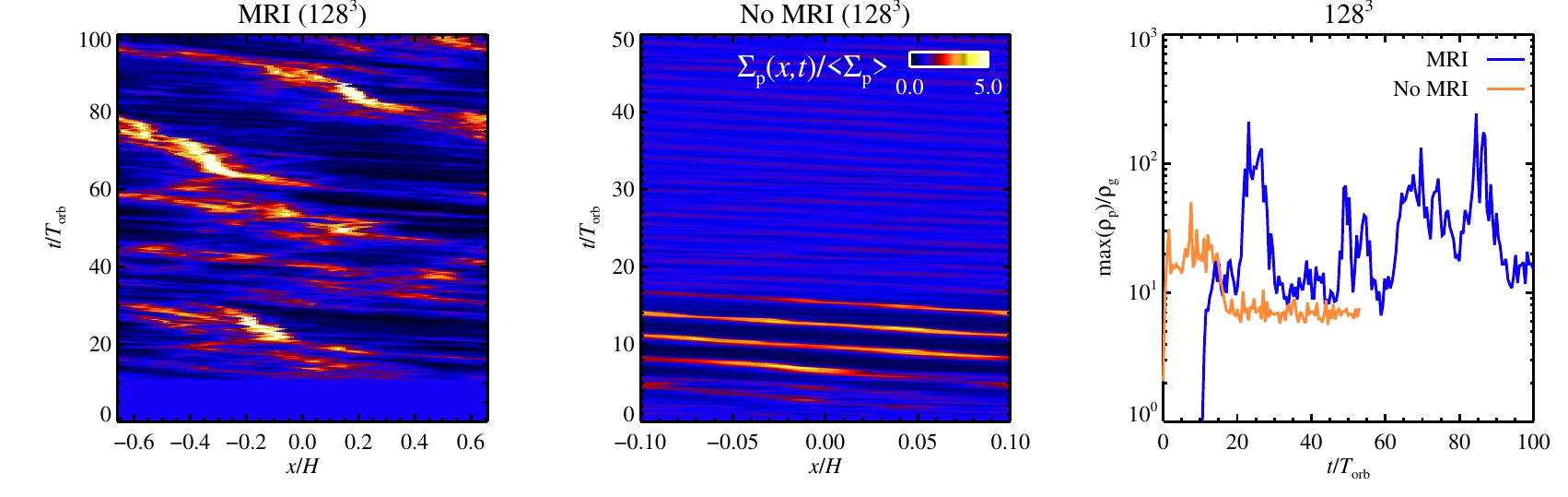}
  \end{center}
  \caption{Comparison of azimuthally averaged boulder column density for models
    with magnetic fields (left plot) and without magnetic fields (middle plot)
    for $128^3$ grid points and marginally coupled particles with
    $\varOmega_{\rm K}\tau_{\rm f}=1$. The model with no magnetic fields
    initially shows some clumping and overdensities, but these die out after 20
    orbits after which the maximum solids-to-gas ratio stays below $10$ for the
    duration of the simulation with no signs of the high density events that
    are obvious in the MRI run.}
  \label{f:Sigmapx_compare}
\end{figure}

\begin{figure}[!t]
  \begin{center}
    \includegraphics[width=\linewidth]{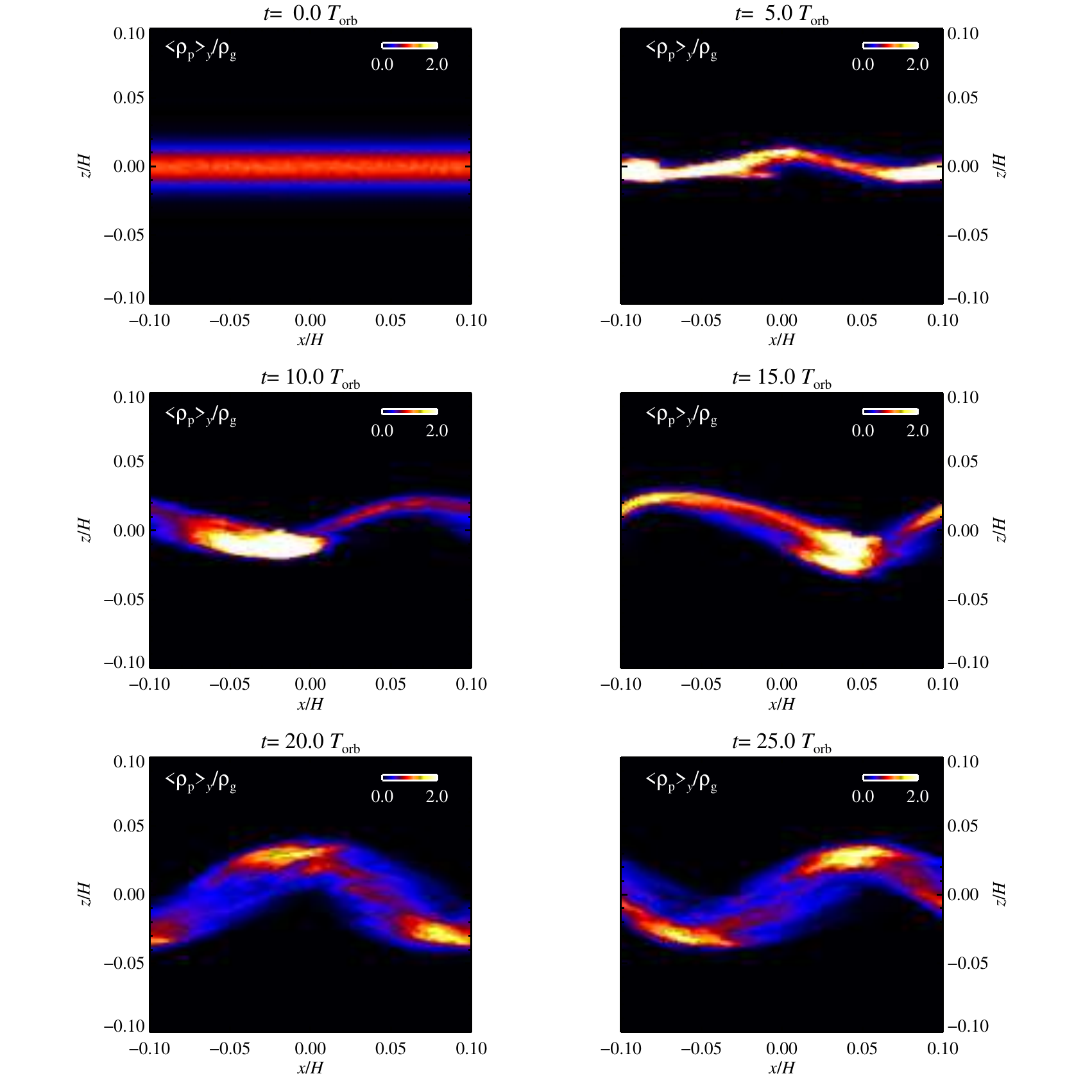}
  \end{center}
  \caption{Time evolution of the boulder density averaged over the azimuthal
    $y$-direction in a 3-D simulation with $128^3$ grid points and $10^6$
    marginally coupled particles with $\varOmega_{\rm K}\tau_{\rm f}=1$.  A
    Kelvin-Helmholtz instability develops from the initial sedimentation,
    driven by the shear in the radial drift velocity of the
    gas\cite{Nakagawa+etal1986}. The whole particle layer eventually
    participates in a single standing wave where particles drift and oscillate
    around the midplane (see main text for details).}
  \label{f:noMRI}
\end{figure}

We show in Fig.\ \ref{f:Sigmapx_compare} a comparison between two 3-D
simulations: one with magnetic fields and one without. Both have a resolution
of $128^3$ grid points, marginally coupled particles with $\varOmega_{\rm
K}\tau_{\rm f}=1$ and a moderate radial pressure support of $\Delta v=-0.05
c_{\rm s}$.  The box size is $L_x=L_y=L_z=1.32 H$ for the MRI run and
$L_x=L_y=L_z=0.2 H$ for the run with no magnetic fields. After an initial
period where strong density enhancements form, the solids-to-gas ratio never
rises above 10 in the simulation without magnetic fields. In Fig.\
\ref{f:noMRI} we show the time evolution of particle density averaged over the
azimuthal direction for the moderate pressure support non-magnetic run
presented in the middle panel of Fig.\ \ref{f:Sigmapx_compare}. The particle
density reaches a configuration where a single standing wave dominates the box,
with no significant overdensities.
This state may come about because of the streaming instability. It was observed
by Johansen \& Youdin\cite{JohansenYoudin2007} that marginal coupling produces
a very strongly turbulent state (with vertical diffusion coefficients similar
to those of the MRI). This state was nevertheless very clumpy in the
simulations of Johansen \& Youdin\cite{JohansenYoudin2007}, but the inclusion
of vertical gravity in the current work may mean that the correlation times are
so short (due to vertical oscillations) that clumping is suppressed. The
simulations by Johansen \& Youdin\cite{JohansenYoudin2007} also showed that
a higher background solids-to-gas ratio leads to a dramatic increase in
growth rate of the streaming instability, and a decrease in the unstable
wavelengths. We show in the next section (\S\ref{s:moddrift_e0.03}) that a
slight increase in the global solids-to-gas ratio indeed leads to
significant particle clumping.

We have tested the effect of self-gravity in the 3-D non-magnetic models with
moderate radial pressure support as well, with the four different particle
sizes we used in the models described in the main letter. We have not been able
to form accreting self-gravitating clusters, like those we observe in discs
with MRI turbulence, for any reasonable column density $\tilde{G}<1$ at either
$64^3$ or $128^3$ resolution.  Global magnetorotational turbulence thus appears
to overall promote rather than impede gravitational collapse of marginally
coupled boulders.

\subsubsection{Higher solids-to-gas ratio}
\label{s:moddrift_e0.03}

\begin{figure}[!t]
  \begin{center}
    \includegraphics[width=\linewidth]{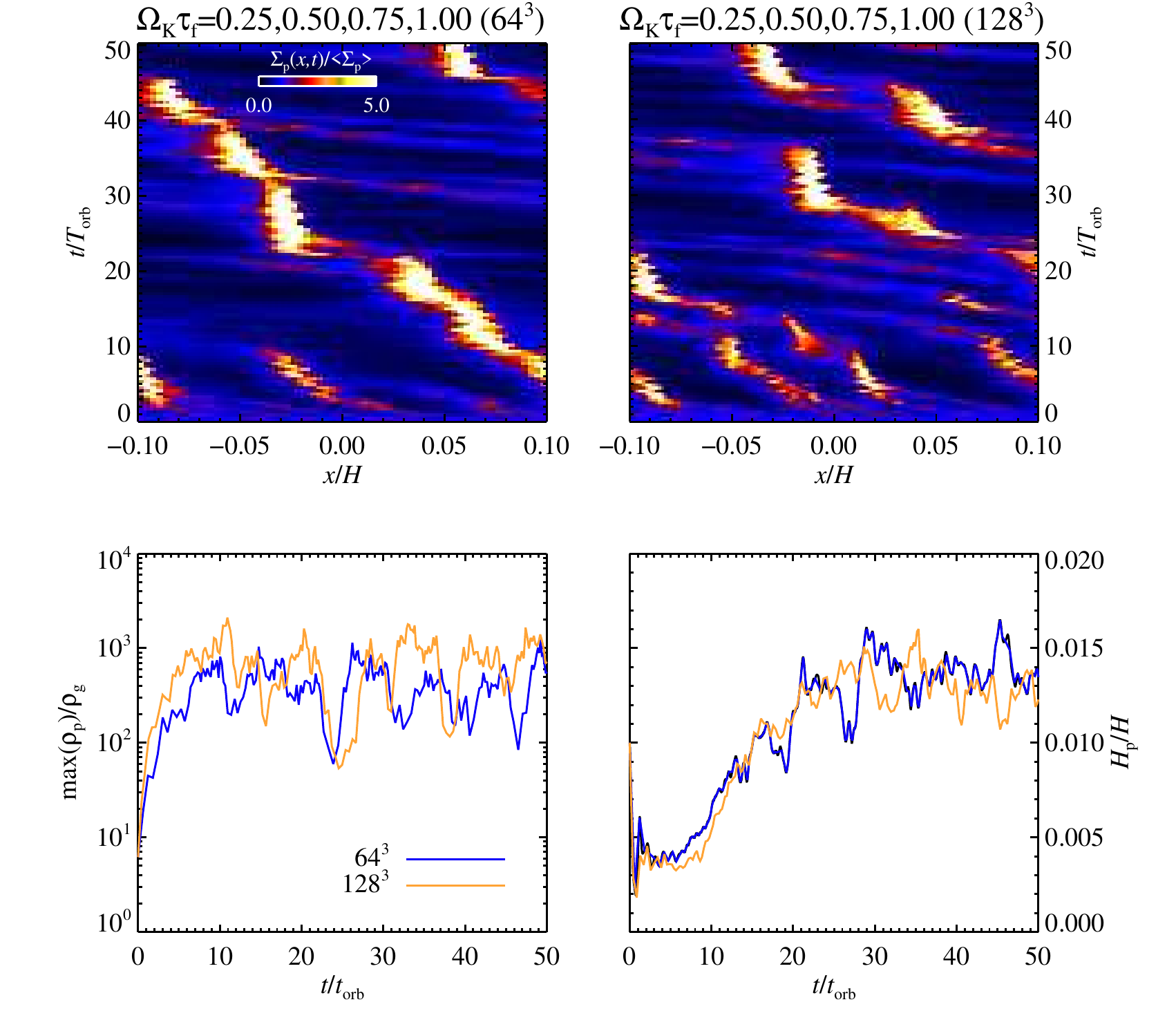}
  \end{center}
  \caption{Particle column density for a global solids-to-gas ration of $0.03$.
    The particle density peaks at more than three orders of magnitude higher
    than the gas density.}
  \label{f:noMRI_moddrift}
\end{figure}

We performed simulations with increasingly higher solids-to-gas ratio with the
expectancy that the turbulent state would get weaker and overdensities larger.
This behaviour was indeed confirmed. In Fig.\ \ref{f:noMRI_moddrift} we show
the evolution of the mid-plane layer for moderate radial pressure support and a
background solids-to-gas ratio of $0.03$. Extremely high overdensities, up to
three orders of magnitude higher than the gas, now occur.
We show in Table~\ref{t:moddv_noMRI} the gravitational collapse for the
moderate pressure support case. The accretion rate is around six times higher
than in the case of low radial pressure support (see
Table~\ref{t:lowdv_noMRI}), but that is probably due to the increased global
solids-to-gas ratio in the moderate pressure support case.

One possible way to enhance the global solids-to-gas ratio is by radial drift
augmentation\cite{StepinskiValageas1996,YoudinShu2002,YoudinChiang2004}. This
would be especially likely in the case of a magnetically active and warm outer
disc with a strong radial drift. Drifting into the dead inner regions of the
disc the boulder velocity field would converge at the rim of the dead zone,
leading to particle pileups there. Another effect, which to our knowledge has
not yet been explored, is that the accreting surface layers must lose all their
solids to the dead zone. Considering an active layer with a width of one gas
scale height $H$ and an orbital distance $r$, the probability for a small dust
grains to accrete all the way to the star without diffusing into the dead zone
is only $(H/r)^2 \ll 1$. Once a grain is lost into the dead zone it is very
unlikely that it will diffuse back into the active layers because the turbulent
diffusion in the dead zone is assumed to be very low. A third possibility for
augmenting the solids-to-gas ratio is by photoevaporation of the gaseous part
of the disc\cite{ThroopBally2005}
\begin{table}[t!]
  \begin{center}
    \begin{tabular}{cccccccc}
      \hline
      \hline
      Resolution & $N_{\rm par}/10^6$ & $\Delta t_{\rm grav}$
                 & $\tilde{G}$ & $Q$
                 & $N_{\rm clusters}$ & $\dot{M}_{\rm cluster}$  \\
      (1) & (2) & (3) & (4) & (5) & (6) & (7) \\
      \hline
      \,\,\,$64^3$ & \,\,\,\,\,\,$0.125$ & $5.0$ & $0.05$  & $32.0$ & $1$ &
            $0.092$ \\
           $128^3$ &             $1.0$   & $5.0$ & $0.025$ & $64.0$ & $3$ &
           $0.004$ \\
      \hline
    \end{tabular}
    \caption{Resolution study for runs with no magnetic fields, $\Delta
    v=-0.05 c_{\rm s}$ and $\epsilon_0=0.03$.
      Col.\ (1): Mesh resolution.
      Col.\ (2): Number of superparticles in millions.
      Col.\ (3): Number of orbits with self-gravity.
      Col.\ (4): Minimum self-gravity parameter where gravitationally bound
                 clusters form (MMSN has $\tilde{G}\approx0.05$ at $r=5$ AU).
      Col.\ (5): Corresponding Toomre $Q\approx1.6 \tilde{G}^{-1}$.
      Col.\ (6): Number of clusters at the end of the simulation.
      Col.\ (7): Accretion rate of the most massive cluster in Ceres masses per
                 orbit.}
  \label{t:moddv_noMRI}
  \end{center}
\end{table}

\subsection{Heating and cooling}\label{s:heating}

Drag and collisions dissipate the kinetic energy of the solids. In the extreme
case, all this heat is released locally in the gas.  We consider here the case
of ineffective radiative cooling and an ideal gas equation of state, rather
than the isothermal equation of state assumed in the rest of the paper. The
only means to transport energy in this model is by turbulent heat conduction.
In reality collisional cooling may transfer most of the energy into deformation
of the colliding bodies, but we consider here the effect of the most extreme
case of local heating on gravitational collapse.

We have ignored viscous heating under the assumption that the gas background
has found an equilibrium between radiative loses and viscous heating. The
extra gas shear and compression in the collapsing particle clusters is
negligible due to the pressure support of the gas, so any extra viscous
heating in regions of collapse is not important. We have found no traces of the
gravitationally contracting clusters in the gas density or in the gas velocity
field (gas is allowed to exert and feel self-gravity in the simulations that
include heating). 

\begin{figure}[!t]
  \begin{center}
    \includegraphics[width=0.49\linewidth]{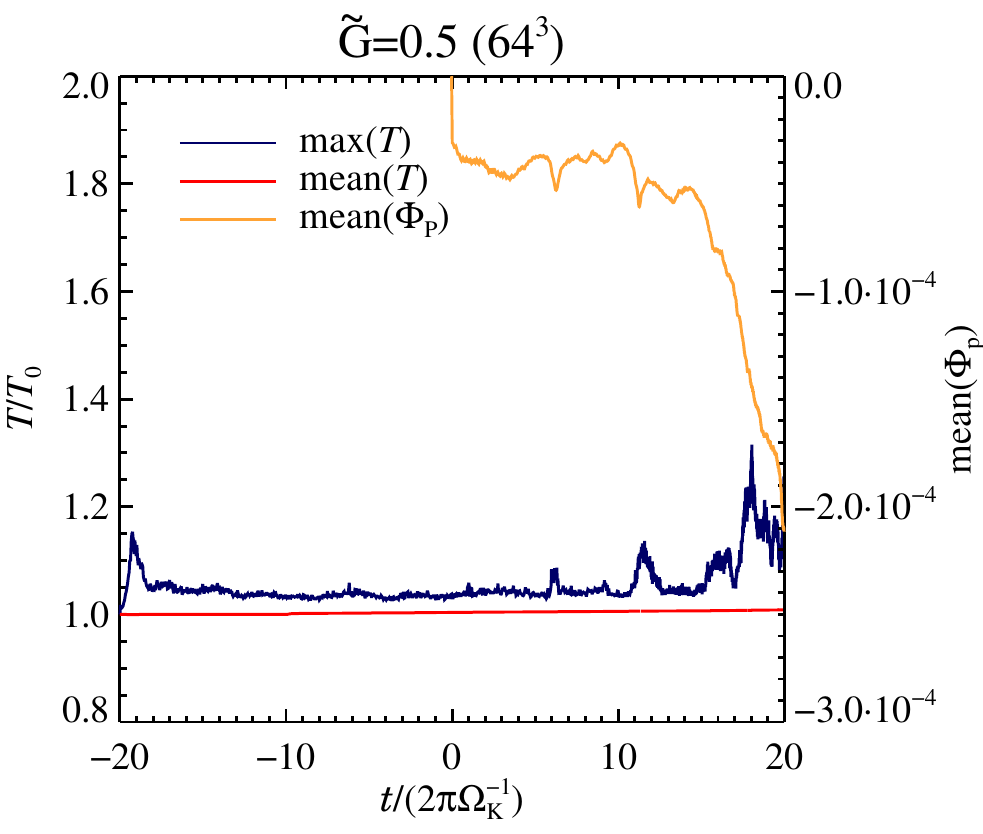}
    \includegraphics[width=0.49\linewidth]{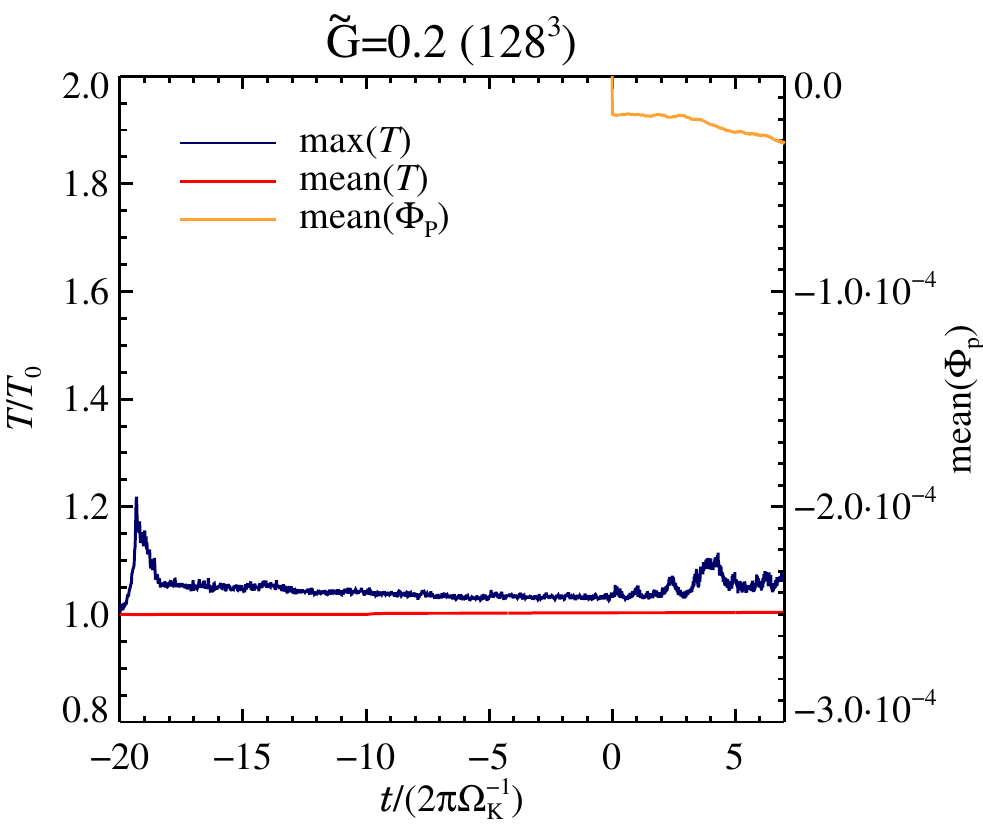}
  \end{center}
  \caption{The maximum and mean temperature of the gas as a function of time in
    our adiabatic model, for models with $64^3$ zones and $\tilde{G} = 0.5$
    (left), and $128^3$ zones and $\tilde{G} = 0.2$. Self-gravity is turned on
    at $t=0$; the first collapse, seen in the fall in the mean potential of the
    particles, begins after approximately 10 orbits for the $64^3$ run. Peaks
    in the maximum temperature coincide well with times where gravitational
    energy is released (the peak at $t=-20$ is due to compression in the
    initial random velocity field). The energy is transferred to the gas by
    drag and inelastic collisions.  The mean temperature rises negligibly
    because gravitational collapse occurs in only a small fraction of the total
    gas volume.}
  \label{f:TTmax_t}
\end{figure}
We show in Fig.\ \ref{f:TTmax_t} the gas temperature as a function of time for
a run with $64^3$ grid points and $\tilde{G}=0.5$ (a 25\% higher column density
was needed for collapse in this case, which we attribute simply to the
stochastic nature of the collapse) and for a run with $128^3$ and the usual
$\tilde{G}=0.2$.  Self-gravity is turned on at a time $t=0$. We let gas exert
and feel self-gravity in the simulations with heating, although this causes no
real difference in the behaviour.  This effect is otherwise not included in the
simulations (see \S\ref{s:disc_mass}). The maximum temperature rises to no more
than 30\% over the initial value at times of gravitational collapse (seen in
the mean gravitational potential energy of the particles). Peaks in the maximum
temperature coincide well with gravitational energy release events. The peak at
$t=-20$ happens due to compression in the initially random velocity field.  

\begin{figure}[!t]
  \begin{center}
    \includegraphics[width=0.49\linewidth]{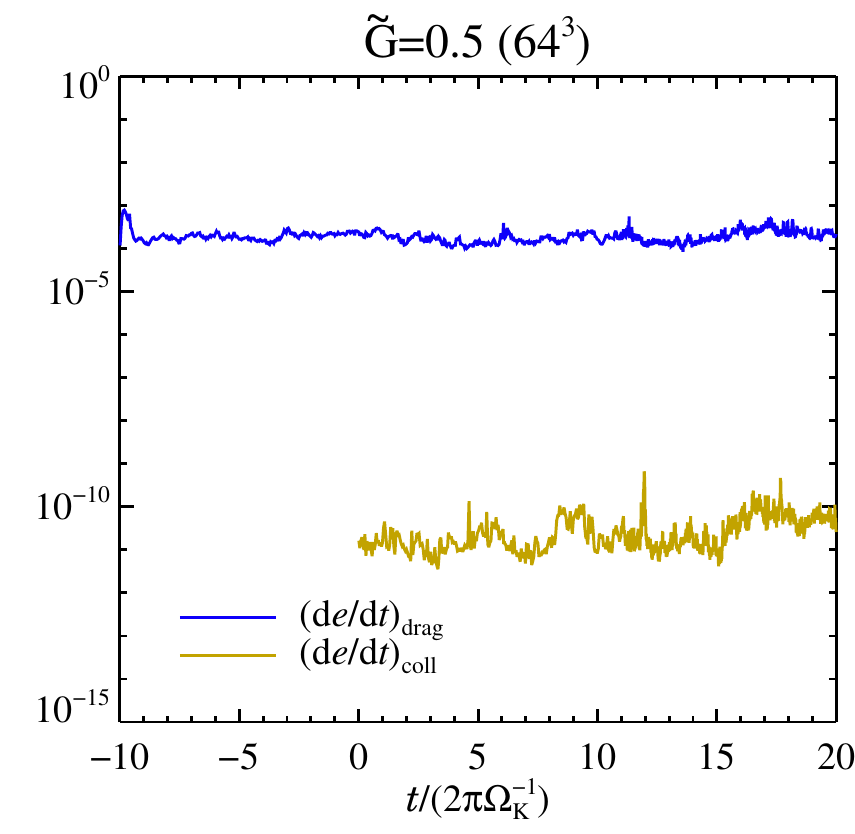}
    \includegraphics[width=0.49\linewidth]{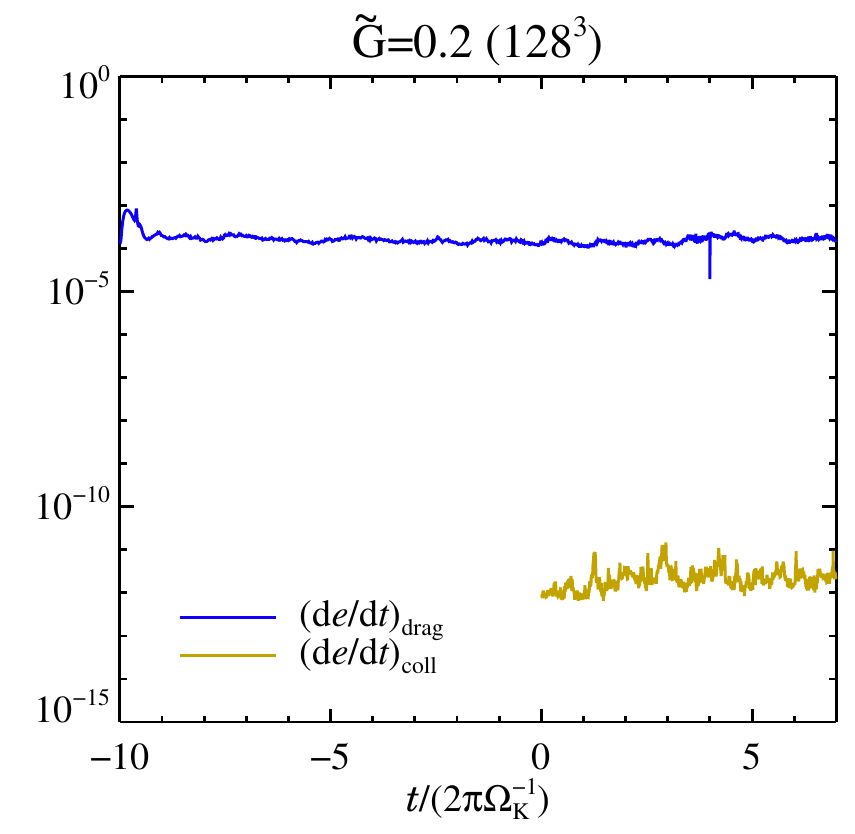}
  \end{center}
  \caption{Volume-averaged heating rate by drag (dark/blue) and by
    inelastic collisions (bright/khaki) in the same runs shown in Fig.\
    \ref{f:TTmax_t}. Drag force dissipates six
    orders of magnitude more kinetic energy than collisions do, although
    collisional dissipation dominates in regions where the bulk density of
    solids is at least two orders of magnitude higher than the gas (see
    \S\ref{s:collisional_cooling}).}
  \label{f:dedragp_decollp_t}
\end{figure}

We show in Fig.\ \ref{f:dedragp_decollp_t} the dissipation rate of drag and
collisions. Drag dominates the volume-averaged energy budget, with a heating
rate at least six orders of magnitude higher, because collisions are only
important in the few grid points where the solids-to-gas ratio greatly exceeds
unity.
Assuming a temperature of around $80\,{\rm K}$ in the non-heated state, the
temperature reaches a maximum of approximately $104\,{\rm K}$ in the $64^3$
simulation, still too low to even melt the icy component of the boulders.

\subsection{Varying radial pressure support}\label{s:vary_drift}

Cuzzi et al.\cite{Cuzzi+etal1993} gives a detailed review of the values
expected for the radial pressure support parameter.  We have assumed throughout
this work a radial pressure support of either $(\dpa \ln P/\dpa \ln r)
(H/r)=-0.04$ or $(\dpa \ln P/\dpa \ln r) (H/r)=-0.1$ (see
\S\ref{s:radial_drift}). The value of $H/r$ directly yields the local
temperature when folded with the mass of the central object and the orbital
radius in the disc. We have run test simulations of colder discs as well, with
$H/r=0.02$, corresponding to a four times lower temperature at a given location
than models with $H/r=0.04$.
\begin{figure}[!t]
  \begin{center}
    \includegraphics[width=0.5\linewidth]{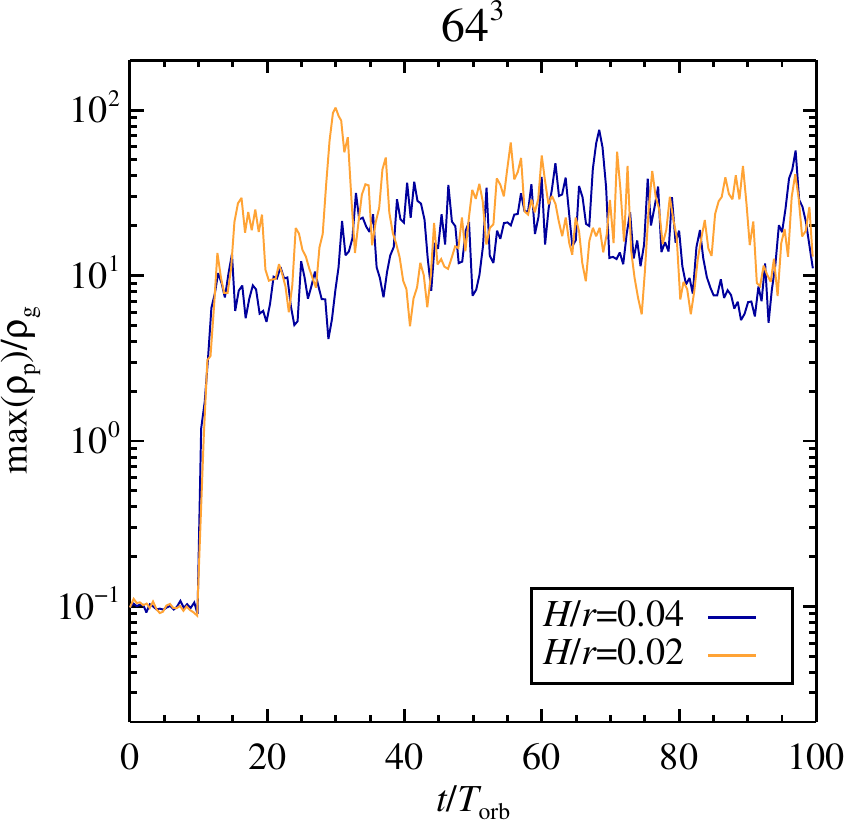}
  \end{center}
  \caption{Maximum particle density versus time for standard run with
    $H/r=0.04$ (blue) and a colder disc with $H/r=0.02$ (yellow).
    The curves appear statistically indistinguishable. A lower disc aspect ratio
    decreases the radial drift of the solids, while making the unstable wave
    lengths of the streaming instability smaller, but apparently neither of
    these effects significantly changes the particle overdensities in
    the considered cases.}
  \label{f:rhopmax_t_Hr_compare}
\end{figure}
In Fig.\ \ref{f:rhopmax_t_Hr_compare} we compare the evolution of the maximum
particle density between the standard run with $H/r=0.04$  and the cold run
with $H/r=0.02$. The two curves are statistically indistinguishable.   Lowering
the disc aspect ratio reduces radial drift and  decreases the unstable
wavelengths of the streaming instability, but this apparently does not change
the evolution of the solids, at least for moderate changes in  temperature.
\begin{figure}[!t]
  \begin{center}
    \includegraphics[type=pdf,ext=.pdf,read=.pdf,width=\linewidth]{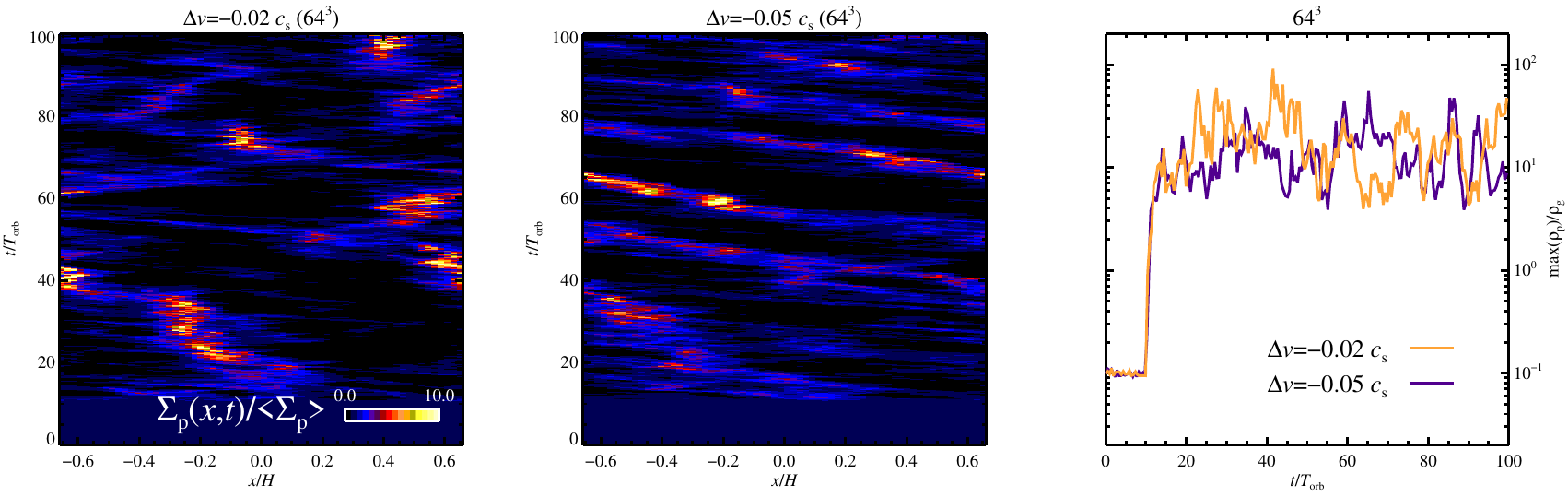}
  \end{center}
  \caption{Comparison of the topography of the particle layer for the standard
    run with $\Delta v=-0.02 c_{\rm s}$ (left) with a run with 2.5 times
    stronger radial pressure support $\Delta v=-0.05 c_{\rm s}$ (middle), both
    at $64^3$ grid points and 125,000 particles.  The increased radial drift is
    clear in the tilted bands, but the streaming instability still produces
    strong overdensities in the particle layer. The right plot shows the
    maximum particle density as a function of time -- the peaks in the particle
    density are reduced by around 30\% in the case of the stronger radial
    pressure support.}
  \label{f:Hr0.1_compare}
\end{figure}

\begin{table}[t!]
  \begin{center}
    \begin{tabular}{cccccccccc}
      \hline
      \hline
      Resolution & $N_{\rm par}/10^6$ & $\Delta t_{\rm grav}$
                 & $\alpha$ & $\tilde{G}$ & $Q$
                 & $N_{\rm clusters}$ & $\dot{M}_{\rm cluster}$  \\
      (1) & (2) & (3) & (4) & (5) & (6) & (7) & (8) \\
      \hline
      \,\,\,$64^3$  & \,\,\,\,\,\,$0.125$ & $10.0$ & $0.002$
          & $0.7$ & $2.3$ & $1$ & $2.7$ \\
      $128^3$ & $1.0$    & $7.0$ & $0.001$
          & $0.4$ & $4.0$ & $4$ & $2.6$ \\
      \hline
    \end{tabular}
    \caption{Resolution study for $\Delta v=-0.05 c_{\rm s}$.
      Col.\ (1): Mesh resolution.
      Col.\ (2): Number of superparticles in millions.
      Col.\ (3): Number of orbits with self-gravity.
      Col.\ (4): Measured turbulent viscosity.
      Col.\ (5): Minimum self-gravity parameter where gravitationally bound
                 clusters form (MMSN has $\tilde{G}\approx0.05$ at $r=5$ AU).
      Col.\ (6): Corresponding Toomre $Q\approx1.6 \tilde{G}^{-1}$.
      Col.\ (7): Number of clusters at the end of the simulation.
      Col.\ (8): Accretion rate of the most massive cluster in Ceres masses per
                 orbit.
\label{t:highdv}}
  \end{center}
\end{table}
A higher radial pressure support can occur for a given value of $H/r$ if the
radial pressure support $|\dpa \ln P/\dpa \ln r|$ exceeds unity. We next
examine the consequences of a higher radial pressure support on the maximum
density of solids. We show in Fig.\ \ref{f:Hr0.1_compare} the topography of the
sedimented particle layer and the maximum particle density of two $64^3$
simulations: the standard run with $(\dpa \ln P/\dpa \ln r) H/r = -0.04$ and a
run with higher radial pressure support and $(\dpa \ln P/\dpa \ln r) H/r =
-0.1$. The same comparison at $128^3$ is shown in the main text. The increased
radial pressure support, $\Delta v=-0.05 c_{\rm s}$, is very clear in the
latter case from the tilted particle bands, but the streaming instability still
produces radial overdensities.  We have checked the column density required for
gravitational collapse in models with $\Delta v=-0.05 c_{\rm s}$ and found that
approximately a factor two higher value of the column density of gas is needed
for the collapse (see Table~\ref{t:highdv}) compared to the standard model with
$\Delta v =-0.02 c_{\rm s}$.

\subsection{Other processes}

In order to isolate the effect of self-gravity on the dynamics of the solids we
have ignored two potentially important collisional effects: coagulation and
collisional fragmentation. In this section we estimate time-scales and effects
of these physical phenomena.

The time-scale of collisions is
\begin{equation}
  \tau_{\rm coll} =
    \frac{\tau_{\rm f}}{(c_{\rm p}/c_{\rm s})(\rho_{\rm p}/\rho_{\rm g})} \, ,
\end{equation}
where $c_{\rm p}$ is the collision speed of the boulders and $\tau_{\rm f}$ is
the friction time. Assuming spherical boulders gives the relation
\begin{equation}
  \frac{a_\bullet}{\dot{a}_\bullet} = 3 \frac{m_\bullet}{\dot{m}_\bullet}
\end{equation}
between the time-scale for radius doubling and the time-scale for mass
doubling, which in turn is just the collisional time-scale. We consider regions
with solid-to-gas ratios characteristic of the average midplane in a turbulent
disc, the peak density in the turbulent midplane, and an intermediate density
in a gravitationally contracting cluster, as described in Table~\ref{t:colls}.
The same table gives the time-scales for collisional fragmentation and for
coagulation for these different regions. 

The average state has a coagulation time of close to 100 orbits, comparable to
the time-scale for radial drift in. To cross the drift barrier effectively a few
coagulation times are needed, so the growing solids are lost into the inner
disc before they can decouple from the gas. A more advanced model would take
into account the sweep up of small grains as the boulders drift, but
essentially the same result is reached\cite{KlahrBodenheimer2006}.
\begin{table}[t!]
  \begin{center}
    \begin{tabular}{l|rrr}
      State & $\langle\rho_{\rm p}\rangle/\langle\rho_{\rm g}\rangle$
            & $\tau_{\rm frag}/T_{\rm orb}$ & $\tau_{\rm coag}/T_{\rm orb}$ \\
      \hline
      Average midplane &    $0.5$ & 30     & 90\\
      Peak midplane    &  $100.0$ &  0.15  &  0.45\\
      Contracting cluster & $1000.0$ &  0.015 &  0.045
    \end{tabular}
  \end{center}
  \caption{Estimated time-scales, in units of orbits, for collisional
    fragmentation and for coagulation in three different states of the
    mid-plane layer.}
  \label{t:colls}
\end{table}

The size of the boulders could increase on the time-scale of only one orbit in
the overdense particle clusters that occur in transient high pressure regions
and due to the streaming instability. Coagulation could go even faster in the
gravitationally contracting regions. This may be how planetesimals eventually
form: by coagulation in an overdense environment set by the self-gravity of the
solids. The strength of the scenario presented in the main text, though, is
that it does not rely on coagulation being efficient in same way as pure
coagulation models do\cite{Weidenschilling1997,DullemondDominik2005}.
Coagulation is needed to build up marginally coupled solids---metre-sized
boulders at 5 AU or centimetre-sized pebbles at 40 AU---but does not directly
drive gravitational collapse.

Collisional fragmentation may also be prevalent in  gravitationally contracting
clusters. In that case one can speculate that the mode of radius growth is by
sweeping up of collisional fragments by the few lucky boulders that avoid
catastrophic collisions with equal-sized bodies.

\end{document}